\documentclass[aps,prd,reprint,preprintnumbers,superscriptaddress,amsmath,amssymb,nofootinbib]{revtex4-2}
\pdfoutput=1
\usepackage{graphicx,orcidlink}
\usepackage{enumitem}
\usepackage[dvipsnames]{xcolor}
\usepackage{hyperref}
\usepackage{scalefnt}
\usepackage[capitalize]{cleveref}
\usepackage{multirow,subcaption,cancel}
\usepackage{acronym}
\usepackage{slashed}
\hypersetup{colorlinks=true,linkcolor=blue,citecolor=blue,urlcolor=blue}
\graphicspath{{../}{./}}
\def\gev{\,\text{Ge\hspace{-0.1em}V}}
\def\mev{\,\text{Me\hspace{-0.1em}V}}

\newcommand{\abbrev}[1]{{\scalefont{.9}#1}}
\newcommand{\eqindentspace}{em}
\newcommand{\eqindent}[1]{\hspace{#1\eqindentspace}}
\newcommand{\MSbar}{\ensuremath{\overline{\abbrev{\text{MS}}}}}

\newcommand{\ren}{\text{\abbrev{R}}}
\newcommand{\ep}{\epsilon}
\newcommand{\alphas}{\alpha_\text{s}}
\newcommand{\api}{a_\text{s}}

\newcommand{\lmut}{L_{\mu \tau}}

\newcommand{\ctr}{T_\text{R}}
\newcommand{\nc}{N_\text{c}}

\newcommand{\nf}{n_\text{f}}

\newcommand{\wrt}{w.r.t.}
\newcommand{\order}[1]{\ensuremath{{\cal O}(#1)}}
\newcommand{\EulerGamma}{\gamma_\text{E}}

\newcommand{\FVE}{\abbrev{FVE}}
\newcommand{\QCD}{\abbrev{QCD}}
\newcommand{\LO}{\abbrev{LO}}
\newcommand{\NLO}{\abbrev{NLO}}
\newcommand{\NNLO}{\abbrev{NNLO}}
\newcommand{\OPE}{\abbrev{OPE}}
\newcommand{\SM}{\abbrev{SM}}
\newcommand{\BSM}{\abbrev{BSM}}
\newcommand{\SFTX}{\abbrev{SFTX}}
\newcommand{\GFSFTX}{\abbrev{GF+SFTX}}
\newcommand{\GF}{\abbrev{GF}}
\newcommand{\HQE}{\abbrev{HQE}}
\newcommand{\HQET}{\abbrev{HQET}}
\newcommand{\VIA}{\abbrev{VIA}}
\newcommand{\RG}{\abbrev{RG}}
\newcommand{\RGE}{\abbrev{RGE}}
\newcommand{\UV}{\abbrev{UV}}

\begin{document}

\preprint{SI-HEP-2026-08,~~P3H-26-024,~~ZU-TH 14/26,~~TTK-26-05}

\title{Heavy-Meson Bag Parameters using Gradient Flow} 

\author{Matthew Black\orcidlink{0000-0002-8952-1755}}\email{matthew.black@ed.ac.uk}
\affiliation{Higgs Centre for Theoretical Physics, School of Physics and Astronomy, University of Edinburgh, Edinburgh EH9 3JZ, United Kingdom}
\affiliation{Theoretische Physik 1, Center for Particle Physics Siegen, Naturwissenschaftlich-Technische Fakult\"at, Universit\"at Siegen, 57068 Siegen, Germany}

\author{Robert V.~Harlander\orcidlink{0000-0002-8715-2458}}
\affiliation{Institute for Theoretical Particle Physics and Cosmology, RWTH Aachen University, 52056 Aachen, Germany}

\author{Jonas T. Kohnen\orcidlink{0009-0001-2904-5945}}
\affiliation{Institute for Theoretical Particle Physics and Cosmology, RWTH Aachen University, 52056 Aachen, Germany}

\author{\\Fabian Lange\orcidlink{0000-0001-8531-5148}}
\affiliation{Physik-Institut, Universität Zürich, Winterthurerstrasse 190, 8057 Zürich, Switzerland}
\affiliation{PSI Center for Neutron and Muon Sciences, 5232 Villigen PSI, Switzerland}

\author{Antonio Rago\orcidlink{0000-0003-1192-5538}}
\affiliation{IMADA and Quantum Theory Center, University of Southern Denmark, Odense, Denmark}

\author{Andrea Shindler\orcidlink{0000-0003-3693-8300}}
\affiliation{Institute for Theoretical Particle Physics and Cosmology, RWTH Aachen University, 52056 Aachen, Germany}
\affiliation{Nuclear Science Division, Lawrence Berkeley National Laboratory, Berkeley, California 94720, USA}
\affiliation{Department of Physics, University of California, Berkeley, CA 94720, USA}

\author{Oliver Witzel\orcidlink{0000-0003-2627-3763}}
\affiliation{Theoretische Physik 1, Center for Particle Physics Siegen, Naturwissenschaftlich-Technische Fakult\"at, and Cluster of Excellence ``Color meets Flavor'', Universit\"at Siegen, 57068 Siegen, Germany}

\date{\today}

\begin{abstract}
We demonstrate the use of the
gradient flow combined with the short flow-time expansion
(\GFSFTX{}) as a renormalization procedure
for four-quark operator matrix
elements and associated bag parameters relevant to neutral heavy-meson mixing
($\Delta Q=2$) and heavy-meson lifetimes ($\Delta Q=0$).  Using six RBC/UKQCD
2+1-flavor domain-wall fermion ensembles, we calculate for a charm-strange system with
physical quark masses 
flowed bag parameters and match them to the
\MSbar\ scheme using perturbative \SFTX{} coefficients up to
next-to-next-to-leading order in \QCD.  We employ a multiscale matching
strategy and a renormalization-group improved flow-time evolution which allows
for a reliable estimate of systematic uncertainties.  For a fictitious neutral
$D_s$ meson, we obtain the $\Delta Q=2$ \MSbar\ bag parameter ${\cal
B}^\text{\MSbar}_1(3\gev)=0.7673(123)$, consistent with
existing short-distance $D^0$ mixing determinations.  For the $\Delta Q=0$
lifetime-ratio operator basis, we find the \MSbar\ results
$B^\text{\MSbar}_1(3\gev)=1.0524(97)$, $B^\text{\MSbar}_2(3\gev)=0.9621(71)$,
$\epsilon^\text{\MSbar}_1(3\gev)=-0.2275(76)$, and
$\epsilon^\text{\MSbar}_2(3\gev)=-0.0005(8)$. We provide conversion formulae 
to reexpress these results for an arbitrary choice of evanescent operators.
These results demonstrate that \GFSFTX{} can
deliver precise
determinations of dimension-six four-quark operators and
establish a framework for future lattice computations including more complex
operator bases, where the challenge of power-divergent mixing is shifted to the
continuum and handled in the \SFTX.
\end{abstract}
\maketitle

\section{Introduction}

The physics of heavy hadrons involves short- and long-distance contributions from \QCD. Using effective field theories such as the heavy-quark expansion (\HQE), they can be factorized into perturbative Wilson coefficients and matrix elements of composite operators which need to be evaluated using nonperturbative methods. In particular, the mixing and decay processes of heavy mesons, which are at the focus of this paper, are mediated by four-quark operators of mass-dimension six, carrying a net heavy-quark flavor quantum number of $\Delta Q=2$ and $\Delta Q=0$, respectively. The relevant hadronic matrix elements are typically parametrized by so-called bag parameters. Their combination with the perturbative Wilson coefficients requires a common renormalization scheme, which presents one of the main obstacles in the theoretical description of these processes.

The calculation of bag parameters for neutral meson mixing is rather well established using numerical lattice \QCD\ simulations~\cite{Gamiz:2009ku,BMW:2011zrh,SWME:2015oos,RBC:2014ntl,Aoki:2014nga,Boyle:2024gge,ETM:2012vvy,Carrasco:2013zta,Carrasco:2014uya,Carrasco:2015pra,Laiho:2011np,Bazavov:2017weg,Bazavov:2016nty,Dowdall:2019bea}.
Both perturbative~\cite{VandeWater:2005uq,Bailey:2012wb} and nonperturbative~\cite{Martinelli:1994ty,Sturm:2009kb,RBC:2014ntl,Boyle:2017skn} renormalization procedures have been used in the past.
While percent-level precision~\cite{FLAG:2024oxs} has been reached for the Standard Model (\SM) 
operator describing neutral kaon-mixing, results for beyond Standard Model
(\BSM) operators are in tension between the \abbrev{ETM} Collaboration (\abbrev{ETMC})~\cite{Carrasco:2015pra} and the determinations by the \abbrev{SWME} \cite{SWME:2015oos} and
\abbrev{RBC/UKQCD}~\cite{Garron:2016mva,Boyle:2024gge} Collaborations.
\abbrev{SWME} uses a perturbative renormalization scheme, whereas both
\abbrev{ETMC} and \abbrev{RBC/UKQCD} perform a nonperturbative renormalization
requiring gauge fixing, referred to as regularization-independent
momentum-subtraction (\abbrev{RI-MOM}). It has been suggested that
the difference between \abbrev{ETMC} and \abbrev{RBC/UKQCD} originates from
differently chosen kinematics in their \abbrev{RI-MOM} schemes
\cite{Boyle:2017skn}.
Currently most precise results for the \SM\ mass-difference operator by the
Fermilab/\abbrev{MILC}~\cite{Bazavov:2016nty} and
\abbrev{HPQCD}~\cite{Dowdall:2019bea} Collaborations exhibit some tension as well; they both
rely on perturbative renormalization using the \MSbar{}-\abbrev{NDR}
scheme.\footnote{Joint work by \abbrev{RBC/UKQCD} and \abbrev{JLQCD} using
\abbrev{RI-(S)MOM} renormalization is ongoing
\cite{Boyle:2021kqn}.} An independent theoretical approach to meson mixing would
certainly help to clarify this situation. It is the aim of this paper to provide
such an approach by considering the short-distance contribution to $D$ meson
mixing, allowing us to compare to few-percent-level results which have been
obtained by \abbrev{ETMC} \cite{Carrasco:2014uya,Carrasco:2015pra} and
Fermilab/\abbrev{MILC} \cite{Bazavov:2017weg}.

Concerning the theoretical description of heavy-meson decays, one faces two major additional challenges. First, so-called ``eye diagrams'' occur whose lattice calculation is known to be numerically expensive. Subsequently, the renormalization of the $\Delta Q=0$ operators introduces a mixing with operators of lower mass dimension, which on the lattice implies power divergences in $1/a$. While different strategies to calculate eye diagrams have been already pursued~\cite{Giusti:2019kff,RBC:2022ddw}, the mixing of operators with lower mass dimensions remains one of the outstanding challenges. Recently, a gauge-invariant renormalization scheme has been proposed~\cite{Spanoudes:2025ryj}, and for the special case of four-quark operators using the static heavy-quark approximation an approach using position-space renormalization has been suggested~\cite{Lin:2024mws}.
Nevertheless, to date no lattice-\QCD{} determination of the corresponding bag parameters with full error budget exists, despite some early works~\cite{DiPierro:1998ty,DiPierro:1999tb,Becirevic:2001fy} as well as preliminary work~\cite{Lin:2022fun} in the static limit.

To a good approximation, both the issue of the eye diagrams as well as the mixing with lower-dimensional operators can be eliminated by considering lifetime ratios rather than absolute lifetimes. However, even in this case, no determination of the bag parameters is available whose accuracy would be at the level of current experimental precision.

We evaluate bag parameters with the help of the gradient
flow~\cite{Narayanan:2006rf,Luscher:2009eq,Luscher:2010iy,Luscher:2013cpa} in
combination with the short flow-time expansion
(\SFTX)~\cite{Luscher:2011bx,Suzuki:2013gza,Luscher:2013vga}.\footnote{
Preliminary work towards a similar application of this topic was presented by the \abbrev{WHOT QCD} collaboration at Lattice~2019~\cite{Taniguchi:2019talk}.}
The \GF{} provides a
gauge-invariant, \UV{}-finite definition of composite operators, while the
\SFTX{} allows one to convert the corresponding matrix elements to the \MSbar\
scheme.  The window condition for the involved length (or mass) scales implied
by this procedure is found to be comfortably fulfilled in our case.  We claim
that the \GFSFTX{} procedure is suitable for high-precision determinations of
various hadronic quantities since it decreases discretization effects in the
lattice calculation and has an easily-attainable matching condition to the
\MSbar\ scheme.  This is supported by recent applications of this procedure to
the determination of renormalized strange and charm quark
masses~\cite{Black:2025gft} and parton distribution
functions~\cite{Shindler:2023xpd,Harlander:2025qsa,Francis:2025rya,Francis:2025pgf,Edwards:2026ixr}, as well as earlier
work on the determination of thermodynamical quantities based on the
energy-momentum tensor~\cite{Iritani:2018idk,Taniguchi:2020mgg}, for example.

We will apply the \GFSFTX\ method to $D_s$ mesons which provide an ideal
heavy-light laboratory for establishing the method.
This has the advantage that in a fully relativistic setup both charm and strange quark can be tuned to their physical masses avoiding extrapolations in the mass or the use of effective actions.
In addition, heavy-meson matrix elements warrant the use of a massive renormalization scheme to ensure a sufficient separation of scales, which typically poses additional challenges; alternative approaches to this challenge have been discussed in Refs.~\cite{Boyle:2016wis,DelDebbio:2024hca}.
Moreover, heavy-light mesons exhibit a rich phenomenological structure, and further insight on the validity of the heavy-quark expansion for charm quarks is highly desired.
Sum-rules techniques have been used by Ref.~\cite{Black:2024bus} to compute results for $D$ mesons and thus provide a comparison point for our results.
They use the heavy-quark effective theory (\HQET) before matching to \QCD\ for both $D$ and $B$ mesons which allows for an additional evaluation on the applicability of \HQET{} for the charm quark, since it is much lighter than the bottom quark.

It is well known that the bag parameters in the \MSbar\ scheme still have a residual scheme ambiguity related to the choice of so-called evanescent operators which are an intrinsic artifact of dimensional regularization. Any dependence on this choice will cancel in the combination of the bag parameters with the corresponding Wilson coefficients, which, therefore, need to be evaluated in the same scheme. We will present our final results for one specific scheme choice in this paper, but also provide conversion formulae to virtually arbitrary other schemes in Appendix~\ref{app:evscheme}.

The remainder of this paper is organized as follows.  We will first motivate our
calculation with a discussion of the phenomenology of heavy mesons
in~\cref{sec:pheno}, before introducing the \GFSFTX\ formalism
in~\cref{sec:GFR}.  The \GFSFTX\ method is based on three major components, all
of which will then be provided in this paper. The first one is
understanding the perturbative matching matrix between flowed and
unflowed operators, whose calculation will be presented in \cref{sec:matching};
the second component is the lattice evaluation of matrix elements of flowed
operators and their extrapolation to the continuum -- details for this will be
described in~\cref{sec:lattice,sec:data}; and the third component is the
extrapolation of the product of the first two components to vanishing flow
time. For this latter part, we have developed a robust procedure which takes
into account various systematic uncertainties  (see \cref{sec:tau0}) where also
other sources of systematic effects are considered. We give our final results in the context of phenomenology in~\cref{sec:phenores}, then~\cref{sec:concl} presents
our conclusions.
Preliminary results have been presented in Refs.\,\cite{Black:2023vju,Black:2024iwb,Black:Thesis24} and we highlight phenomenological importance of the obtained results for heavy meson lifetimes in Ref.~\cite{Black:2026rbz}.

\section{Heavy Meson Phenomenology}\label{sec:pheno}

Charm and bottom quarks are of special phenomenological interest. Due to their large mass, a plethora of decay modes exist which may be affected by physics beyond the \SM.
Studying the decays of heavy-light mesons containing one heavy quark $Q$ is particularly promising due to the precise experimental results available, building on a long history of measurements from dedicated flavor factories as well as general-purpose collider programs.

From a theory perspective, the hadron's decay rate is largely dominated by the free weak decay of the heavy quark $Q$, while bound-state effects are suppressed by the heavy-quark mass.
They can be consistently taken into account using the {heavy quark expansion} (\HQE)~\cite{Khoze:1983yp,Shifman:1984wx}, which is an {operator product expansion} (\OPE)~\cite{Wilson:1969zs,Wilson:1973jj} based on the hierarchy $\Lambda_\text{\QCD}\ll m_Q$, where $\Lambda_\text{\QCD}$ is the \QCD\ confinement scale and $m_Q$ is the heavy quark mass.
In the following general discussion, $H$ denotes a (neutral) heavy-light $D$ or $B$ meson, i.e., the heavy quark $Q$ is either a charm or bottom quark, whereas the light quark $q$ can be up, down, or strange.
Focusing on dimension-six four-quark operators, we discuss two interesting sectors of heavy quark phenomenology: the mixing of neutral mesons with their antiparticles (via $\Delta Q=2$ operators), and spectator effects in the decay of heavy mesons (via $\Delta Q=0$ operators).

\subsection{Neutral Heavy Meson Mixing\label{sec:mixing}}

\begin{figure}[tb]
    \centering
    \includegraphics[width=0.49\columnwidth]{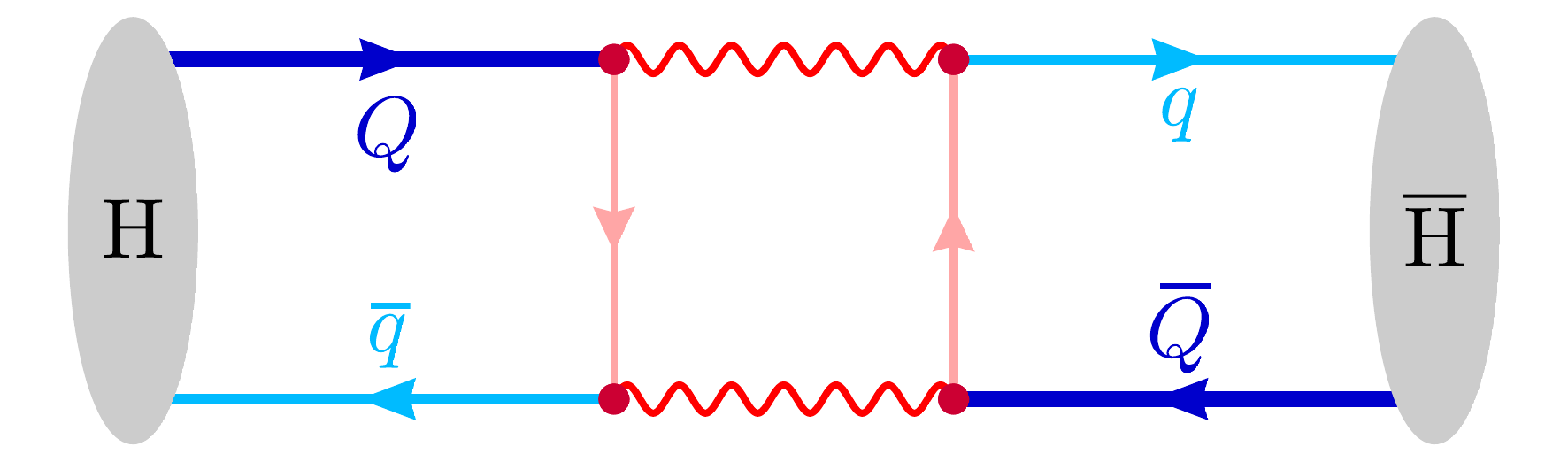}
    \includegraphics[width=0.49\columnwidth]{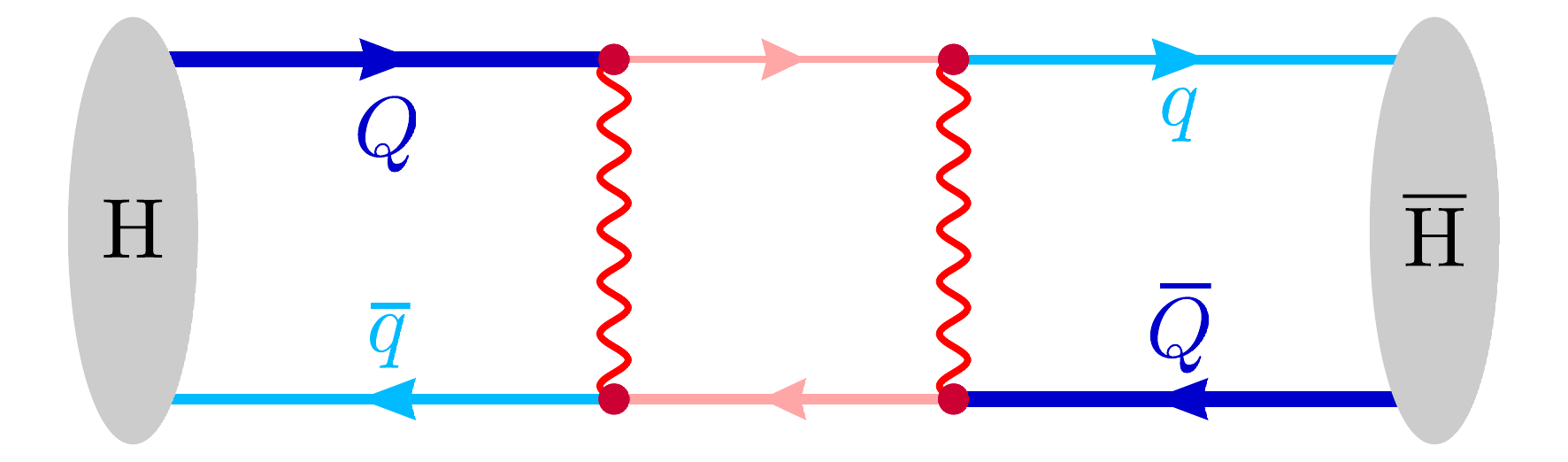}\\
    \includegraphics[width=0.49\columnwidth]{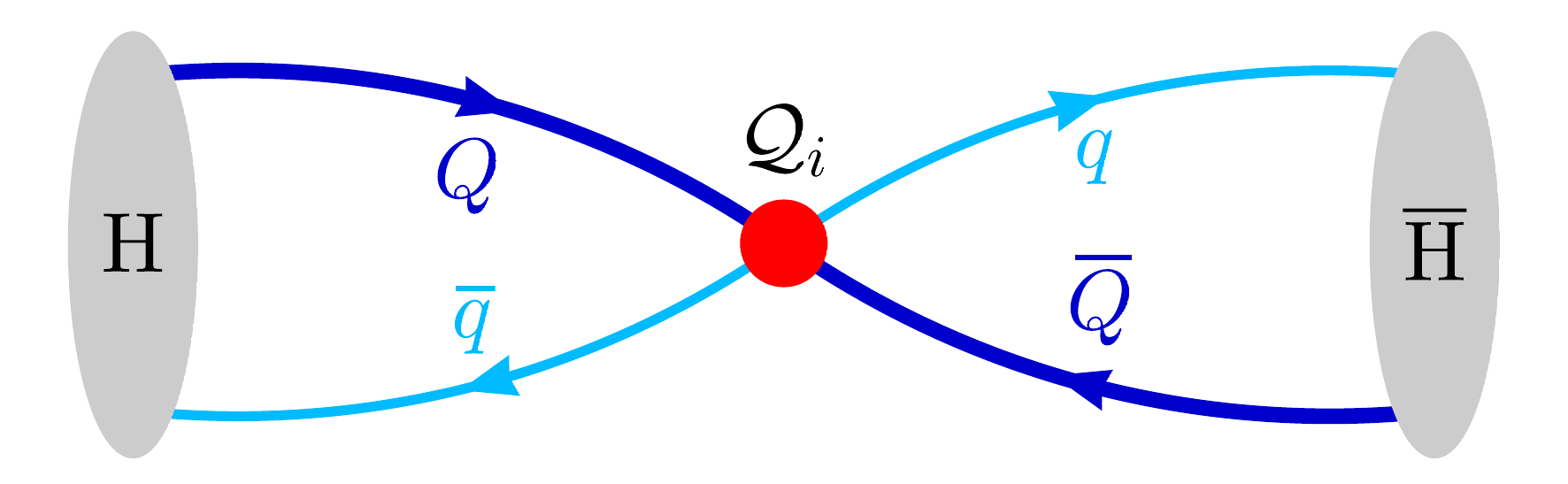}
    \caption{\SM{} box diagrams contributing to neutral meson mixing (top).
    After integrating out the heavy electroweak degrees of freedom, this is
    represented by a point-like $\Delta Q=2$ four-quark operator (bottom). The diagrams were produced with the help of \textsc{FeynGame}~\cite{Harlander:2020cyh,Bundgen:2025utt}.}
    \label{fig:mixingBox}
\end{figure}
For a neutral meson $H$ (with quark content $Q\bar{q}$), the box diagrams shown in~\cref{fig:mixingBox} induce the mixing with its antiparticle $\overline{H}$ (with quark content $\bar{Q}q$).
The time evolution of this mixing is described by
\begin{equation}
    \begin{aligned}
        {\rm i}\frac{\rm d}{{\rm d}t}\begin{pmatrix} |H(t)\rangle \\ |\overline{H}(t)\rangle \end{pmatrix} &= \Bigg(\hat{M} - \frac{\rm i}{2}\hat{\Gamma}\Bigg)\begin{pmatrix} |H(t)\rangle \\ |\overline{H}(t)\rangle \end{pmatrix} .
        \label{eq:time_evol}
    \end{aligned}
\end{equation}
Diagonalizing the matrices $\hat{M}$ and $\hat{\Gamma}$ leads to the heavy ($H_{\rm H}$) and light ($H_{\rm L}$) mass eigenstates of the system. The differences
in the mass and decay rates of $H_\text{H}$ and $H_\text{L}$ are
determined by the off-diagonal elements of the dispersive matrix $\hat{M}$ and absorptive matrix $\hat{\Gamma}$, respectively,
\begin{align}\label{eq:mesmix1}
    \Delta M &= 2|M_{12}|, &
    \Delta\Gamma &= 2|\Gamma_{12}|\cos\phi_{12},
\end{align}
where $\phi_{12} = \arg(-{M_{12}}/{\Gamma_{12}})$.
The mass differences $\Delta M$ are well measured by experiments; in particular,
after a rich history of
measurements~\cite{ALEPH:1996ofb,DELPHI:1997pdr,DELPHI:2002aou,OPAL:1997pdn,OPAL:1997tfr,OPAL:2000qeg,CDF:1997oyi,CDF:1998trv,CDF:1999jbn,CDF:1999jfn,CDF:1999hlp,L3:1998igq,BaBar:2001qlo,BaBar:2001bcs,BaBar:2002epc,BaBar:2002jxa,BaBar:2005laz,Belle:2002lms,Belle:2002qxn,Belle:2004hwe,Belle-II:2023bps,LHCb:2012mhu,LHCb:2013fep,LHCb:2016gsk,CDF:2006imy,LHCb:2011vae,LHCb:2013fep,LHCb:2013lrq,LHCb:2014iah,LHCb:2020qag,LHCb:2021moh,LHCb:2023sim,CMS:2020efq,BELLE:2007dgh,BaBar:2007kib,CDF:2007bdz,LHCb:2012zll,CDF:2013gvz,Belle:2014yoi,LHCb:2016zmn,LHCb:2018zpj,LHCb:2019mxy,LHCb:2021ykz,LHCb:2021dcr,LHCb:2022gnc,LHCb:2024hyb,LHCb:2024yxi,LHCb:2025kch,Belle:2024dux},
they have reached subpercent level precision for $B_d$ mesons and subpermille
level precision for $B_s$ mesons, while $D^0$ mixing, which occurs at much lower
frequencies, is measured with 10\% level accuracy. The world averages of the
mass differences as determined by the heavy-flavor averaging group
(\abbrev{HFLAV})~\cite{HFLAV:2024ctg} are
\begin{equation}
    \begin{aligned}
        \Delta M_{B_d} &= 0.5069 \pm 0.0019\,{\rm ps}^{-1}, \\
        \Delta M_{B_s} &= 17.765 \pm 0.006\,{\rm ps}^{-1}, \\
        \Delta M_{D^0} &= 9.4\pm1.1\,{\rm ns}^{-1}.
    \end{aligned}
\end{equation}
Theoretically, the off-diagonal elements $M_{12}$ and $\Gamma_{12}$ in~\cref{eq:mesmix1} can be expressed as
\begin{align}\label{eq:mixHeff}
    2M_H\Bigg(M_{12}-\frac{\rm i}{2}\Gamma_{12}\Bigg) &= \langle H|{\cal H}^{\Delta Q=2}_{\rm eff}|\overline{H}\rangle \\
    &+ \sum_n \frac{\langle H|{\cal H}^{\Delta Q=1}_{\rm eff}|n\rangle\langle n|{\cal H}^{\Delta Q=1}_{\rm eff}|\overline{H}\rangle}{M_H - E_n + {\rm i}\epsilon}, \nonumber
\end{align}
where $M_H$ refers to the average mass of the meson $H$,
and $\mathcal{H}^{\Delta Q=1}_{\rm eff}$, $\mathcal{H}^{\Delta Q=2}_{\rm eff}$ denote weak effective Hamiltonians consisting of $\Delta Q=1$ and $\Delta Q=2$ operators.
The sum over $n$ represents all possible states which the meson $H$ can decay into with energy $E_n$.  While the first term only contributes to $\hat{M}$,
the second term contributes to both $\hat{M}$ and $\hat{\Gamma}$ in \cref{eq:mixHeff}. The interactions incorporated in the second term are inherently long distance, and their proper calculation presents a significant challenge (for an overview in the context of $D$ mesons, see Ref.~\cite{Lenz:2020awd}, and for a proposed strategy to tackle this problem with lattice \QCD{}, see Ref.~\cite{DiCarlo:2025mvt}). For the mass differences $\Delta M$ of $B$ mesons, these long-distance effects are negligible~\cite{Artuso:2015swg,Lenz:2020awd}, however they are in fact the dominant contribution for $D$ mesons.
It also follows that these long-distance interactions generate the decay-rate difference $\Delta\Gamma$ for all heavy mesons.

Nevertheless, in the context of $\Delta M$, our focus in this paper is on the $\Delta Q=2$ effective Hamiltonian. Its matrix elements are formed from short-distance operators inserted between clearly-defined meson states which are well suited to be extracted on the lattice.
In the \SM{}, this Hamiltonian yields a contribution from a single $\Delta Q=2$ operator,\footnote{The \SM{} prediction of $\Delta\Gamma$ inherits two more four-quark operators of the $\Delta Q=2$ Hamiltonian by applying the \HQE{} to the long-distance term~\cite{Artuso:2015swg}.} namely
\begin{align}\label{eq:DB2_ops}
    {\cal Q}_1 &= (\bar{Q}\gamma_\mu(1-\gamma_5)q)(\bar{Q}\gamma_\mu(1-\gamma_5)q) ,
\end{align}
whose Wilson coefficient is known to next-to-leading order (\NLO{})~\cite{Inami:1980fz,Buras:1990fn}.
First estimates of the matrix element of this operator used the vacuum insertion approximation (\VIA), which assumes that the matrix element of any four-quark operator is saturated by the intermediary vacuum state between two quark bilinears, i.e.
\begin{equation}\label{eq:VIA}
    \langle H|{\cal O}|\overline{H}\rangle \stackrel{\text{\VIA}}{=} \langle H|\bar{Q}^iC_{ij}\Gamma_1q^j|0\rangle\cdot\langle0|\bar{Q}^kC_{kl}\Gamma_2q^l|\overline{H}\rangle,
\end{equation}
where $\Gamma_1$ and $\Gamma_2$ are the Dirac structures of each bilinear in the four-quark operator and $C_{ij}$ and $C_{kl}$ represent generic color structures such as $\delta_{ij}$ or $T^A_{ij}$ for fundamental color indices $i,j,k,l$ and adjoint index $A$.
For example, if we consider the \VIA{} for operator ${\cal Q}_1$ from~\cref{eq:DB2_ops}, first we can identify the matrix element of the two (identical) quark bilinears as
\begin{equation}\label{eq:current}
    \langle H(\vec{p})|\bar{Q}\gamma^\mu(1-\gamma_5)q|0\rangle = -{\rm i}f_H p^\mu,
\end{equation}
which is just the matrix element of the axial-vector current.
Then, the matrix element of ${\cal Q}_1$ in the \VIA{} is given by
\begin{equation}\label{eq:Q1VIA}
    \langle H|{\cal Q}_1|\bar{H}\rangle \stackrel{\text{\VIA}}{=} \frac{8}{3}
    f_H^2 M_H^2,
\end{equation}
where the additional factor $8/3$ accounts for spin and color algebra.
Beyond the \VIA{}, an additional hadronic quantity is introduced which parametrizes the deviation of the true matrix element from~\cref{eq:VIA}.
This is the bag parameter, ${\cal B}_1^H(\mu)$, which carries a dependence on the renormalization scale $\mu$.
The full matrix element of ${\cal Q}_1$ is then given as
\begin{align} \label{eq:Q1_ME}
    \langle H|{\cal Q}_1|\overline{H}\rangle(\mu) &= \frac83\, f_{H}^2\,M_{H}^2\,{\cal B}^{H}_1(\mu),
\end{align}
which is parametrized in terms of the decay constant $f_{H}$, the mass $M_{H}$, and the bag parameter ${\cal B}^{H}_1(\mu)$.
By comparing~\cref{eq:Q1VIA,eq:Q1_ME}, one can see that in the \VIA{}, ${\cal B}_1^H(\mu)=1$.
Since the meson mass $M_H$ is precisely measured~\cite{PDG:2024cfk} and the decay constant is well predicted from nonperturbative methods~\cite{FLAG:2024oxs}, it is convenient to focus our efforts on calculating the bag parameter in order to contribute to the precise determination of the four-quark matrix element.

The bag parameter ${\cal B}^H_1(\mu)$ has been determined for both neutral $B$ and $D$ mesons using sum rule techniques in \HQET{}~\cite{Grozin:2016uqy,Kirk:2017juj,Kirk:2018kib,King:2019lal}, and also from lattice \QCD{} calculations~\cite{Carrasco:2014uya,Carrasco:2015pra,Bazavov:2017weg,Carrasco:2013zta,Aoki:2014nga,Gamiz:2009ku,Bazavov:2016nty,Dowdall:2019bea,Boyle:2021kqn}.
Beyond mass-dimension six, the matrix elements of the dimension-seven four-quark operators have been explored in lattice \QCD{} by the \abbrev{HPQCD} collaboration~\cite{Davies:2019gnp} using tree-level matching between lattice and the \MSbar{} scheme.

\subsection{Heavy Meson Lifetimes}\label{sec:lifetimes}
\begin{figure}[tb]
  \centering
   \includegraphics[width=0.49\columnwidth]{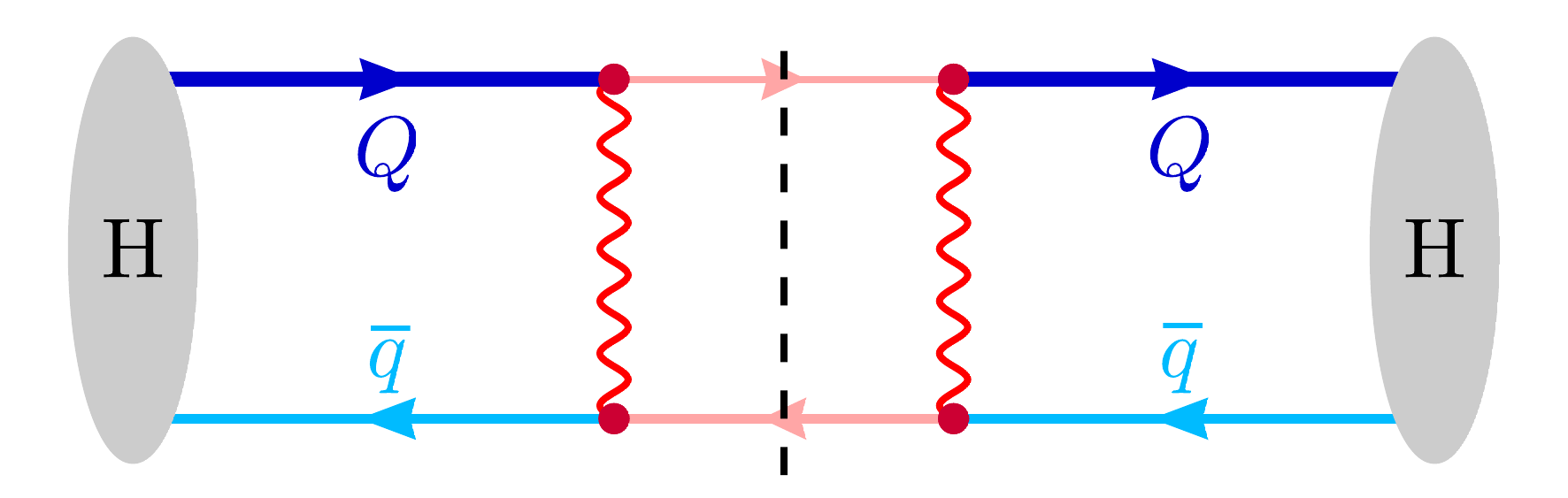}
   \includegraphics[width=0.49\columnwidth]{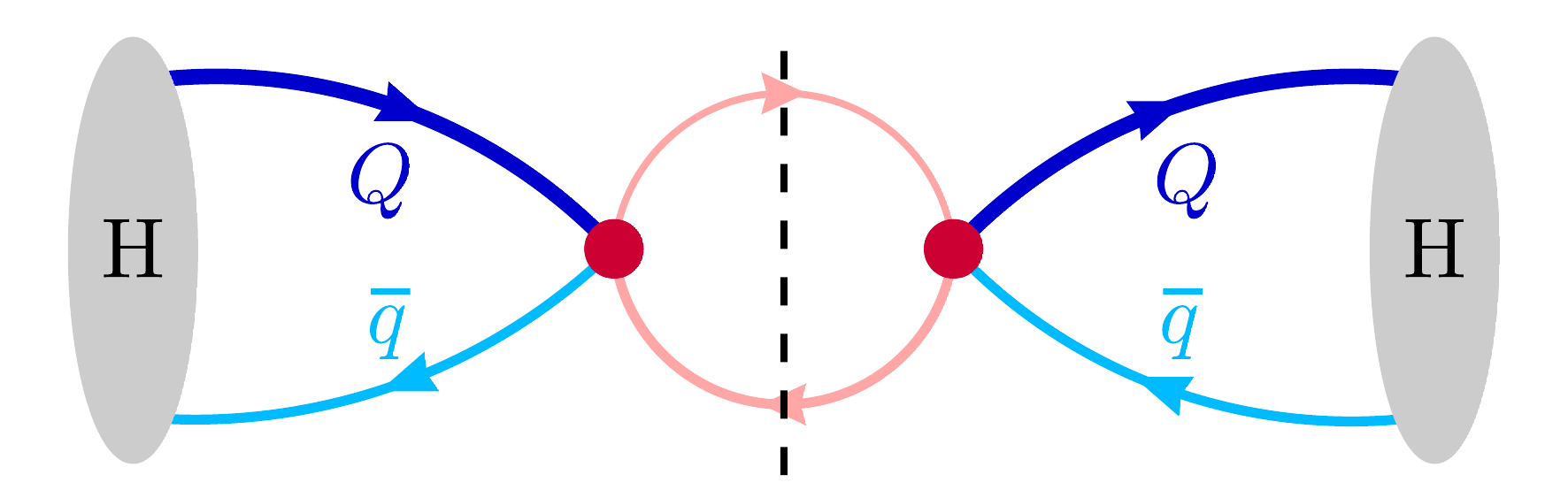}\\
   \includegraphics[width=0.49\columnwidth]{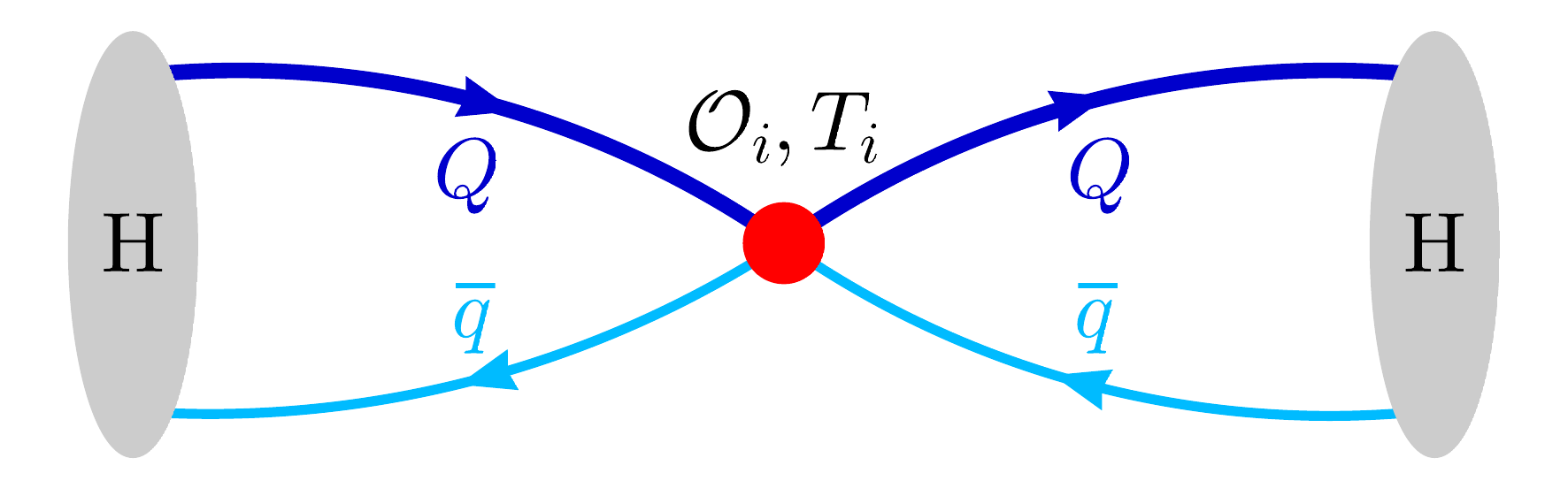}
   \caption{\label{fig:QuarkDiagram}
   Imaginary part of a $\Delta Q=0$ box diagram in the \SM{} (top left).
   After integrating out the $W$ bosons, this is described by a double insertion of the resulting $\Delta Q=1$ effective weak Hamiltonian (top right) which can be matched to local $\Delta Q=0$ four-quark operators of the HQE (bottom).}
\end{figure}

\begin{figure}[tb]
  \centering
   \includegraphics[width=0.49\columnwidth]{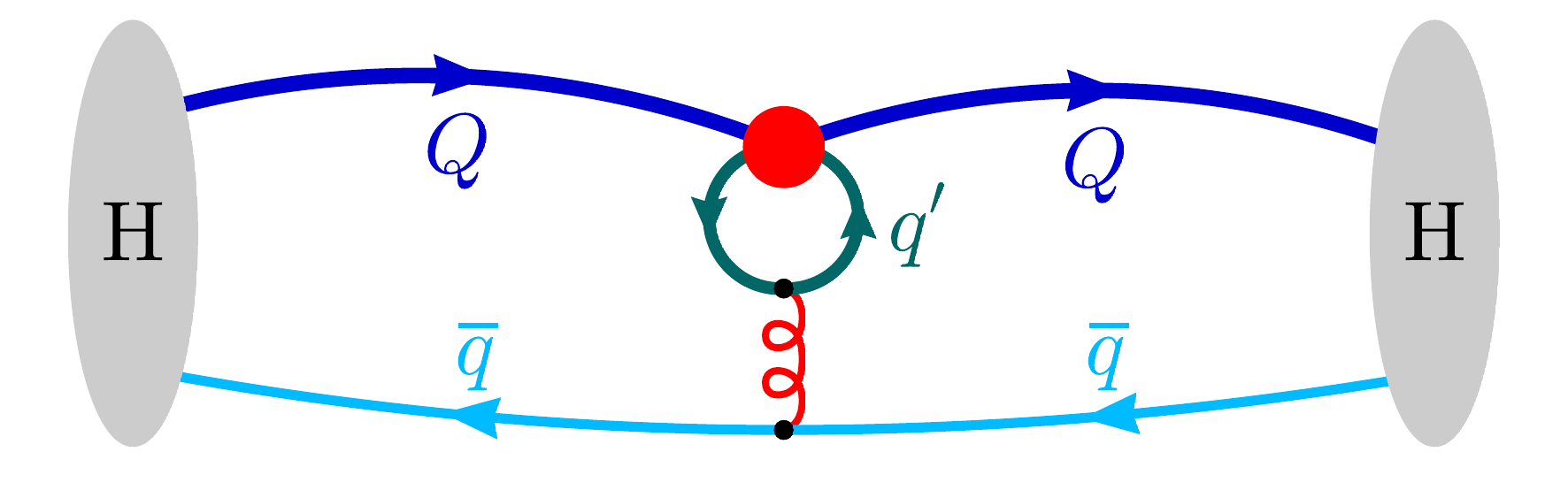}
   \includegraphics[width=0.49\columnwidth]{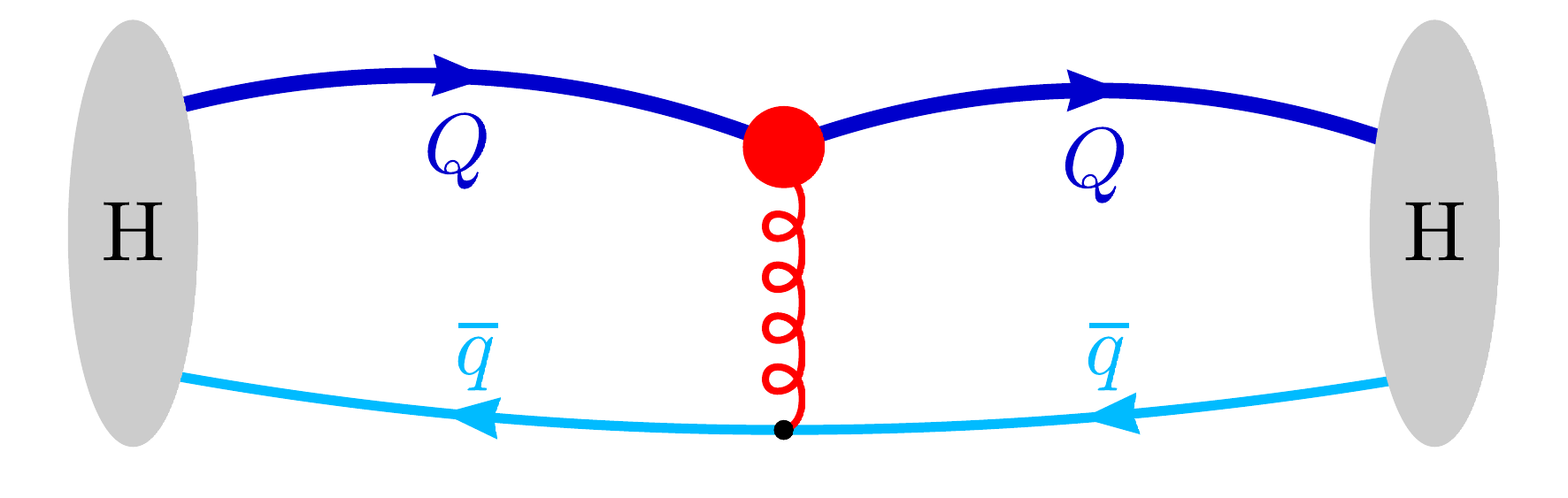}
   \caption{\label{fig:eye} Eye diagram (left) and contributions from
   lower-dimensional operators (right). The quark $q'$ forming the ``eye'' can be any light quark flavor. }
\end{figure}

The lifetimes of heavy mesons have received much attention from various experiments over the years, leading to high-precision results for $B$~mesons~\cite{DELPHI:1995hxy,DELPHI:1995pkz,DELPHI:1996dkh,DELPHI:2000aij,DELPHI:2003hqy,DELPHI:2000gjz,ALEPH:1996geg,ALEPH:1997rqk,ALEPH:2000kte,ALEPH:2000cjd,L3:1998pnf,OPAL:1995bfe,OPAL:1997zgk,OPAL:1997ufs,OPAL:1998msi,OPAL:2000qeg,SLD:1997wak,CDF:1997axv,CDF:1998pvs,CDF:1998htf,CDF:2002ixx,CDF:2010ibe,CDF:2010gif,CDF:2011utg,CDF:2011kjt,CDF:2012nqr,D0:2004ije,D0:2004jzq,D0:2008nly,D0:2011ymu,D0:2014ycx,D0:2016nbv,BaBar:2001mmd,BaBar:2002nat,BaBar:2002jxa,BaBar:2002war,BaBar:2005laz,Belle:2004hwe,ATLAS:2014nmm,ATLAS:2016pno,ATLAS:2020lbz,CMS:2015asi,CMS:2017ygm,CMS:2019bbr,CMS:2020efq,LHCb:2012zwr,LHCb:2013dzm,LHCb:2013cca,LHCb:2013odx,LHCb:2014iah,LHCb:2014qsd,LHCb:2014bqh,LHCb:2014wet,LHCb:2016crj,LHCb:2016tuh,LHCb:2017knt,LHCb:2017hbp,LHCb:2019sgv,LHCb:2019nin,LHCb:2021wte,LHCb:2021awg}
as well as for $D$~mesons~\cite{E691:1987poq,E687:1993lxk,E687:1993xsd,E791:1998zjs,E791:1999bzz,CLEO:1999xvl,SELEX:2000ijz,SELEX:2001miq,FOCUS:2002sso, FOCUS:2005gui,LHCb:2017knt,Belle-II:2021cxx,Belle-II:2023eii}.
Concerning $B$ mesons, \abbrev{HFLAV}~\cite{HFLAV:2024ctg} combines these results to find
\begin{align}
    \tau(B^0) &= 1.517 \pm 0.004 \,\text{ps}, \\
    \tau(B^+) &= 1.638 \pm 0.004 \,\text{ps}, \, \frac{\tau (B^+)}{\tau (B^0)} = 1.076 \pm 0.004, \nonumber \\
    \tau(B_s^0) &= 1.516 \pm 0.006 \,\text{ps}, \, \frac{\tau (B_s^0)}{\tau (B^0)} = 1.0021 \pm 0.0034, \nonumber
\end{align}
while the $D$-meson measurements are combined by the Particle Data Group (\abbrev{PDG})~\cite{PDG:2024cfk} to give
\begin{align}
    \tau (D^0) &= 0.4103 \pm 0.0010 \,\text{ps},  \\
    \tau (D^+) &= 1.033 \pm 0.005 \,\text{ps},\hspace{0.8em} \frac{\tau (D^+)}{\tau (D^0)} = 2.518 \pm 0.014, \nonumber \\
    \tau (D_s^+) &= 0.5012 \pm 0.0022 \,\text{ps}, \frac{\tau (D_s^+)}{\tau (D^0)} = 1.2215 \pm 0.006. \nonumber
\end{align}
The decay rate of a heavy meson $H$ is given by $\hat{\Gamma}_{11}$ of \cref{eq:time_evol}, which, similar to \cref{eq:mixHeff}, can be written as the sum of the decay of $H$ to all possible final states $n$,
\begin{equation}
    \Gamma_H = \frac{1}{2M_H} \sum_n \int_{\rm PS} (2\pi)^4\delta^{(4)}(p_{H}-p_n)|\langle n|{\cal H}_{\rm eff}^{\Delta Q=1}|H\rangle|^2,
\end{equation}
where ${\cal H}_{\rm eff}^{\Delta Q=1}$ is the weak effective Hamiltonian describing the decay of the heavy quark $Q$, and $\int_{\rm PS}$ denotes the phase-space integral for each final state $n$.
Via the optical theorem, this expression can be related to the imaginary part of the forward-scattering amplitude. The corresponding ``transition operator''
\begin{equation}\label{eq:transitionop}
    {\cal T}^{\Delta Q=0} = {\rm i}\,\int d^4x\,{\rm T}\Big\{{\cal H}_{\rm eff}^{\Delta Q=1}(x)\,{\cal H}_{\rm eff}^{\Delta Q=1}(0)\Big\}
\end{equation}
has $\Delta Q=0$, because it mediates between two identical external states. In turn, the total decay rate of $H$ can be reexpressed as
\begin{equation}
    \Gamma_H = \frac{1}{2M_H}\,{\rm Im}\langle H|{\cal T}^{\Delta Q=0}|H\rangle.
\end{equation}
In the limit where the heavy quark $Q$ is infinitely heavy, the decay rate of the heavy hadron is identical to the decay rate of the free heavy quark.
The \HQE\ allows one to systematically include corrections to this infinite mass limit.
By assuming that the heavy quark $Q$ inside the meson interacts with the light degrees of freedom via momentum exchanges $O(\Lambda_\text{\QCD})\ll m_Q$, one can see that the heavy quark's momentum will be dominated by its mass and can thus be decomposed as
\begin{equation}
    p^\mu_Q = m_Qv^\mu + k^\mu,
\end{equation}
where $v^\mu$ is the meson velocity and $k^\mu\sim O(\Lambda_\text{QCD})$ is the residual momentum stemming from nonperturbative interactions with the meson's light degrees of freedom.
The heavy-quark field can therefore be rescaled as
\begin{equation}
    Q(x) = {\rm e}^{-{\rm i}m_Qv\cdot x}Q_v(x),
\end{equation}
such that the field $Q_v(x)$ contains only low frequencies of order $k$.
From here on, when discussing heavy quark fields, we are referring to this rescaled field $Q_v(x)$.
This allows us to perform a second \OPE\ upon the transition operator ${\cal T}^{\Delta Q=0}$ of~\cref{eq:transitionop} via the small parameter $\Lambda_\text{\QCD}/m_Q\ll 1$,\footnote{The question of suitability of the HQE for a charm quark is challenging and remains open; see e.g.~Ref.~\cite{King:2021xqp}.} which expresses the decay rate as
\begin{equation}\label{eq:gammaH}
    \Gamma_H = \Gamma_3 + \cdots + 16\pi^2\left[\Gamma_6\frac{\overline{\langle{\cal O}_6\rangle}}{m_Q^3} + {\Gamma}_7\frac{\overline{\langle{\cal O}_7\rangle}}{m_Q^4}+\cdots \right],
\end{equation}
where, for each mass-dimension $d$, $\Gamma_d$ are short-distance coefficients computed in perturbation theory, and $\overline{\langle{\cal O}_d\rangle}\equiv\langle H|{\cal O}_d|H\rangle/2M_H$ are the matrix elements of the operators of the $\Delta Q=0$ effective Hamiltonian.
The free decay of the heavy quark $Q$ is described by the leading term $\Gamma_3$, while corrections arise starting at $O(1/m_Q^2)$, induced by two-quark operators of mass-dimension five, contained in the first ellipse of~\cref{eq:gammaH}; see Ref.~\cite{Albrecht:2024oyn} for an overview.
The four-quark operators ${\cal O}_d$, which are of interest here, first arise at mass-dimension six, and carry a relative factor of $16\pi^2$ from loop enhancement compared to the two-quark operators; see e.g.~Ref.~\cite{Piscopo:2021ogu}.
Their Wilson coefficients ${\Gamma}_{6/7}$ are known to \NLO{}~\cite{Guberina:1979xw,Shifman:1986mx,Uraltsev:1996ta,Neubert:1996we,Beneke:2002rj,Franco:2002fc,Lenz:2013aua} and leading order (\LO{})~\cite{Gabbiani:2003pq,Gabbiani:2004tp} respectively.
In the \SM{}, the $\Delta Q=0$ four-quark operators appearing at mass-dimension six are\footnote{The dimension-seven four-quark operators have yet to be studied beyond the \VIA{} and are beyond the scope of this work.}
\begin{equation}
    \label{eq:DB0_ops}
    \begin{aligned}
        {\cal O}_1 &= (\bar{Q}\gamma_\mu(1-\gamma_5)q)(\bar{q}\gamma^\mu(1-\gamma_5)Q), \\
        T_1 &= (\bar{Q}\gamma_\mu(1-\gamma_5)T^Aq)(\bar{q}\gamma^\mu(1-\gamma_5)T^AQ), \\
        {\cal O}_2 &= (\bar{Q}(1-\gamma_5)q)(\bar{q}(1+\gamma_5)Q), \\
        T_2 &= (\bar{Q}(1-\gamma_5)T^Aq)(\bar{q}(1+\gamma_5)T^AQ),
    \end{aligned}
\end{equation}
where $T^A$ are the generators of SU(3) color. Feynman diagrams illustrating the line of reasoning from the \SM\ description of the heavy-meson decay to the \HQE\ formulation are shown in \cref{fig:QuarkDiagram}. When considering absolute lifetimes, however, so-called eye diagrams also contribute; see \cref{fig:eye}.
On the lattice, their signal-to-noise ratio is
notoriously challenging. In addition, the calculation of absolute lifetimes involves contributions from lower-dimensional operators, indicated by the right diagram in \cref{fig:eye}, implying power divergences on the lattice. They also require significantly increased efforts on the perturbative side due to the mixing with unphysical operators.
However, both of these types of contributions cancel when considering lifetime ratios in the limit of isospin (or SU(3)$_{\rm F}$) symmetry; see e.g.~Ref.~\cite{Black:2024bus}.
Furthermore, the bag parameters contributing to absolute lifetimes have been computed in \HQET{} within the sum-rule framework of Refs.~\cite{King:2021jsq,King:Thesis22}.
In those studies, the eye-diagram contributions in \HQET{} were found to be $O(0.2\%)$, and we assume that their magnitude does not change significantly when going from \HQET{} to full \QCD{}.
For the sake of this first study, we will therefore focus on the assumption of SU(3)$_{\rm F}$ symmetry, thus avoiding the associated technical problems. Nevertheless, our method should be applicable also for absolute lifetimes, as we will discuss in more detail below.

Continuing with the operators of \cref{eq:DB0_ops}, in analogy to \cref{eq:Q1_ME}, their matrix elements are parametrized in terms of the decay constant $f_H$, quark and hadron masses $m_q$, $m_Q$, and $M_H$, and the bag parameters $B^H_i, \epsilon^H_i$ (for $i=1,2$) describing the deviation from the \VIA{}:
\begin{equation}
    \label{eq:DB0_ME}
    \begin{aligned}
        \langle H|{\cal O}_1|H\rangle &= f_{H}^2\,M_{H}^2\,{B}^{H}_1, \\
        \langle H|{\cal O}_2|H\rangle &= \frac{M_{H}^2}{(m_Q+m_q)^2}f_{H}^2\,M_{H}^2\,{B}_2^{H},  \\
        \langle H|T_1|H\rangle &= f_{H}^2\,M_{H}^2\,\epsilon_1^{H},  \\
        \langle H|T_2|H\rangle &= \frac{M_{H}^2}{(m_Q+m_q)^2}f_{H}^2\,M_{H}^2\,\epsilon_2^{H}.
    \end{aligned}
\end{equation}
The bag parameters $B_i^H$ of the color-singlet operators ${\cal O}_i$ share a similar notation to the $\Delta Q=2$ operators, motivated by the similarity of the operators, while the color-octet operators $T_i$ have their bag parameters written instead as $\epsilon_i^H$.
A clear distinction can be understood by comparing their values in the \VIA{}, where
\begin{equation}\label{eq:B_ep_VIA}
    B_i^H \stackrel{\rm VIA}{=} 1, \qquad\qquad
    \epsilon_i^H \stackrel{\text{VIA}}{=} 0.
\end{equation}

Similarly to the case of neutral meson mixing, \HQET{} sum rules can be applied to predict the bag parameters of the $\Delta Q=0$ operator basis and have yielded results for $B$ and $D$ meson lifetimes and lifetime ratios~\cite{Kirk:2017juj,Kirk:2018kib,King:2021xqp,King:2021jsq,King:Thesis22,Black:2024bus}.
However, as discussed above, there are no complete determinations of these quantities from lattice \QCD{}.

\subsection{Renormalization of Bag Parameters}\label{sec:ren_bag}

In a general prescription, renormalized operators are given by
\begin{equation}\label{eq:Brenorm}
    {\cal O}^\text{ren}_i = \sum_j Z_{ij}{\cal O}_j,
\end{equation}
where $Z_{ij}$ is the renormalization matrix.
Importantly for lattice calculations, this renormalization must occur at finite lattice spacing (defining operators at a scale $a\mu$ in e.g.~the \abbrev{RI-MOM} scheme) in order to properly define a continuum limit.
In simple cases, this mainly introduces additional cutoff effects which must be controlled and can be particularly delicate for heavy-quark processes~\cite{DelDebbio:2024hca}.
For more complicated operator bases -- such as those for lifetime calculations including eye contractions -- lattice symmetries may allow for mixing with lower-dimensional operators whose coefficients $\propto 1/a^n$ such that the continuum limit is not meaningful unless these contributions are properly accounted for.
A renormalization scheme that both improves the approach to the continuum limit and avoids power-divergent subtractions at finite $a$ is highly desirable.
The \GF\ formalism indeed provides such a scheme, as we will review in the following.

\section{Gradient Flow Renormalization}\label{sec:GFR}

\subsection{Flow equations}\label{sec:flow_equations}

The \QCD{} \GF\ formalism~\cite{Luscher:2010iy,Luscher:2011bx} extends
the \QCD{} fields by an auxiliary parameter, the flow time
$\tau$ which has mass-dimension $[\tau]=-2$. The \textit{flowed fields} $\chi$, $\bar{\chi}$, ${\cal A}$ are connected to
the regular \QCD{} fields $\psi_f$, $\bar\psi_f$, $A$ via the initial conditions
\begin{equation}
\begin{aligned}
\label{eq:bound}
\chi_{f}(\tau=0,x)&=\psi_{f}(x)\,,\\
\bar{\chi}_{f}(\tau=0,x)&=\bar{\psi}_{f}(x)\,,\\
{\cal A}_\mu^A(\tau=0,x)&=A_\mu^A(x)\,.
\end{aligned}
\end{equation}
We denote color indices in the adjoint representation by capital letters $A,B,C$, while $f$ denotes the flavor, e.g. $\psi_f\in\{Q,q\}$.

At positive flow time, the behavior of the flowed fields is described by the
flow equations
\begin{equation}
\begin{aligned}
\label{eq:floweq}
\partial_\tau
{\cal A}_\mu^A &= D^{AB}_\nu G_{\nu\mu}^B+\kappa\,D_\mu^{AB}\partial_\nu {\cal A}_\nu^B\,,\\
\partial_\tau \chi &= \Delta\chi+{\rm i}\kappa\,g_0\,\partial_\mu {\cal A}_\mu^AT^A\chi\,,\\
\partial_\tau \bar{\chi} &=
\bar{\chi}\overleftarrow{\Delta}-{\rm i}\kappa\,g_0\,\bar{\chi}\partial_\mu {\cal A}_\mu^AT^A\,,
\end{aligned}
\end{equation}
where
\begin{equation}
\label{eq:trianglenot}
\Delta=D^{\rm F}_\mu D^{\rm F}_\mu\,,\qquad\overleftarrow{\Delta}
=\overleftarrow{D}^{\rm F}_\mu\overleftarrow{D}^{\rm F}_\mu\,,
\end{equation}
and
\begin{equation}
\begin{aligned}
\label{eq:flcovder}
D_\mu^{AB}&=\partial_\mu\delta^{AB}-g_0\,f^{ABC}{\cal A}_\mu^C\,,\\
D^{\rm F}_\mu&=\partial_\mu-{\rm i}g_0\,{\cal A}_\mu^AT^A\,,\\
\overleftarrow{D}^{\rm F}_\mu&=\overleftarrow{\partial}_\mu+{\rm i}g_0\,{\cal A}_\mu^AT^A\,,
\end{aligned}
\end{equation}
with the bare \QCD{} coupling $g_0$, and $[T^A,T^B] = {\rm i}f^{ABC}T^C$.
The flowed field-strength tensor is given by
\begin{equation}
\label{eq:flfldstr}
G_{\mu\nu}^A=\partial_\mu {\cal A}_\nu^A-\partial_\nu {\cal A}_\mu^A+g_0f^{ABC}{\cal A}_\mu^B{\cal A}_\nu^C\,.
\end{equation}
The parameter $\kappa$ is a flowed gauge parameter \cite{Luscher:2010iy}; in the perturbative calculation (see \cref{sec:matching}), it is advantageous to set
$\kappa=1$, while we set $\kappa$ to zero on the lattice. However, our lattice and perturbative results are separately independent of this choice. Using the flow equations, one can derive a
Lagrangian of flowed \QCD{} \cite{Luscher:2011bx}, which incorporates the
Feynman rules of regular (unflowed) \QCD{} as well as additional flowed Feynman
rules, which are listed in Ref.~\cite{Artz:2019bpr}.

The solutions of the flow equations of~\cref{eq:floweq} lead to an exponential
suppression of high-momentum modes. In particular, composite operators of
flowed fields do not require any operator
renormalization~\cite{Luscher:2011bx,Luscher:2013cpa,Hieda:2016xpq}.
In our choice of normalization, the flowed gluon field renormalizes in the same way as the gauge coupling, while flowed quarks need
to be renormalized in order to result in finite Green's functions.  Throughout
our perturbative calculation, we will adopt the ringed scheme, defined as~\cite{Makino:2014taa}
\begin{equation}\label{eq:gff:gapy}
  \begin{aligned}
    \mathring{\chi}(\tau)&=\mathring{Z}^{1/2}_\chi(\tau)\chi(\tau)\,,\qquad
    \langle\mathring{\bar{\chi}}\slashed{D}^\text{F}\mathring{\chi}\rangle\equiv
    -\frac{2\nc}{(4\pi \tau)^2}\,,
  \end{aligned}
\end{equation}
where $\nc=3$ is the number of colors. The explicit form of $\mathring{Z}_\chi$
is given in \cref{app:ren}. Our results for the perturbative matching
coefficients of the four-quark operators in \cref{sec:pt_results} will be
provided in this scheme. However, as we will see later, the wave function
renormalization factors $\mathring{Z}_\chi$ actually cancel in the ratios of
matrix elements defining the bag parameters. The renormalization of the \QCD{}
parameters (coupling and quark masses) is the same as in regular \QCD{}.

\subsection{Short Flow-Time Expansion}\label{sec:sftx}

With the gradient flow, the Wilson coefficients $C_n$ and operators $\mathcal{O}_n$ of an effective Hamiltonian can be expressed in terms of \textit{flowed operators} $\widetilde{\cal O}_n(\tau)$, composed of flowed fields, and \textit{flowed Wilson coefficients} $\widetilde{C}_n(\tau)$, i.e.
\begin{equation}\label{eq:Heff}
    {\cal H}_{\rm eff} = \sum_n C_n{\cal O}_n = \sum_n \widetilde{C}_n(\tau)\widetilde{\cal O}_n(\tau).
\end{equation}
While Wilson coefficients and matrix elements of operators depend on the choice of the renormalization scheme, the effective Hamiltonian itself (and therefore any physical observable derived from it) is independent of this choice (up to perturbative truncation effects).
This does not change in the flowed theory, where the dependence on the flow time between the operator matrix elements and Wilson coefficients will mutually cancel.
In the short flow-time expansion, the relation between matrix elements of the flowed operators and the regular ones is
\begin{equation}
    \label{eq:sftx}
    \langle\widetilde{\cal O}_n(\tau)\rangle
    = \sum_m\zeta_{nm}(\tau,\mu)\langle{\cal O}^\text{\MSbar}_m\rangle(\mu) + O(\tau),
\end{equation}
where $\zeta_{nm}(\tau,\mu)$ are perturbatively-calculated matching coefficients, and ${\cal O}^\text{\MSbar}$ are the regular operators renormalized in the \MSbar\ scheme. Therefore, by inverting this relation, the matrix elements of the operators in the \MSbar\ scheme can be obtained from the matrix elements of the flowed operators with the help of the inverse matching matrix.

In our analysis, terms of order $\tau$ (and higher) will be neglected in \cref{eq:sftx}, meaning that the rhs\ of \cref{eq:Heff} will have a residual flow-time dependence of $O(\tau)$, implying an upper limit on $\tau$ in its applicability that needs to be quantified. Strict independence on $\tau$ is formally reached only in the limit $\tau\to 0$. However, in a lattice evaluation of the flowed matrix elements, the ``smearing radius'' $\sqrt{8\tau}$ must be larger than the lattice spacings in order to suppress discretization effects coming from the boundary and to sufficiently approach a renormalized trajectory.
Moreover, the perturbative matching coefficients depend on the renormalization scale $\mu$, which needs to be chosen in a regime where the strong coupling $\alphas(\mu)$ assumes perturbative values. The dependence on $\tau$ is logarithmic, which also restricts its values to a finite interval in order to preserve perturbativity. One therefore needs to identify a window $\tau\in[\tau_\text{min},\tau_\text{max}]$ in flow time where all these aspects are taken into account, and which allows for a reliable extrapolation $\tau\to 0$ within well-defined uncertainties. We will show that, in our case, a suitable interval for $\tau$ exists. We will describe in detail our procedure for the $\tau\to 0$ extrapolation, resulting in the bag parameters in the \MSbar\ scheme at a fixed renormalization scale $\mu_0$.

It is worth pointing out that any mixing which occurs in lattice operator renormalization as power divergences in $1/a^n$ will turn into divergences of the form $1/\tau^{n/2}$ in the \GFSFTX\ procedure; see e.g.~\cite{Kim:2021qae}. Such terms are absent for the operator basis presented here; they will appear when including the full operator basis for absolute lifetimes through eye diagrams and penguin operators though. The main strategy for renormalizing these operators using \GFSFTX\ remains the same as described here, however.

\subsection{Application to the four-quark operators}\label{sec:sftx_bag}

Focusing on heavy-meson mixing and lifetime ratios, the set of flowed operators is simply obtained by replacing the regular fields in~\cref{eq:DB2_ops,eq:DB0_ops} by flowed ones. Adopting a unified notation for both mixing and lifetime operators, we write them as
\begin{equation}
  \begin{aligned}\label{eq:flowops}
  \widetilde{{\cal O}}_1&=\mathring{Z}_\chi^2
  \left(\bar{\chi}_\alpha\gamma_\mu (1-\gamma_5)
  \chi_\beta\right)\left(\bar{\chi}_\gamma\gamma_\mu (1-\gamma_5)
  \chi_\delta\right)\,, \\
  \widetilde{T}_1&=\mathring{Z}_\chi^2
  \left(\bar{\chi}_\alpha\gamma_\mu (1-\gamma_5) T^A \chi_\beta\right)
  \left(\bar{\chi}_\gamma\gamma_\mu (1-\gamma_5) T^A \chi_\delta\right)\,, \\
  \widetilde{{\cal O}}_2&=\mathring{Z}_\chi^2
  \left(\bar{\chi}_\alpha (1-\gamma_5)\chi_\beta\right)
  \left(\bar{\chi}_\gamma (1+\gamma_5)\chi_\delta\right)\,, \\
  \widetilde{T}_2&=\mathring{Z}_\chi^2
  \left(\bar{\chi}_\alpha (1-\gamma_5)T^A \chi_\beta\right)
  \left(\bar{\chi}_\gamma (1+\gamma_5)T^A \chi_\delta\right)\,,
  \end{aligned}
\end{equation}
where $\mathring{Z}_\chi$ is the flowed quark renormalization in the ringed scheme as defined in \cref{eq:gff:gapy}.
The indices $\alpha,\beta,\gamma,\delta$ in \cref{eq:flowops} denote the quark
flavor. By choosing the initial conditions $\chi(\tau=0)=\psi$, with
\begin{equation}\label{eq:497:60}
  \begin{aligned}
    \psi_\alpha = \psi_\gamma = Q\,,\qquad
    \psi_\beta = \psi_\delta = q\,,
  \end{aligned}
\end{equation}
$\tilde{\cal O}_1$ corresponds to the flowed version of the mixing operator $\mathcal{Q}_1$ in \cref{eq:DB2_ops}, while setting
\begin{equation}\label{eq:497:61}
  \begin{aligned}
    \psi_\alpha = \psi_\delta = Q\,,\qquad
    \psi_\beta = \psi_\gamma = q\,,
  \end{aligned}
\end{equation}
leads to the flowed versions of the lifetime operators of \cref{eq:DB0_ops}. The lattice evaluation of the corresponding matrix elements relevant for heavy-meson mixing and decay will be described in \cref{sec:lattice,sec:data}. Before that, we describe the perturbative calculation of the corresponding matching coefficients.

\section{Perturbative matching coefficients}\label{sec:matching}

We will calculate matching matrices for the flowed operators of~\cref{eq:flowops} through next-to-next-to-leading order (\NNLO) \QCD. Since we focus on mixing parameters and lifetime ratios, we can assume that the flavor indices of the four quark fields in \cref{eq:flowops} are mutually distinct, i.e.\ we can assume that they denote quarks of four different flavors. For the sake of clarity, the discussion in this section is mostly devoted to the lifetime operators. The treatment of the mixing operator will be discussed at the end of \cref{sec:pt_results}.

According to the color structure of the operators in \cref{eq:DB0_ops,eq:flowops}, the matching matrix is block diagonal, so that we can write the \SFTX\ at the operator level as
\begin{equation}\label{eq:517:38}
  \begin{aligned}
    \begin{pmatrix}
      \widetilde{\cal O}_i\\\
       \widetilde{T}_i
      \end{pmatrix}
      = \zeta_i
      \begin{pmatrix}
        {\cal O}^\text{\MSbar}_i\\\
        T^\text{\MSbar}_i
        \end{pmatrix} \,,\qquad i\in\{1,2\},
    \end{aligned}
\end{equation}
with the $2\times2$ matching matrices $\zeta_1$ and $\zeta_2$.

\subsection{Evanescent operators}\label{sec:evops}

It is well known that the operator bases of \cref{eq:DB2_ops,eq:DB0_ops} are not
closed under renormalization when working in dimensional regularization.
Instead, one needs to include so-called evanescent
operators~\cite{Buras:1989xd,Dugan:1990df,Herrlich:1994kh} which vanish in $D=4$
space-time dimensions due to algebraic identities~\cite{Fierz:1937wjm}, but
are non-zero when working at $D=4-2\ep$. At each perturbative order, additional
evanescent operators have to be introduced. Their choice is not unique and can
affect the finite parts of the final results, thus defining sub-variants of the
\MSbar\ scheme. In the main part of this paper, we will adopt the following set
of evanescent operators for the lifetime calculations, referring to it as
the \textit{minimal basis} (cf.~\cref{eq:497:61} for the notation):
\begin{align}
    \mathcal{E}_{{\cal O}_1}^{(1)}=&\left(\bar{\psi}_\alpha\gamma_{\mu\nu\rho}
    (1-\gamma_5)
    \psi_\beta\right)\left(\bar{\psi}_\gamma\gamma_{\rho\nu\mu}
    (1-\gamma_5)
  \psi_\delta\right) \nonumber \\
  &-4\,{\cal O}_1\,,\hspace*{3.7em}\nonumber \\
  \mathcal{E}_{T_1}^{(1)}=&\left(\bar{\psi}_\alpha\gamma_{\mu\nu\rho}
  (1-\gamma_5) T^A
  \psi_\beta\right)\left(\bar{\psi}_\gamma\gamma_{\rho\nu\mu}
  (1-\gamma_5) T^A
  \psi_\delta\right)\nonumber \\&
  -4\,T_1\,,\nonumber \\
  \mathcal{E}_{{\cal O}_2}^{(1)}=&
  \left(\bar{\psi}_\alpha\gamma_{\mu\nu}
  (1-\gamma_5)\psi_\beta\right)
  \left(\bar{\psi}_\gamma\gamma_{\nu\mu}
  (1+\gamma_5)\psi_\delta\right)\nonumber \\
  &-4\,{\cal O}_2\,,\nonumber \\
  \mathcal{E}_{T_2}^{(1)}=&
  \left(\bar{\psi}_\alpha\gamma_{\mu\nu}
  (1-\gamma_5)T^A\psi_\beta\right)
  \left(\bar{\psi}_\gamma\gamma_{\nu\mu}
  (1+\gamma_5)T^A\psi_\delta\right)\nonumber \\
  &-4\,T_2\,,\label{eq:evopsmin1}
\end{align}
and
\begin{align}
    \mathcal{E}_{{\cal O}_1}^{(2)}=&\left(\bar{\psi}_\alpha
    \gamma_{\mu\nu\rho\sigma\tau}
  (1-\gamma_5)
  \psi_\beta\right)\left(\bar{\psi}_\gamma\gamma_{\tau\sigma\rho\nu\mu}
  (1-\gamma_5)
  \psi_\delta\right) \nonumber \\
  &-16\,{\cal O}_1\,, \nonumber \\
  \mathcal{E}_{T_1}^{(2)}=&\left(\bar{\psi}_\alpha\gamma_{\mu\nu\rho\sigma\tau}
  (1-\gamma_5)T^A
  \psi_\beta\right)\left(\bar{\psi}_\gamma\gamma_{\tau\sigma\rho\nu\mu}
  (1-\gamma_5)T^A
  \psi_\delta\right) \nonumber \\
  &-16\,T_1\,,\nonumber  \\
  \mathcal{E}_{{\cal O}_2}^{(2)}=&
  \left(\bar{\psi}_\alpha\gamma_{\mu\nu\rho\sigma}
  (1-\gamma_5)
  \psi_\beta\right)
  \left(\bar{\psi}_\gamma\gamma_{\sigma\rho\nu\mu}
  (1+\gamma_5)
  \psi_\delta\right) \nonumber \\
  &-16\,{\cal O}_2\,, \nonumber \\
  \mathcal{E}_{T_2}^{(2)}=&
  \left(\bar{\psi}_\alpha\gamma_{\mu\nu\rho\sigma}
  (1-\gamma_5)T^A
  \psi_\beta\right)
  \left(\bar{\psi}_\gamma\gamma_{\sigma\rho\nu\mu}
  (1+\gamma_5)T^A
  \psi_\delta\right) \nonumber \\
  &-16\,T_2\,, \label{eq:evopsmin2}
\end{align}
where
\begin{equation}
  \gamma_{\mu_1\ldots\mu_n}=\gamma_{\mu_1}\cdots\gamma_{\mu_n}\,.
\end{equation}
The $\mathcal{E}_{{\cal O}_i}^{(n)},\mathcal{E}_{T_i}^{(n)}$ are required for the calculation of the $\zeta_i$, $i\in\{1,2\}$ at $n$-loop level, respectively, cf.~\cref{eq:zetas}.
In our calculation, we actually allow for a more general choice of evanescent operators, as described in \cref{app:evscheme}. The results and anomalous dimensions in this general scheme can be found in the ancillary file; see \cref{app:anc} and Ref.~\cite{SupplementalMaterial}.

A priori, the evanescent operators contribute to the \SFTX{} of the physical
operators. The matching matrix of the full operator basis can be schematically written as
\begin{equation}\label{eq:gff:john}
  \begin{aligned}
  \begin{pmatrix}
    \widetilde{\mathcal{O}}(\tau)\\
    \widetilde{\mathcal{E}}(\tau)
  \end{pmatrix}&=
  \begin{pmatrix}
    \zeta_{\mathcal{O}\mathcal{O}}(\tau,\mu)&\zeta_{\mathcal{O}\mathcal{E}}(\tau,\mu)\\
    \zeta_{\mathcal{E}\mathcal{O}}(\tau,\mu)&\zeta_{\mathcal{E}\mathcal{E}}(\tau,\mu)
  \end{pmatrix}
  \begin{pmatrix}
    \mathcal{O}^\ren(\mu)\\
    \mathcal{E}^\ren(\mu)
  \end{pmatrix}\,,
  \end{aligned}
\end{equation}
where
\begin{equation}
  \begin{pmatrix}
    \mathcal{O}^\ren\\
    \mathcal{E}^\ren
  \end{pmatrix}=
  \begin{pmatrix}
    \mathcal{Z_{\mathcal{O}\mathcal{O}}}&\mathcal{Z_{\mathcal{O}\mathcal{E}}}\\
    \mathcal{Z_{\mathcal{E}\mathcal{O}}}&\mathcal{Z_{\mathcal{E}\mathcal{E}}}
  \end{pmatrix}
  \begin{pmatrix}
    \mathcal{O}\\
    \mathcal{E}
  \end{pmatrix}\,.
\end{equation}
Here, $\mathcal{O}$ and $\mathcal{E}$ refer to the collection of physical and
evanescent operators, respectively.  The sub-matrix
$\mathcal{Z}_{\mathcal{E}\mathcal{O}}$ contains a non-minimal finite part
which is defined such that the renormalized evanescent operators vanish at
$D=4$\ \cite{Dugan:1990df,Buras:1989xd,Herrlich:1994kh} and thus do not
contribute to physical observables. Explicit results are given in
\cref{app:ren}. In the \GF{}, this
renormalization condition can be implemented by requiring
that~\cite{Harlander:2022tgk}
\begin{equation}\label{eq:gff:isis}
  \begin{aligned}
    \zeta_{\mathcal{E}\mathcal{O}}(\tau,\mu)&=\mathcal{O}(\ep)\,.
  \end{aligned}
\end{equation}
This follows from the fact that matrix elements of flowed operators are free
from \UV{} divergences; one can therefore write
$\langle\widetilde{\mathcal{E}}(\tau)\rangle = \order{\ep}$. Combining
\cref{eq:gff:john,eq:gff:isis} then also enforces
$\langle\mathcal{E}^\ren(\mu)\rangle=\order{\ep}$.
The results for the finite renormalization matrix in a general scheme of evanescent operators are provided in the ancillary file; see \cref{app:anc} and Ref.~\cite{SupplementalMaterial}.

\subsection{Method of projectors}\label{sec:projectors}

Our procedure to determine the matching matrices $\zeta_i$ of \cref{eq:517:38} relies on the method of
projectors~\cite{Gorishnii:1983su,Gorishnii:1986gn}. For each of the operators
of \cref{eq:DB0_ops} (assuming four different quark flavors; see above) as well as each of the evanescent operators of \cref{sec:evops}, we
construct a projector $P_i$, consisting of external states and derivatives
\wrt\ masses and external momenta, such that
\begin{equation}
  \label{eq:projgeneral}
  P_{i}[\mathcal{O}_j]=\delta_{ij}\,.
\end{equation}
Setting all masses and external momenta to zero before loop integration eliminates all higher-order contributions on the lhs of this relation.
Applying these projectors on the flowed operators thus directly gives the individual components of the matching
matrix~\cite{Harlander:2018zpi},
\begin{equation}
  \label{eq:calcmix}
  \zeta_{mn}=P_{n}[\widetilde{\mathcal{O}}_m]\,.
\end{equation}
For the calculation of the resulting Feynman diagrams, we use an updated
version of the setup described in Ref.~\cite{Artz:2019bpr} which consists of
\texttt{qgraf}~\cite{Nogueira:1991ex,Nogueira:2006pq} for generating the
Feynman diagrams, \texttt{tapir/exp}~\cite{Harlander:1998cmq,Seidensticker:1999bb,Gerlach:2022qnc}\footnote{We recently replaced \texttt{q2e}~\cite{Harlander:1998cmq,Seidensticker:1999bb} by
\texttt{tapir}~\cite{Gerlach:2022qnc}; in fact, we used this project to check
our implementation of \texttt{tapir} against the \texttt{q2e} version.} for
identifying the momentum topologies and inserting the Feynman rules, and
\texttt{FORM}\ \cite{Vermaseren:2000nd,Ruijl:2017dtg} for the calculation of
the Dirac and color traces~\cite{vanRitbergen:1998pn}.  We adopt anticommuting $\gamma_5$ throughout the
calculation; this is justified, because all traces contain an even number of
$\gamma_5$ (see Ref.~\cite{Harlander:2022tgk}).
The resulting scalar integrals are then reduced to the one and six master integrals of Ref.~\cite{Harlander:2018zpi} at one and two loops, respectively, by solving integration-by-parts identities~\cite{Tkachov:1981wb,Chetyrkin:1981qh,Laporta:2000dsw,Artz:2019bpr} with \texttt{Kira}$\otimes$\texttt{FireFly}~\cite{Maierhofer:2017gsa,Klappert:2020nbg,Klappert:2019emp,Klappert:2020aqs}.

\subsection{Results}\label{sec:pt_results}

The results for $\zeta_1$ have already been presented in Ref.~\cite{Harlander:2022tgk} through \NNLO. We re-evaluated the matrix as a
check for our calculation of $\zeta_2$ and include them here for the sake of completeness. For convenience of the reader, we quote the results for
the inverse matching matrix rather than the matrix itself, as it is this quantity that is required to relate the lattice matrix elements to the perturbative matching coefficient; cf.~\cref{eq:sftx}. We use the definitions
\begin{equation}
\label{eq:transcendentals}
  \begin{aligned}
    \zeta(2)=\text{Li}_2(1) &= \frac{\pi^2}{6} = 1.64493\ldots\\
    \text{Li}_2(1/4) &= 0.267652\ldots\,,
  \end{aligned}
\end{equation}
where $\text{Li}_2(z)=\sum_{k=1}^\infty z^k/k^2$ is the dilogarithm.
The flowed quark fields are renormalized in the ringed scheme, and we define the evanescent operators as in \cref{sec:evops}. Furthermore, we denote the number of quark flavors by $\nf$, and define $\api(\mu)=\frac{g^2(\mu)}{4\pi^2}$ with $g(\mu)$ the strong coupling in the $\MSbar$ scheme. The coefficient $c_\chi^{(2)}$ appearing in the results arises from the conversion to the ringed scheme and is only known numerically; its value can be found in \cref{eq:chi2,eq:chi2coeffs}. The matching coefficients depend on
\begin{equation}
\label{eq:lmut}
  \begin{aligned}
    \lmut = \ln 2\mu^2 \tau + \EulerGamma\,,
  \end{aligned}
\end{equation}
with $\EulerGamma = 0.577215\ldots$ being the Euler-Mascheroni constant.

Finally, for $i=1$, we find\footnote{The results of
Ref.\,\cite{Harlander:2022tgk} are equivalent to ours, even though the choice of
basis is slightly different. The difference amounts to the choice of $B_{1,0}$
and $B_{2,0}$ as defined in \cref{eq:evops2} and \cref{eq:evopscoeffs2}. This
choice does not impact the physical part of the matching matrix though (see
\cref{app:evscheme}).
}
\begin{widetext}
\begin{align}
    (\zeta_1^{-1})_{11} =& 1+\api\,\Bigl(-\frac{1}{3}
  +\frac{8}{3}\ln2+2\ln3\Bigr)
  +\api^2\,\Biggl\{-\frac{1}{8}c_\chi^{(2)}+\frac{2413}{144}
  -\frac{35}{72}\nf
  +\left(\frac{119}{72}-\frac{1}{6}\nf\right)\zeta(2)+\frac{70}{3}\ln2
  +\frac{22}{9}\ln^22 \nonumber \\
  &\eqindent{1}+8\ln2\ln3-\frac{83}{3}\ln3
  +3\ln^23
-\frac{107}{6}\text{Li}_2\left(1/4\right) \nonumber \\
&+\lmut\biggl[-\frac{3}{8}+\frac{1}{18}\nf+\left(\frac{22}{3}-\frac{4}{9}\nf\right)\ln2+\left(\frac{11}{2}-\frac{1}{3}\nf\right)\ln3\biggr]
+\frac{1}{4}\lmut^2\Biggr\}+\mathcal{O}(\api^3)\,,  \nonumber \\
    (\zeta_1^{-1})_{12} =&-\api\,\Bigl(\frac{3}{4}
  +\frac{3}{2}\lmut\Bigr)+\api^2\,\Biggl\{-\frac{301}{48}+\frac{133}{288}\nf+\left(-\frac{67}{96}+\frac{1}{8}\nf\right)\zeta(2)-\frac{451}{24}\ln2+\frac{201}{16}\ln3
-\frac{27}{16}\text{Li}_2\left(1/4\right) \nonumber \\
&+\lmut\biggl[-\frac{245}{32}+\frac{5}{12}\nf-4\ln2-3\ln3\biggr]
+\lmut^2\biggl[-\frac{39}{16}+\frac{1}{8}\nf\biggr]\Biggr\}+\mathcal{O}(\api^3)\,,  \nonumber \\
    (\zeta_1^{-1})_{21} =& -\api\,\Bigl(\frac{1}{6}
  +\frac{1}{3}\lmut\Bigr)+\api^2\,\Biggl\{-\frac{733}{216}+\frac{133}{1296}\nf+\left(-\frac{67}{432}+\frac{1}{36}\nf\right)\zeta(2)-\frac{451}{108}\ln2+\frac{67}{24}\ln3
-\frac{3}{8}\text{Li}_2\left(1/4\right) \nonumber \\
&+\lmut\biggl[-\frac{317}{144}+\frac{5}{54}\nf-\frac{8}{9}\ln2-\frac{2}{3}\ln3\biggr]
+\lmut^2\biggl[-\frac{13}{24}+\frac{1}{36}\nf\biggr]\Biggr\}+\mathcal{O}(\api^3)\,, \nonumber \\
    (\zeta_1^{-1})_{22} =&  1+\api\,\Bigl(\frac{17}{12}
  +\frac{8}{3}\ln2+2\ln3+\frac{1}{2}\lmut\Bigr)
  +\api^2\,\Biggl\{-\frac{1}{8}c_\chi^{(2)}+\frac{6913}{288}
  -\frac{625}{864}\nf
  +\left(\frac{181}{96}-\frac{5}{24}\nf\right)\zeta(2)+\frac{2419}{72}\ln2 \nonumber \\
  &\eqindent{1}+\frac{22}{9}\ln^22+8\ln2\ln3-\frac{1385}{48}\ln3
  +3\ln^23
-\frac{829}{48}\text{Li}_2\left(1/4\right) \nonumber \\
&+\lmut\biggl[\frac{281}{96}-\frac{1}{12}\nf+\left(\frac{26}{3}-\frac{4}{9}\nf\right)\ln2+\left(\frac{13}{2}-\frac{1}{3}\nf\right)\ln3\biggr]
+\lmut^2\biggl[\frac{17}{16}-\frac{1}{24}\nf\biggr]\Biggr\}+\mathcal{O}(\api^3)\,, \label{eq:resu:irus}
\end{align}
and for $i=2$
\begin{align}
  (\zeta_2^{-1})_{11}=&1+\api\,\Bigl(\frac{4}{3}
  +\frac{8}{3}\ln2+2\ln3+2\lmut\Bigr)
  +\api^2\,\Biggl\{-\frac{1}{8}c_\chi^{(2)}+\frac{2291}{72}
  -\frac{10}{9}\nf
  +\left(\frac{7}{4}-\frac{1}{3}\nf\right)\zeta(2)+\frac{143}{3}\ln2
  +\frac{22}{9}\ln^22 \nonumber \\
  &\eqindent{1}+8\ln2\ln3-\frac{251}{6}\ln3
  +3\ln^23
-\frac{47}{2}\text{Li}_2\left(1/4\right) \nonumber \\
&+\lmut\biggl[\frac{29}{2}-\frac{1}{2}\nf+\left(\frac{38}{3}-\frac{4}{9}\nf\right)\ln2+\left(\frac{19}{2}-\frac{1}{3}\nf\right)\ln3\biggr]
+\lmut^2\biggl[\frac{19}{4}-\frac{1}{6}\nf\biggr]\Biggr\}+\mathcal{O}(\api^3)\,, \nonumber \\
(\zeta_2^{-1})_{12}=&\frac{1}{2}\api+\api^2\,\Biggl\{\frac{17}{192}-\frac{1}{144}\nf+\frac{5}{96}\zeta(2)+\frac{77}{24}\ln2+\frac{1}{16}\ln3
-\frac{25}{16}\text{Li}_2\left(1/4\right)-\frac{19}{32}\lmut\Biggr\}+\mathcal{O}(\api^3)\,, \nonumber \\
(\zeta_2^{-1})_{21}=&\frac{1}{9}\api+\api^2\,\Biggl\{-\frac{91}{864}-\frac{1}{648}\nf+\frac{5}{432}\zeta(2)+\frac{77}{108}\ln2+\frac{1}{72}\ln3
-\frac{25}{72}\text{Li}_2\left(1/4\right)+\frac{17}{144}\lmut\Biggr\}+\mathcal{O}(\api^3)\,, \nonumber \\
(\zeta_2^{-1})_{22}=&1+\api\,\Bigl(\frac{19}{24}+\frac{8}{3}\ln2+2\ln3-\frac{1}{4}\lmut\Bigr)
+\api^2\,\Biggl\{-\frac{1}{8}c_\chi^{(2)}+\frac{21223}{1152}-\frac{845}{1728}\nf
  +\left(\frac{935}{576}-\frac{7}{48}\nf\right)\zeta(2)+\frac{209}{9}\ln2 \nonumber \\
  &\eqindent{1}+\frac{22}{9}\ln^22
  +8\ln2\ln3-\frac{275}{12}\ln3
  +3\ln^23
  -\frac{797}{48}\text{Li}_2\left(1/4\right) \nonumber \\
  &+\lmut\biggl[-\frac{427}{192}+\frac{1}{8}\nf
    +\left(\frac{20}{3}-\frac{4}{9}\nf\right)\ln2
+\left(5-\frac{1}{3}\nf\right)\ln3\biggr]
+\lmut^2\biggl[-\frac{5}{16}+\frac{1}{48}\nf\biggr]\Biggr\}+\mathcal{O}(\api^3)\,. \label{eq:resu:itys}
\end{align}
\end{widetext}
Note that the \NLO{} result for both $\zeta_1^{-1}$ and $\zeta_2^{-1}$ was recently obtained in Ref.~\cite{Crosas:2026ofx} using the 't Hooft-Veltman scheme for $\gamma_5$. A direct comparison is not possible without a dedicated calculation.

In principle, $\zeta_{11}^{-1}$ is both an element of the matching matrix for the $\Delta Q=0$ operators of \cref{eq:DB0_ops}, as well as the matching coefficient for the only relevant $\Delta Q=2$ operator given in \cref{eq:DB2_ops}. However, for the latter case, it is more common
to define~\cite{Buras:1989xd,Buras:2006gb}
\begin{equation}\label{eq:791:66}
  \begin{aligned}
    \mathcal{Q}_{\pm} = \frac{1}{2}\left(\mathcal{Q}_1 \pm \mathcal{Q}_2\right)\,,
  \end{aligned}
\end{equation}
with $\mathcal{Q}_1$ from \cref{eq:DB2_ops}, and
\begin{align}\label{eq:797:68}
    \mathcal{Q}_2 &= (\bar{Q}^i\gamma_\mu(1-\gamma_5)q^j)
    (\bar{q}^j\gamma^\mu(1-\gamma_5)Q^i)\\& =
    2\,(\bar{Q}\gamma_\mu (1-\gamma_5)T^Aq)(\bar{q}\gamma^\mu(1-\gamma_5)T^AQ)
    +\frac{1}{\nc}\mathcal{Q}_1\,,\notag
\end{align}
where $i,j$ are color indices, and we used the SU($\nc$) identity
\begin{equation}\label{eq:807:26}
  \begin{aligned}
    T^A_{ij}T^A_{kl} = \frac{1}{2}\left(\delta_{il}\delta_{jk}-\frac{1}{\nc}\delta_{ij}\delta_{kl}\right)\,.
  \end{aligned}
\end{equation}
In $D=4$ space-time dimensions, it is $\mathcal{Q}_1=\mathcal{Q}_2=\mathcal{Q}_+$ due to Fierz identities, while $\mathcal{Q}_-$ is evanescent. This also implies $\widetilde{\mathcal{Q}}_1=\widetilde{\mathcal{Q}}_+$, since flowed operators are \UV\ finite. Nevertheless, this change of basis affects the higher orders in the matching relation, which we now write as
\begin{equation}\label{eq:805:75}
  \begin{aligned}
    \mathcal{Q}^\text{\MSbar}_+ =
    \zeta^{-1}_{++}\tilde{\mathcal{Q}}_+=
    \zeta^{-1}_{++}\tilde{\mathcal{Q}}_1
    \,.
  \end{aligned}
\end{equation}
Choosing this non-mixing basis, corresponding to the evanescent operators of \cref{eq:evopscoeffsnonmixing}, we find
\begin{widetext}
\begin{equation}
  \begin{aligned}\label{eq:zetapp}
    \zeta_{++}^{-1}&=1+\api\,\Bigl(-\frac{5}{4}
  +\frac{8}{3}\ln2+2\ln3-\frac{1}{2}\lmut\Bigr)
  +\api^2\,\Biggl\{-\frac{1}{8}c_\chi^{(2)}+\frac{62527}{3600}
  -\frac{85}{288}\nf
  +\left(\frac{409}{288}-\frac{1}{8}\nf\right)\zeta(2)+\frac{367}{24}\ln2 \\
  &\eqindent{1}+\frac{22}{9}\ln^22+8\ln2\ln3-\frac{397}{16}\ln3
  +3\ln^23
-\frac{883}{48}\text{Li}_2\left(1/4\right) \\
&+\lmut\biggl[-\frac{83}{32}+\frac{7}{36}\nf+\left(6-\frac{4}{9}\nf\right)\ln2+\left(\frac{9}{2}-\frac{1}{3}\nf\right)\ln3\biggr]
+\lmut^2\biggl[-\frac{9}{16}+\frac{1}{24}\nf\biggr]\Biggr\}+\mathcal{O}(\api^3)\,.
  \end{aligned}
\end{equation}
\end{widetext}

In Ref.~\cite{Suzuki:2020zue} the \NLO{} result was obtained within the dimensional-reduction scheme for $\gamma_5$; direct comparison requires further calculation beyond the scope of this work.

The calculation of all the matching coefficients has been carried out in general
$R_\xi$ gauge, the independence of the final result on the gauge parameter $\xi$
providing a strong check.
The full result including the dependence on the choice of evanescent operators is provided in the ancillary file; see \cref{app:anc} and Ref.~\cite{SupplementalMaterial}.

\section{Flowed matrix elements}\label{sec:lattice}

In this section, we describe the evaluation of the flowed matrix elements required for the calculation of the bag parameters.

\subsection{Simulation details}\label{sec:simulation}

Our work is based on six ensembles of gauge field configurations generated by the \abbrev{RBC/UKQCD} collaboration \cite{Allton:2008pn,Aoki:2010dy,Blum:2014tka,Boyle:2018knm}. These 2+1 flavor ensembles feature the effect of degenerate up/down (light) and strange quarks in the sea sector, and simulations are performed using the Iwasaki gauge action \cite{Iwasaki:1983iya} and Shamir domain-wall fermions (\abbrev{DWF}) \cite{Kaplan:1992bt,Shamir:1993zy,Furman:1994ky}. Details of the ensembles are summarized in~\cref{tab:confs}.
\begin{table*}[tp]
    \centering
    \begin{tabular}{l@{~~~}c@{~~~}c@{~~~}c@{~~~}c@{~~~}c@{~~~}cccccccc}
        \hline\hline & L/a & T/a & $a^{-1}[\gev]$ & $M_\pi[\mev]$ & $am_s^{\rm sea}$ & $am_s^\text{val}$ & $am_c$
                      & $N_\text{src}\times\text{N}_{\text{conf}}$ & $\sigma$ & $N_\sigma$ & $\Delta T_{\rm min}$ \\\hline\hline
        C1 & 24 & 64 & 1.7848(50) & 339.8(1.2) & 0.04\phantom{144} & 0.03224 & 0.64& $32\times101$ & 4.5 & 400 & 20 \\
        C2 & 24 & 64 & 1.7848(50) & 430.6(1.4) & 0.04\phantom{144} & 0.03224 & 0.64 & $32\times101$ & 4.5 & 400 & 20 \\[1.2ex]
        M1 & 32 & 64 & 2.3833(86) & 303.6(1.4) & 0.03\phantom{144} & 0.02477 & 0.45 & $32\times\phantom{0}79$ & 6.5 & 400 & 22 \\
        M2 & 32 & 64 & 2.3833(86) & 360.7(1.6) & 0.03\phantom{144} & 0.02477 & 0.45 & $32\times\phantom{0}89$ & 6.5 & 100 & 22 \\
        M3 & 32 & 64 & 2.3833(86) & 410.8(1.7) & 0.03\phantom{144} & 0.02477 & 0.45 & $32\times\phantom{0}68$ & 6.5 & 100 & 22 \\[1.2ex]
        F1S & 48 & 96 & 2.785(11) & 267.6(1.3) & 0.02144 & 0.02167 & 0.37 & $24\times\phantom{0}98$ & & & 28 
        \\\hline\hline
    \end{tabular}
    \caption{RBC/UKQCD's $N_f=2+1$ Shamir \abbrev{DWF} ensembles with Iwasaki gauge action \cite{Allton:2008pn,Aoki:2010dy,Blum:2014tka,Boyle:2018knm} characterized by inverse lattice spacing ($a^{-1}$), unitary pion mass ($M_\pi$), sea strange ($am_s^{\rm val}$), valence strange ($am_s^\text{val}$), and charm ($am_c$) quark masses. We use two coarse (C1, C2), three medium (M1, M2, M3) and one fine ensemble (F1S). The Jacobi smearing procedure is applied on the C and M ensembles with smearing width $\sigma$ and $N_\sigma$ iterations. $\Delta T_{\rm min}$ indicates the minimum source separation $\Delta T$ considered to have sufficient resolution of the ground state.}
    \label{tab:confs}
\end{table*}

In the valence sector, we simulate strange quarks using the same \abbrev{DWF} action as in the sea sector, but set the bare mass to its physical value determined in Refs.~\cite{Blum:2014tka,Boyle:2018knm}.
For charm quarks, we use stout-smeared~\cite{Morningstar:2003gk} M\"obius \abbrev{DWF}~\cite{Brower:2012vk} optimized for simulating heavy flavors~\cite{Cho:2015ffa}.
We tune the bare charm quark mass to its physical value using the mass of the $D_s$ pseudoscalar meson~\cite{PDG:2024cfk} and $a^{-1}$ from Refs.~\cite{Blum:2014tka,Boyle:2018knm}.

The setup for our measurements is inspired by the determination of bag parameters for neutral $B$-meson mixing~\cite{Boyle:2018knm}.
All propagators are generated with $\mathbb{Z}_2$ wall sources and we choose an equal-time separation for $N_\text{src}$ source positions as specified in~\cref{tab:confs}.
We suppress excited states by applying iterative Gaussian smearing~\cite{Gusken:1989ad} to strange quark sources using a width $\sigma$ and $N_\sigma$ iterations.

The evolution of gauge and propagator fields is performed using a third-order Runge-Kutta algorithm with step size $\delta\tau/a^2=0.01$~\cite{Luscher:2010iy,Luscher:2013cpa}.
Flowed fields are contracted into two- and three-point correlation functions, initially every 10 steps and later every 40 steps.
The maximum flow times in lattice units are set such that all ensembles are flowed by the same physical distance.

\begin{figure}[t]
    \includegraphics[width=0.8\columnwidth]{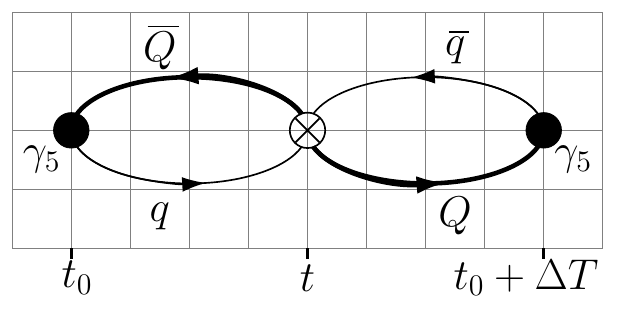}
    \caption{Quark-line diagram to calculate $\Delta Q=2$ operators on the lattice for neutral meson mixing.}
    \label{fig:DB2_diagram}
\end{figure}

\begin{figure}[t]
    \includegraphics[width=0.8\columnwidth]{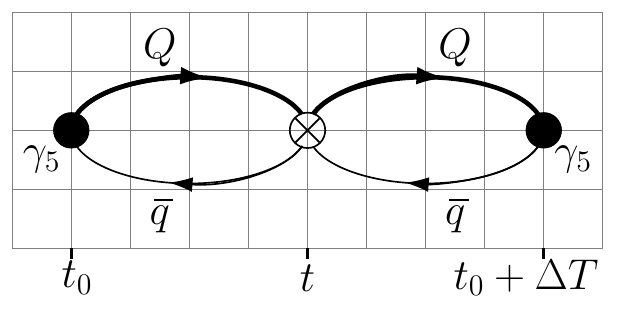}
    \caption{Quark-line diagrams to calculate $\Delta Q=0$ operators on the lattice for valence-quark contributions to lifetimes.}
    \label{fig:DB0_diagram}
\end{figure}

The construction of three-point functions is shown by the quark-line diagrams in~\cref{fig:DB2_diagram,fig:DB0_diagram} where the pseudoscalar meson states are placed at Euclidean times $t_0$ and $t_0+\Delta T$. We contract the four-quark operators at all times $t\in[t_0,t_0+\Delta T]$.
To study the dependence of the signal on the separation of both sources, we perform contractions for multiple choices of $\Delta T$ up to half the temporal extent of the lattice.

We use the \texttt{Grid}~\cite{Grid16} and \texttt{Hadrons}~\cite{HADRONS,Hadrons22} lattice \QCD{} software libraries to carry out our measurements.
The gradient-flow evolution of both the gauge and propagator fields is implemented in
new \texttt{Hadrons} modules\footnote{\href{https://github.com/aportelli/Hadrons/pull/137}{https://github.com/aportelli/Hadrons/pull/137}} including an example run application.\footnote{\href{https://github.com/mbr-phys/HeavyMesonLifetimes}{https://github.com/mbr-phys/HeavyMesonLifetimes}}

Instead of working  with the operators defined in \cref{eq:DB0_ops}, it is more convenient to use a basis for the lattice simulations where the quark pairs on each spin line are color singlets:
\begin{equation}
    \label{eq:DB0_ops_latt}
    \begin{aligned}
        {\cal O}_1 &= (\bar{Q}\gamma_\mu(1-\gamma_5)q)(\bar{q}\gamma_\mu(1-\gamma_5)Q), \\
        {\cal O}_2 &= (\bar{Q}(1-\gamma_5)q)(\bar{q}(1+\gamma_5)Q), \\
        {\cal T}_1 &= (\bar{Q}\gamma_\mu(1-\gamma_5)Q)(\bar{q}\gamma_\mu(1-\gamma_5)q), \\
        {\cal T}_2 &= (\bar{Q}\gamma_\mu(1+\gamma_5)Q)(\bar{q}\gamma_\mu(1-\gamma_5)q).
    \end{aligned}
\end{equation}
Through \cref{eq:807:26}, they are related to the operators defined in \cref{eq:DB0_ops} via
\begin{equation}
  \begin{pmatrix} {\cal O}_1 \\ {\cal O}_2 \\ T_1 \\ T_2 \end{pmatrix} = \begin{pmatrix} 1 & 0 & 0 & 0 \\ 0 & 1 & 0 & 0 \\ -\frac{1}{2\nc} & 0 & -\frac12 & 0 \\ 0 & -\frac{1}{2\nc} & 0 & \frac14 \end{pmatrix}\; \begin{pmatrix} {\cal O}_1 \\ {\cal O}_2 \\ {\cal T}_1 \\ {\cal T}_2 \end{pmatrix}.
\end{equation}

Furthermore, since \QCD{} is parity invariant, we are only interested in the parity-even components of these operators, i.e.\ the terms that involve an even number of $\gamma_5$ matrices. While the basis of \cref{eq:DB0_ops_latt} is used in the simulations, we only show results that have been converted back to the bases defined in \cref{eq:DB0_ops}.

\section{Data Analysis}\label{sec:data}

In the following, we will provide a detailed description of our analysis strategy. In order to test the coherence and reliability of our framework, all numerical analyses described below have been performed using two independent setups: one based on the bootstrap procedure~\cite{Efron:1979bxm}, and the other based on the $\Gamma$ method~\cite{Wolff:2003sm,Schaefer:2010hu,Ramos:2018vgu,Joswig:2022qfe,Bruno:2022mfy}. Both analyses lead to consistent results.

\subsection{Correlator Fits}
While the matrix elements of four-quark operators themselves can be extracted from the three-point correlation functions, the bag parameters can be extracted by comparing the matrix element to its \VIA{} using the ratio of the three-point function over two two-point functions which individually yield the bilinear matrix elements of the \VIA{} for each operator.
We thus form ratios of correlation functions for each flow time $\tau$, reading
\begin{align}
    R_X(t_0,t,\Delta T;\tau) &= \frac{C^{\rm 3pt}_{X}(t_0,t,t_0+\Delta T;\tau)}{\eta_XC_X(t_0,t;\tau)\,C_X(t_0+\Delta T,t;\tau)} \notag \\
    &\quad\to \widetilde{B}_X(\tau)\,, \label{eq:ratio}
\end{align}
where the arrow indicates the limits $t_0\ll t\ll t_0+\Delta T$ and $\Delta T\to\infty$.
The flowed two- and three-point functions are given by
\begin{equation}\label{eq:corrs}
  \begin{aligned}
    C^\text{3pt}_{X}(t_1,t_2,t_3;\tau) &= \langle0| j_P(t_1)\widetilde{X}(t_2;\tau)j_P^\dagger(t_3)|0\rangle\,,\\
    C_X(t_1,t_2;\tau) &= \langle0| \widetilde{j}_X(t_2;\tau)j_P^\dagger(t_1)|0\rangle\,,
  \end{aligned}
\end{equation}
where $t_1 < t_2 < t_3$ are generic positions, and $j_P=\bar{q}\gamma_5 Q$ and $j_X=\bar{q}\Gamma_X Q$ are zero-momentum currents.
Note that we are using $X$ both as an operator and as a subscript in \cref{eq:corrs}.
The associated Dirac structures are $\Gamma_{{\cal Q}_1}=\Gamma_{\mathcal{O}_1}=\Gamma_{T_1}=\gamma_0\gamma_5$, and $\Gamma_{\mathcal{O}_2}=\Gamma_{T_2}=\gamma_5$.
Furthermore, we use the notation $B_i = B_{\mathcal{O}_i}\,, \epsilon_i = B_{{T}_i}$.
The normalization factor $\eta_X$ is given as $\eta_{{\cal Q}_1}=8/3$ and $\eta_{{\cal O}_i}=\eta_{T_i}=1$.
In all cases, the gradient-flow evolution is applied only at the sink operator.
As discussed in the context of~\cref{eq:gff:gapy}, the flowed-quark wave function renormalization $\mathring{Z}_\chi$ cancels in the ratios of \cref{eq:ratio}, and thus the bag parameters extracted at $\tau>0$ are \UV{} finite without any additional renormalization.

\begin{figure*}[tb]
    \includegraphics[width=0.32\textwidth]{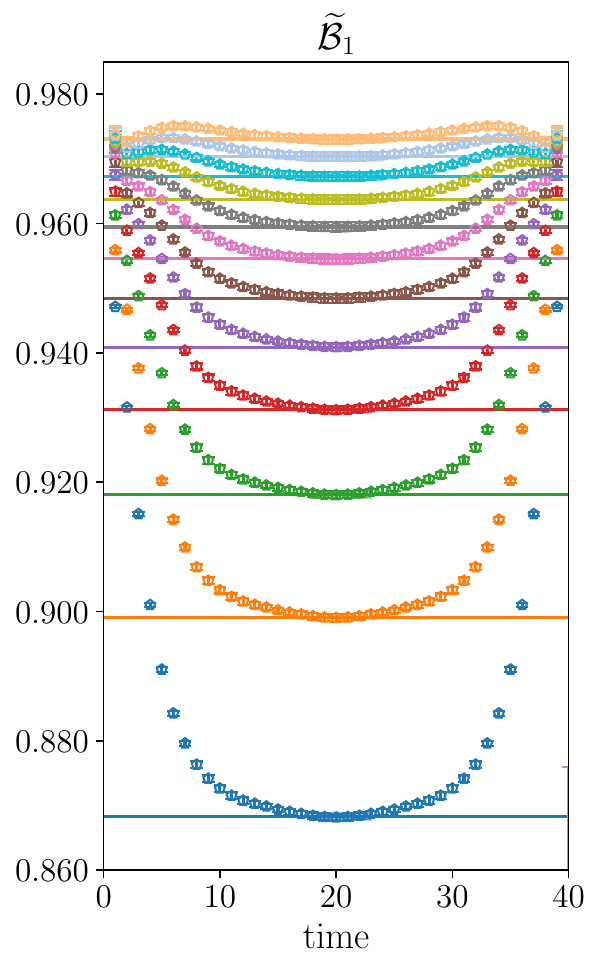}\hfill
    \includegraphics[width=0.32\textwidth]{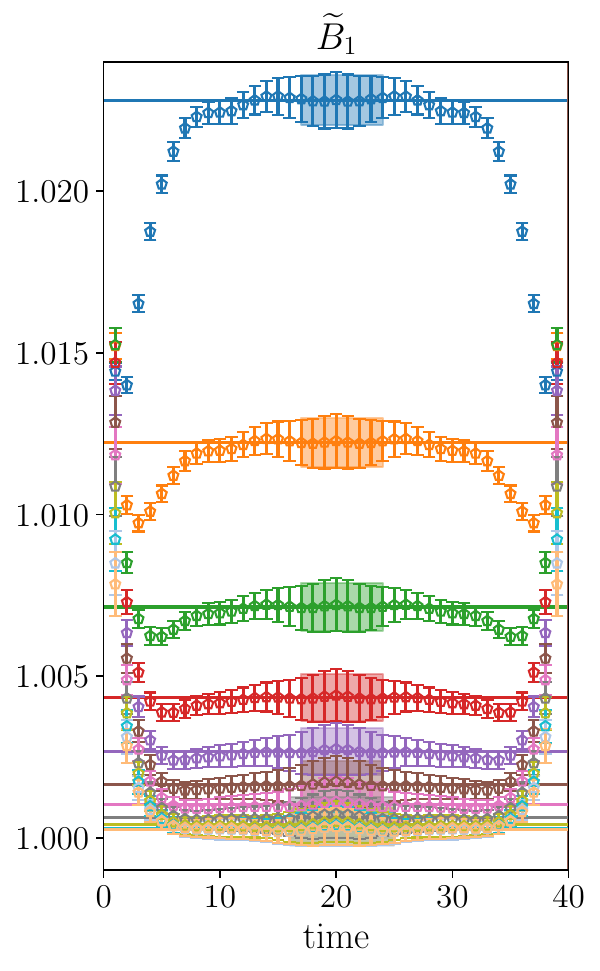}\hfill
    \includegraphics[width=0.32\textwidth]{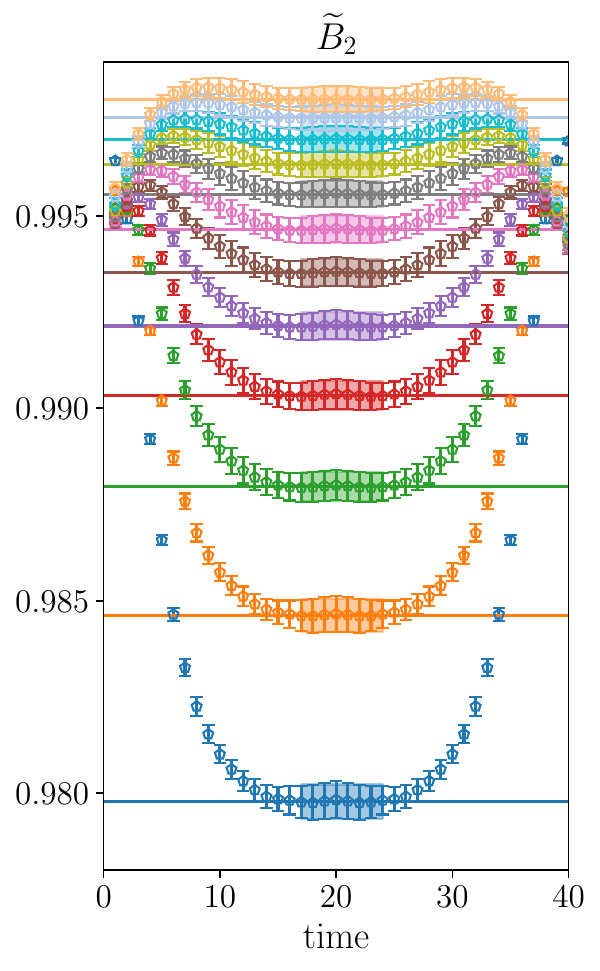}\\
    \includegraphics[width=0.32\textwidth]{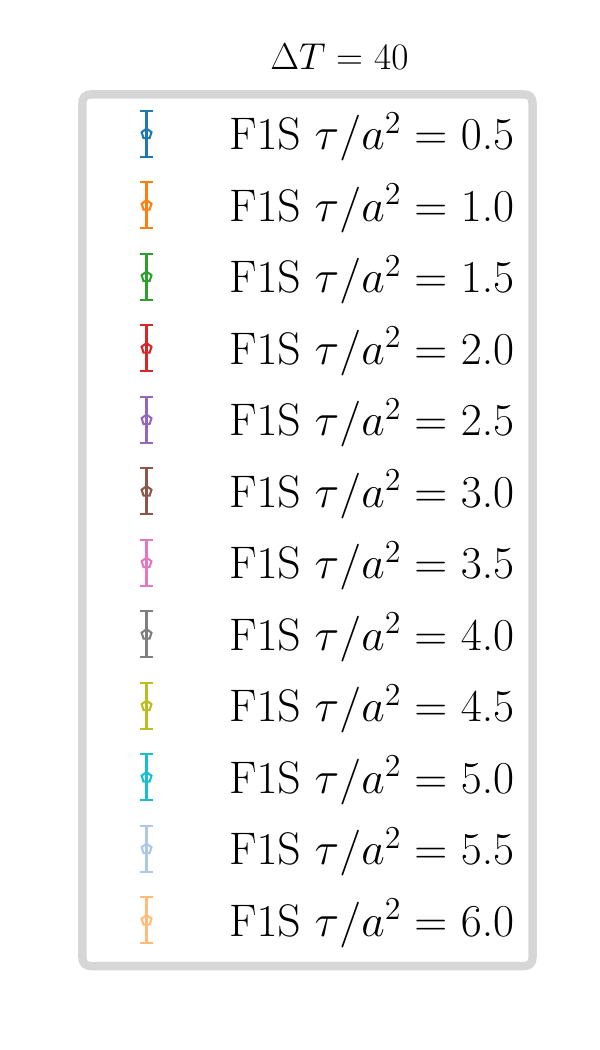}\hfill
    \includegraphics[width=0.32\textwidth]{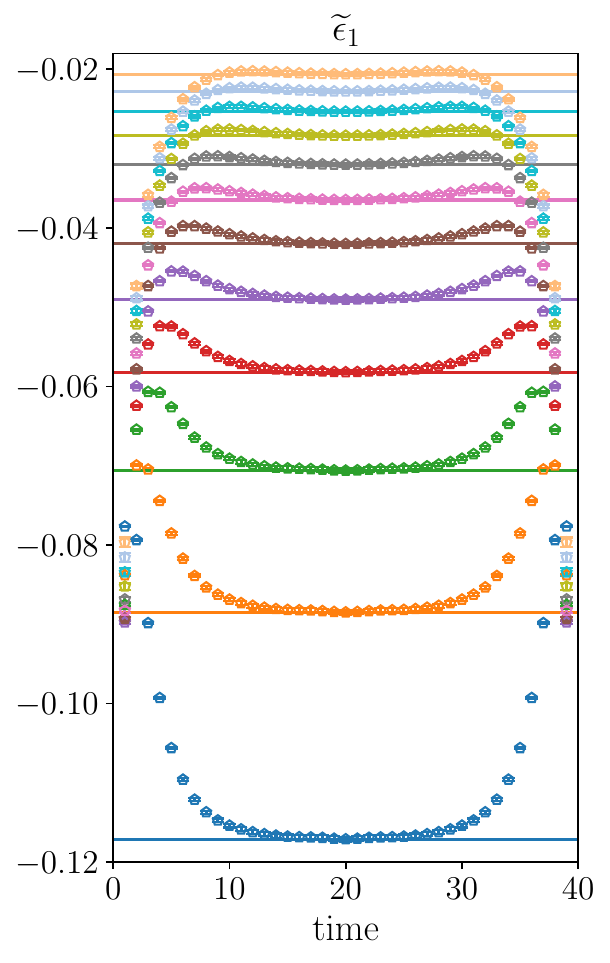}\hfill
    \includegraphics[width=0.32\textwidth]{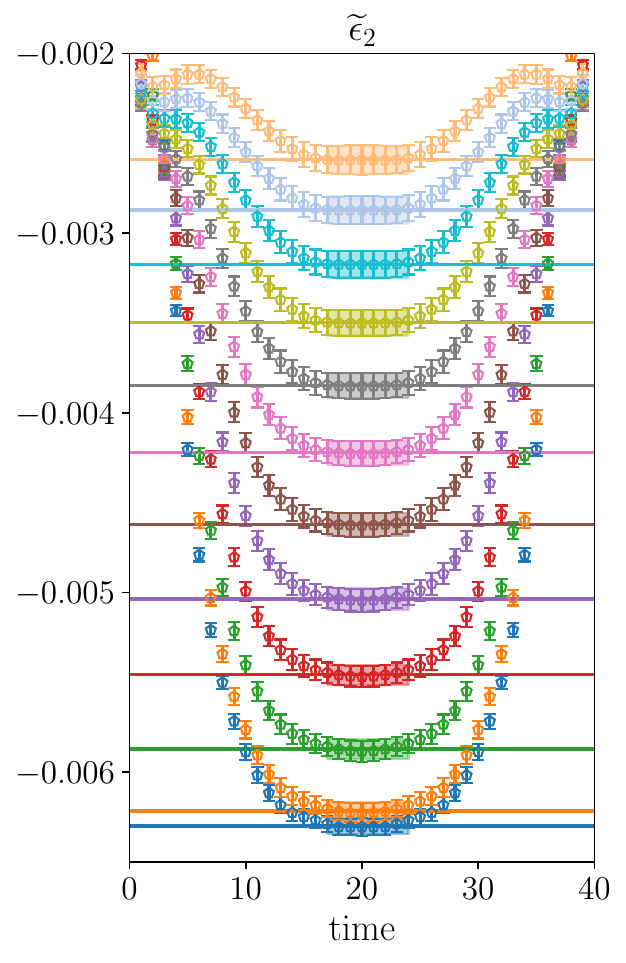}
\caption{Example fits to the ratio $R_X$ (\cref{eq:ratio}) extracting the different bag parameters at various flow times on the F1S ensemble for fixed $\Delta T=40$. }
\label{fig:exampleRatios}
\end{figure*}

\begin{figure*}[p]
    \includegraphics[width=0.45\textwidth]{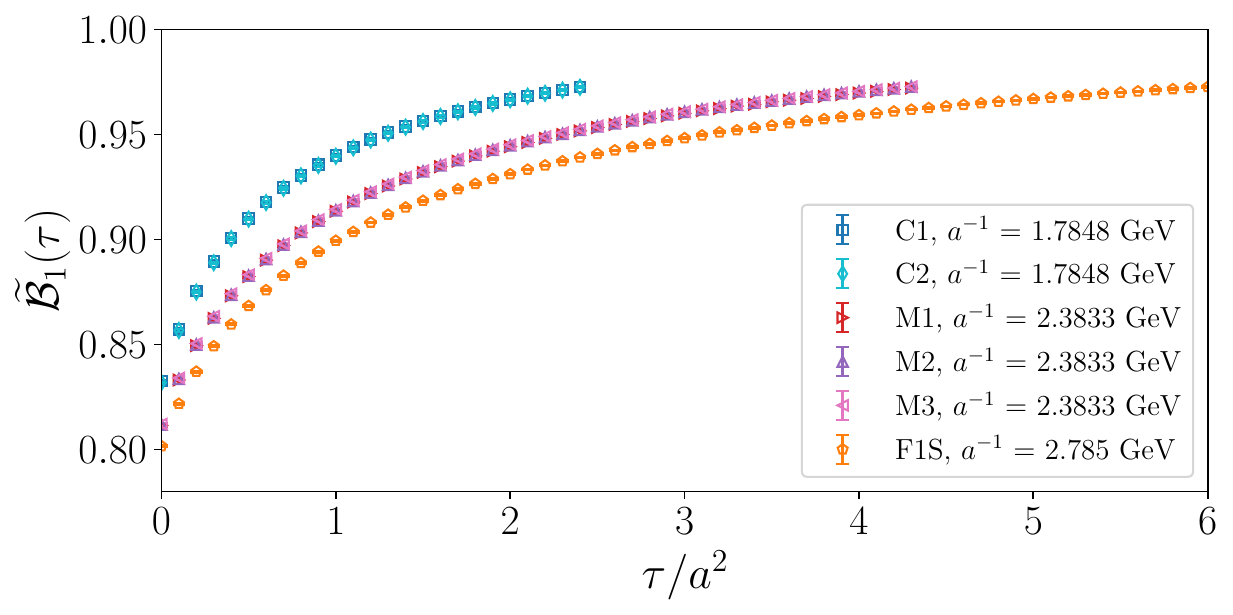}\hfill
    \includegraphics[width=0.45\textwidth]{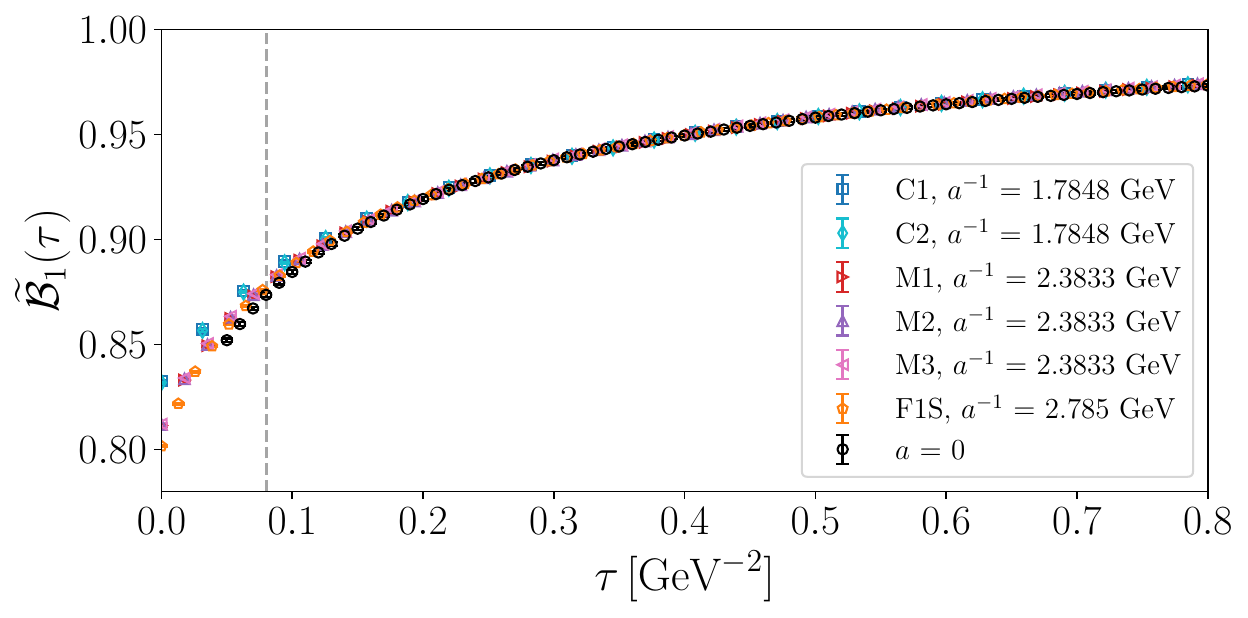}\\
    \includegraphics[width=0.45\textwidth]{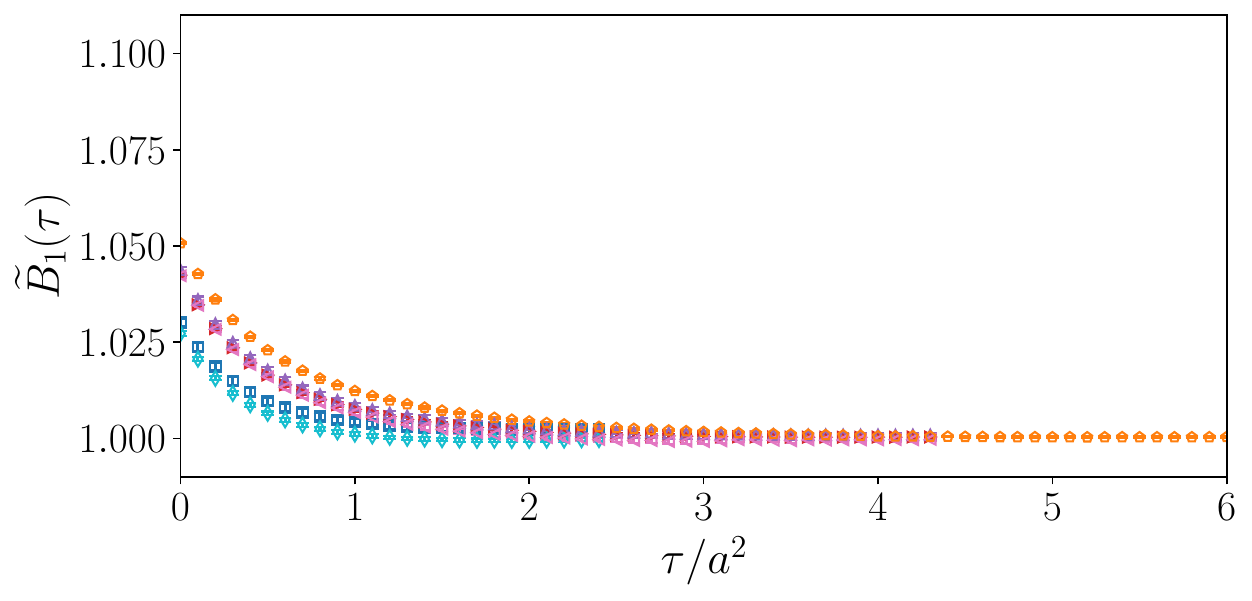}\hfill
    \includegraphics[width=0.45\textwidth]{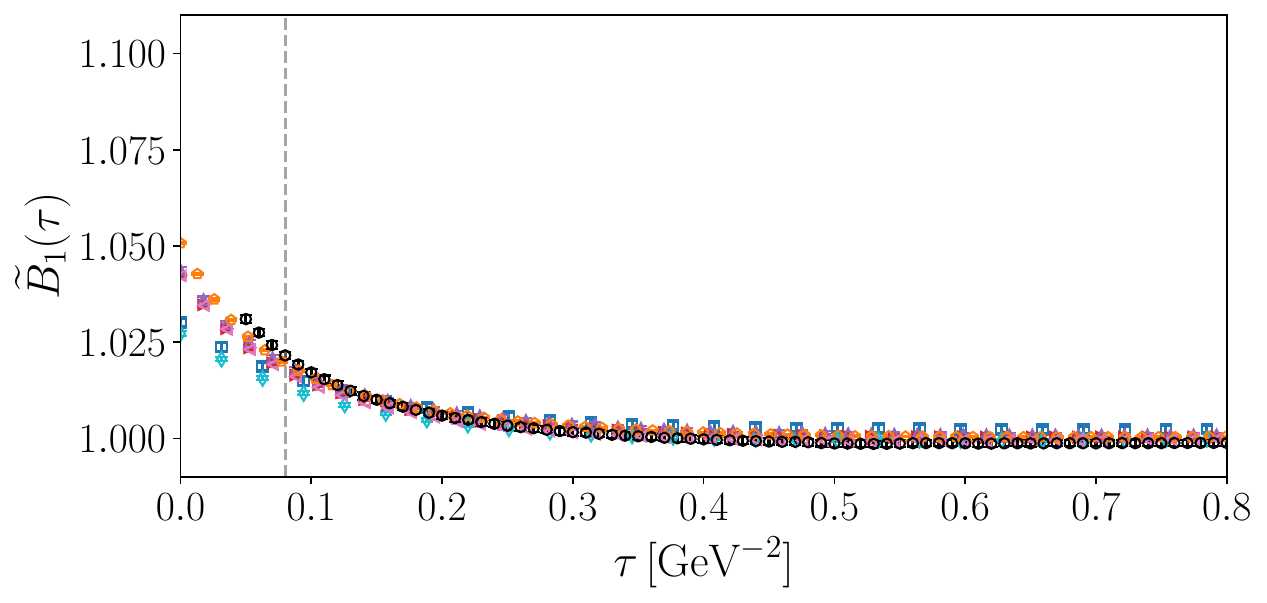}\\
    \includegraphics[width=0.45\textwidth]{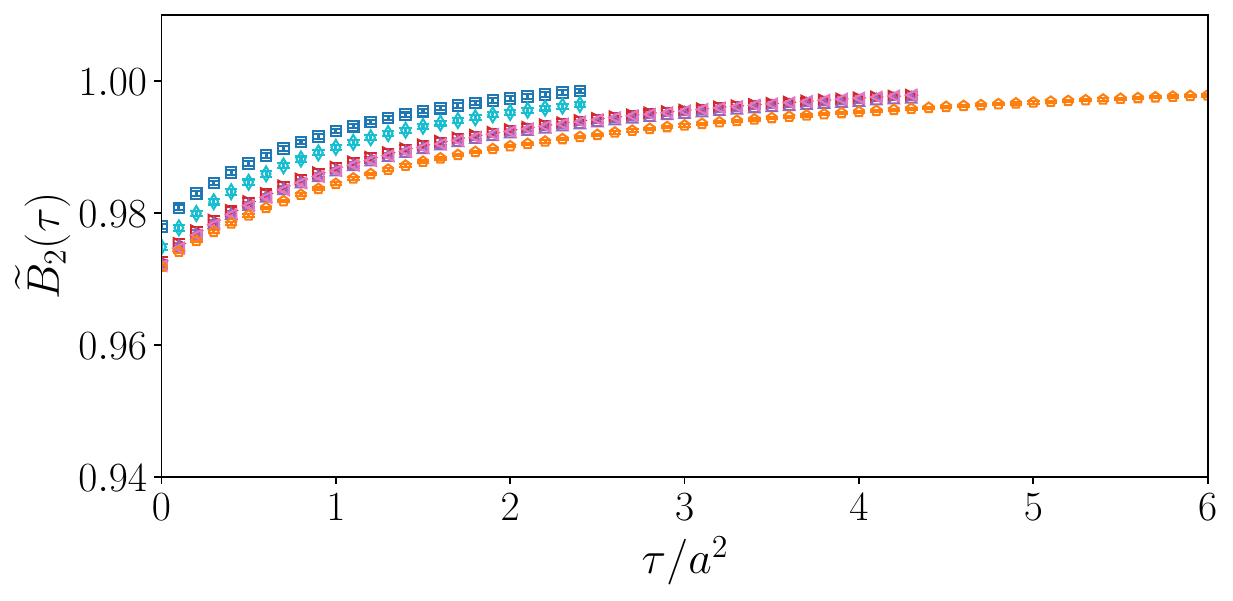}\hfill
    \includegraphics[width=0.45\textwidth]{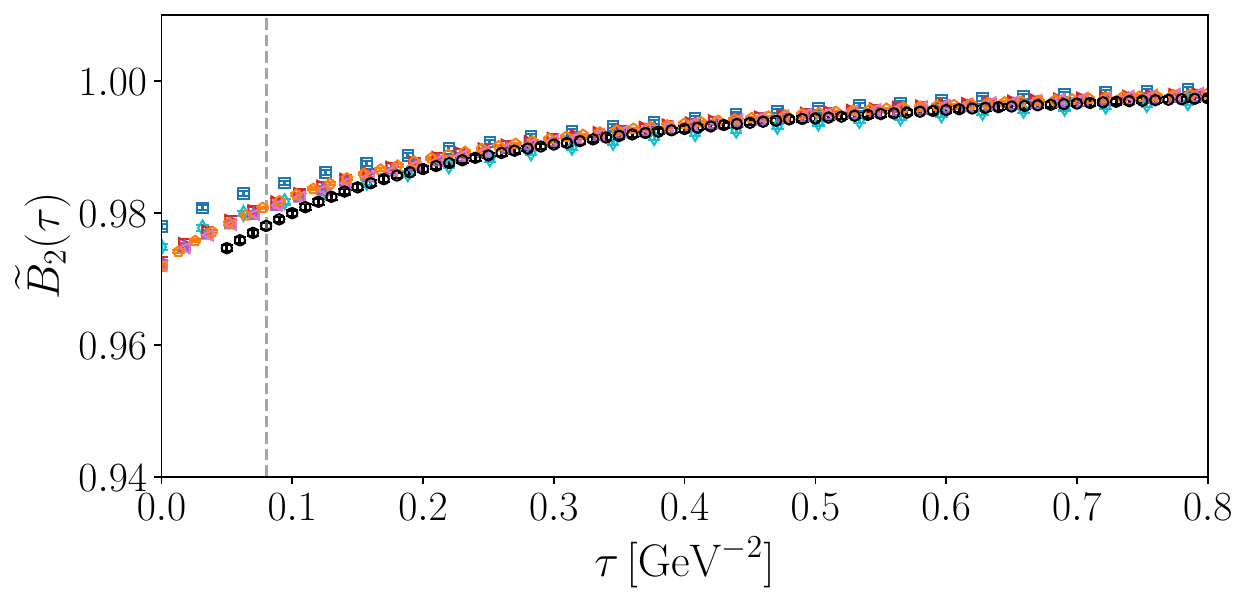}\\
    \includegraphics[width=0.45\textwidth]{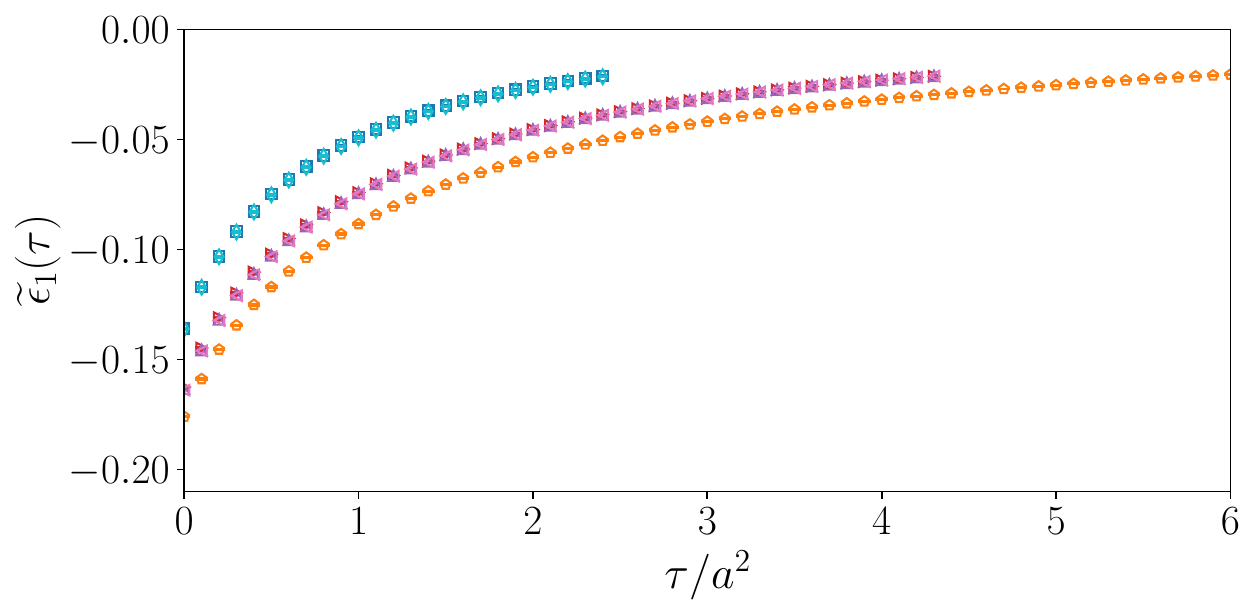}\hfill
    \includegraphics[width=0.45\textwidth]{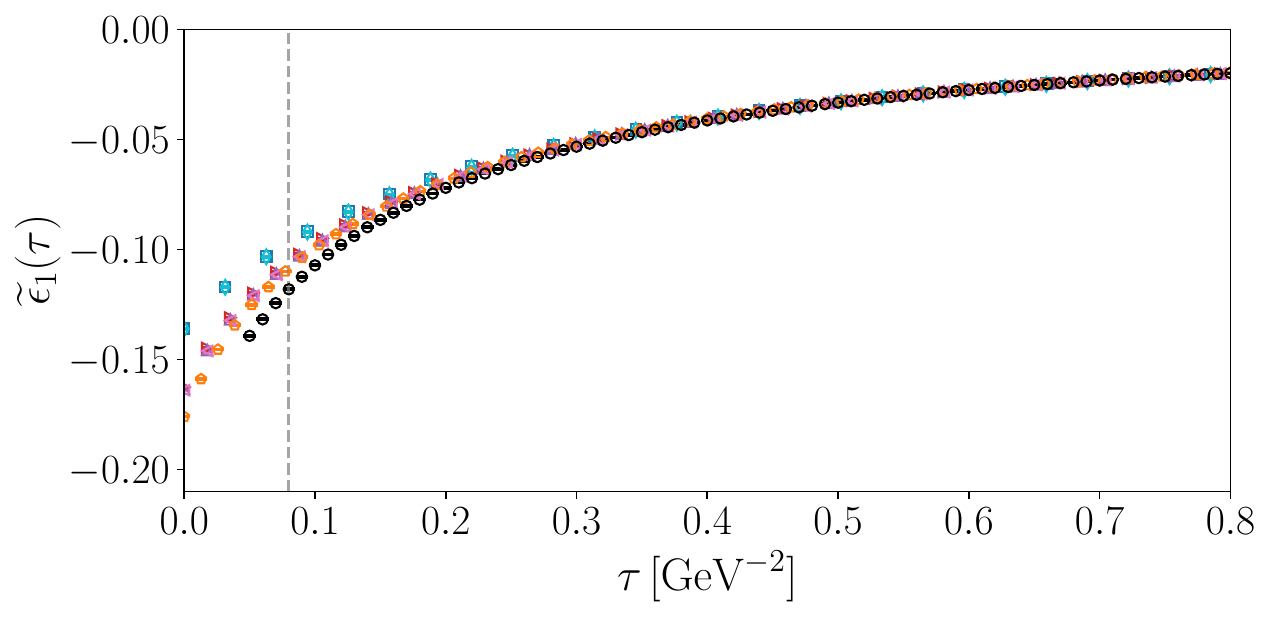}\\
    \includegraphics[width=0.45\textwidth]{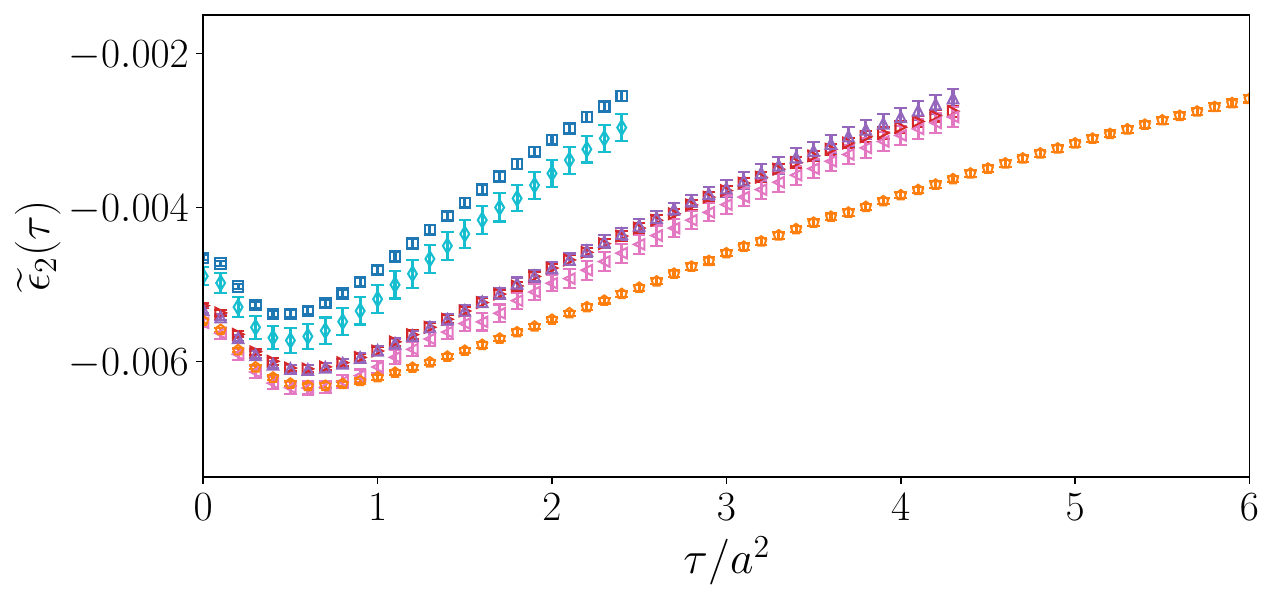}\hfill
    \includegraphics[width=0.45\textwidth]{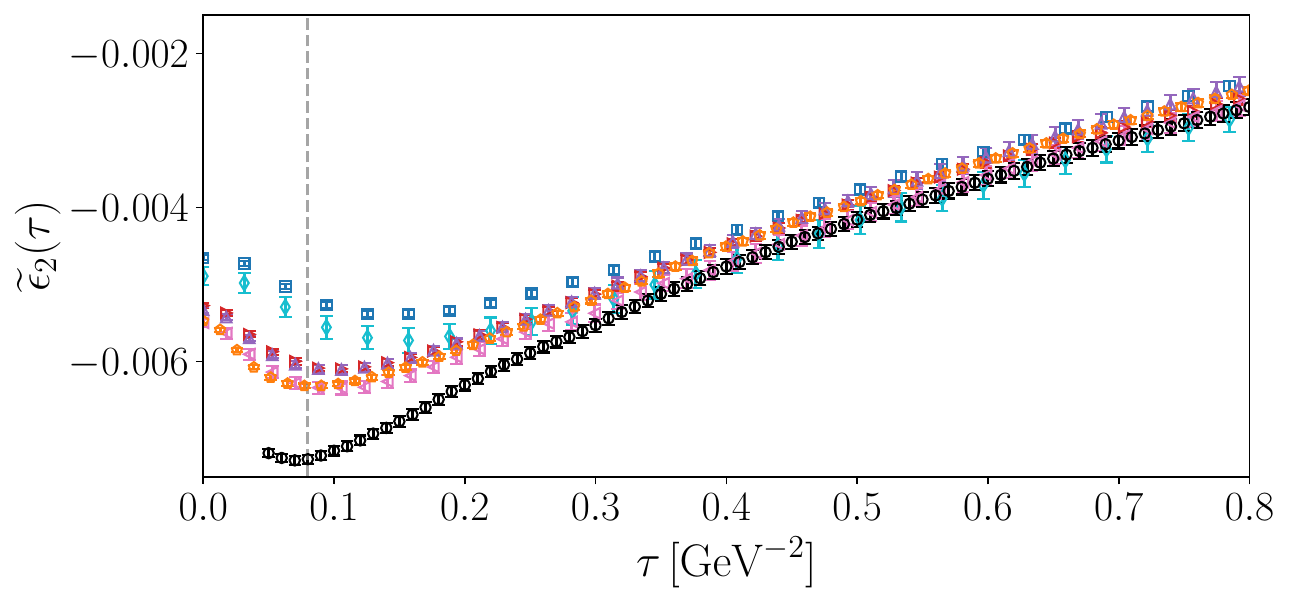}
\caption{Flow-time evolution of the bag parameters. The panels on the left show the bag parameters in lattice units $\tau/a^2$, whereas we show the conversion to physical units $\tau\, [\gev^{-2}]$ in the right panels. The dashed gray lines indicate $\tau_{\rm min}$, the minimum flow time for which the continuum limit is considered reliable, defined relative to our coarsest lattice spacing.}
\label{fig:flowTimeEvolution}
\end{figure*}

In the limit of large Euclidean time separation between sources, these ratios will reach a plateau corresponding to the value of the flowed bag parameter of their inserted operator.
For each flow time, we use \cref{eq:ratio} to extract ground state values of our ratios by performing correlated, simultaneous fits to all separations  $\Delta T\geq\Delta T_{\rm min}$ (listed in~\cref{tab:confs}). Excited states are not statistically resolved.
Examples of the ratio fits at various flow times for each operator are shown in~\cref{fig:exampleRatios}.
The flow-time evolutions of the fitted bag parameters are shown in~\cref{fig:flowTimeEvolution} where the data are plotted first as a function of the lattice flow time $\tau/a^2$ on the left and then in physical flow time [GeV$^{-2}$] on the right.
We see that as the operators are evolved along the gradient flow, the data from different ensembles quickly converge to lie on top of each other at the same physical flow time.

\subsection{Continuum Limits}\label{sec:continuum}

\begin{figure}[t]
    \centering
    \includegraphics[width=0.9\columnwidth]{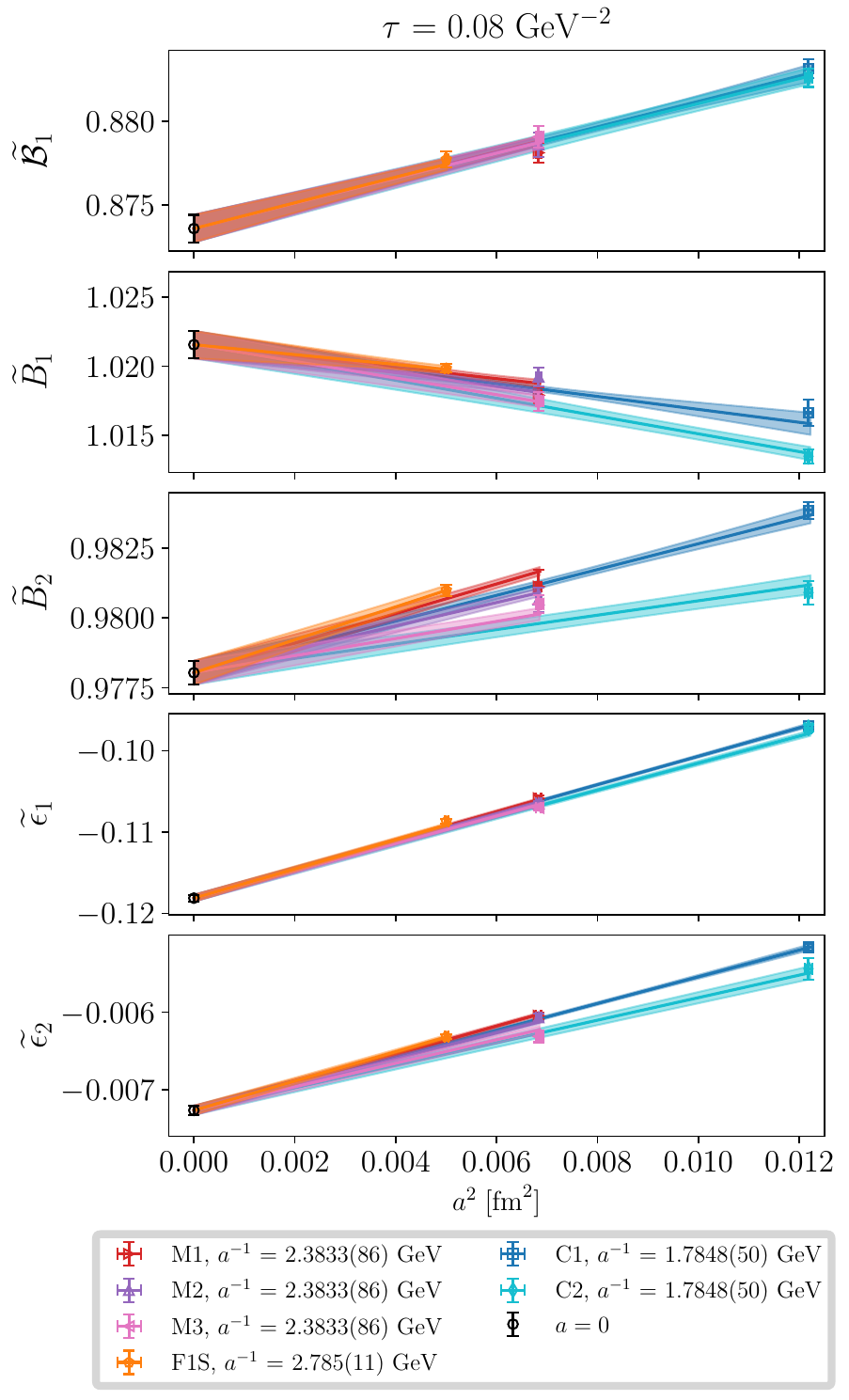}
    \caption{\label{fig:continuum1} Chiral-continuum limit fits for each bag parameter at $\tau=0.08\,{\rm GeV}^{-2}$.
        The black data point is the continuum-extrapolated value at physical pion mass, and the colored bands are the linear $a^2$ relations at each ensemble's pion mass after fitting to~\cref{eq:chiral-cont}.}
\end{figure}

\begin{figure}[t]
    \centering
    \includegraphics[width=0.9\columnwidth]{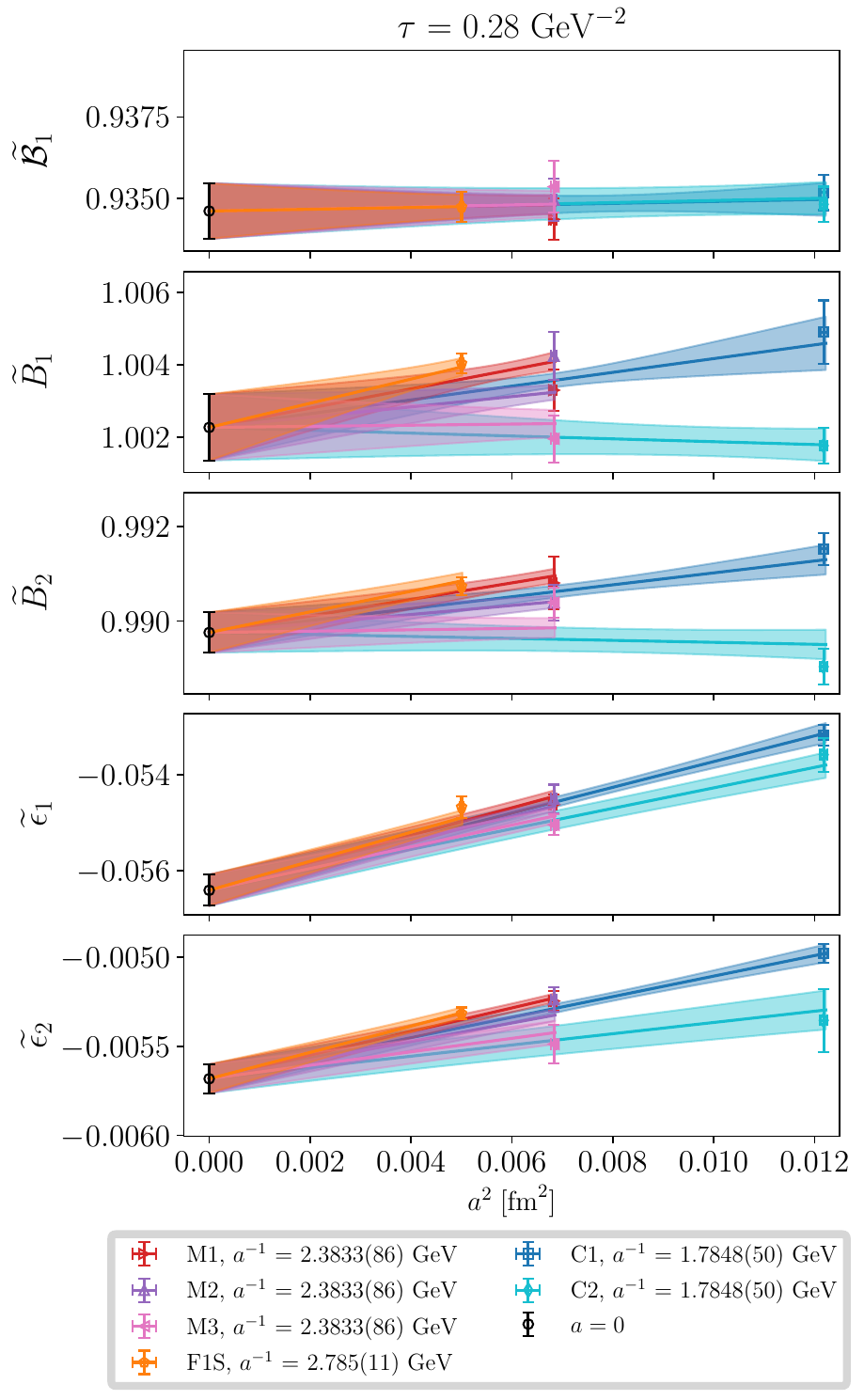}
    \caption{\label{fig:continuum2} Chiral-continuum limit fits for each bag parameter at $\tau=0.28\,{\rm GeV}^{-2}$.
        The black data point is the continuum-extrapolated value at physical pion mass, and the colored bands are the linear $a^2$ relations at each ensemble's pion mass after fitting to~\cref{eq:chiral-cont}.}
\end{figure}

In order to take a continuum limit at fixed values of the flow time in physical units, we first linearly interpolate the values on each ensemble to a common set of flow times $\tau_\text{phys} =\{0.01,\, 0.02,\,\ldots,\, 0.80\}$.\footnote{The flow time is technically a continuous variable and any choice of physical flow times is equally valid.}
The continuum limit is then taken at all values $\tau_\text{phys}$, but the $a\to 0$ extrapolation is reliable only for a sufficiently large ``smearing radius'' $\sqrt{8\tau}$ as discussed in \cref{sec:sftx}.
We find that a suitable choice is $8\tau > 2a^2_\text{max}$, where $a_\text{max}$ is the coarsest lattice spacing in the analysis.
In practice, this translates to $\tau\gtrsim\tau_{\min}\equiv0.08\gev^{-2}$. Smaller values of $\tau$ are considered only in order to check for possible systematic effects.
For increasing values of $\tau>\tau_\text{min}$, the continuum-limit extrapolations quickly become very flat, which may also be attributed to the use of \abbrev{DWF}. For our choice of lattice actions, discretization effects are expected to appear at $O(a^2)$.\footnote{The fermionic gradient flow is classically $O(a)$ improved and does not introduce relevant discretization effects beyond those from the 4D theory for $\tau\gtrsim\tau_{\rm min}$.}
Therefore, we choose a linear ansatz in $a^2$ to perform the $a\to 0$ extrapolation.
Although we study only charm-strange mesons, we observe that small differences due to different pion masses in the sea sector are resolved. We therefore parametrize the pion-mass dependence in a chiral-continuum limit using the ansatz
\begin{align}\label{eq:chiral-cont}
    \widetilde{B}_i^{\rm latt}(a^2,M_\pi^{\rm latt};\tau) = &\widetilde{B}_i(\tau) + C\,a^2 \\
        +& D\,a^2\Big[(M_\pi^{\rm latt})^2 - (M_\pi^{\rm phys})^2\Big], \nonumber
\end{align}
where $\widetilde{B}_i^{\rm latt}(a^2,M_\pi^{\rm latt};\tau)$ are the bag parameters extracted on each ensemble from~\cref{eq:ratio}, $\widetilde{B}_i(\tau)$ denotes the extrapolated estimator in the continuum limit at the physical pion mass, $M_\pi^\text{phys}$~\cite{PDG:2024cfk}, while $C$ and $D$ are dimensionful fit parameters describing the leading discretization effects and their dependence on the sea-pion mass, $M_\pi^{\rm latt}$, which is in physical units.
Example chiral-continuum limits at fixed flow times are shown in~\cref{fig:continuum1,fig:continuum2}.

We then define the \SFTX\ at the level of the bag parameters, i.e.
\begin{equation}
  \begin{aligned}\label{eq:D2match}
    {\cal B}_1(\tau,\mu) &= \zeta^{-1}_{{\cal B}_1}(\tau,\mu)\, \widetilde{\cal B}_1(\tau)
  \end{aligned}
\end{equation}
for the mixing, and
\begin{equation}
  \begin{aligned}\label{eq:D0match}
    \begin{pmatrix} B_1(\tau,\mu) \\ \epsilon_1(\tau,\mu) \end{pmatrix} &= \zeta^{-1}_{B,1}(\tau,\mu)\, \begin{pmatrix} \widetilde{B}_1(\tau) \\ \widetilde{\epsilon}_1(\tau) \end{pmatrix}, \\
    \begin{pmatrix} B_2(\tau,\mu) \\ \epsilon_2(\tau,\mu) \end{pmatrix} &= \zeta^{-1}_{B,2}(\tau,\mu)\, \begin{pmatrix} \widetilde{B}_2(\tau) \\ \widetilde{\epsilon}_2(\tau) \end{pmatrix},
  \end{aligned}
\end{equation}
for the lifetimes, where $\widetilde{\cal B}_1(\tau)$, $\widetilde{B}_i(\tau)$, and $\widetilde{\epsilon}_i(\tau)$ represent the continuum-extrapolated bag parameters from the lattice data.
The corresponding matching matrices are given by
\begin{equation}\label{eq:zetas}
  \begin{aligned}
    \zeta_{{\cal B},1} = \zeta_{++}\zeta_A^{-2}\,,\quad
    \zeta_{B,1} = \zeta_1\zeta_A^{-2}\,,\quad
    \zeta_{B,2} = \zeta_2\zeta_P^{-2}\,,
  \end{aligned}
\end{equation}
with $\zeta_1$, $\zeta_2$, and $\zeta_{++}$ evaluated in \cref{sec:matching} and provided explicitly in \cref{eq:resu:irus,eq:resu:itys,eq:zetapp}.
Finally, $\zeta_P$ and $\zeta_A$ are the \SFTX{} matching coefficients for the pseudoscalar and the axial-vector currents, which are available through \NNLO\ \QCD~\cite{Endo:2015iea,Hieda:2016lly,Borgulat:2023xml}.

\section{Zero-flow-time extrapolation}\label{sec:tau0}

Throughout this paper, we use $\alphas(M_Z)=0.1180$~\cite{PDG:2024cfk} as input and evolve it down to $\alphas(\mu=3\,\gev)=0.2535$ at five-loop level, decoupling the bottom quark at $\mu=m_b$ at four-loop level. For any further change of the renormalization scale, we evolve $\alphas$ to the corresponding perturbative order, i.e.\ one-loop running at \NLO\ and two-loop running at \NNLO. All of these operations are done with the help of \texttt{RunDec}~\cite{Chetyrkin:2000yt,Herren:2017osy}.\footnote{A {\tt Python} wrapper for {\tt RunDec} was also utilized, provided by \url{https://github.com/DavidMStraub/rundec-python}.}

\subsection{Fixed perturbative order}

\begin{figure}[t]
    \centering
    \includegraphics[width=0.9\columnwidth]{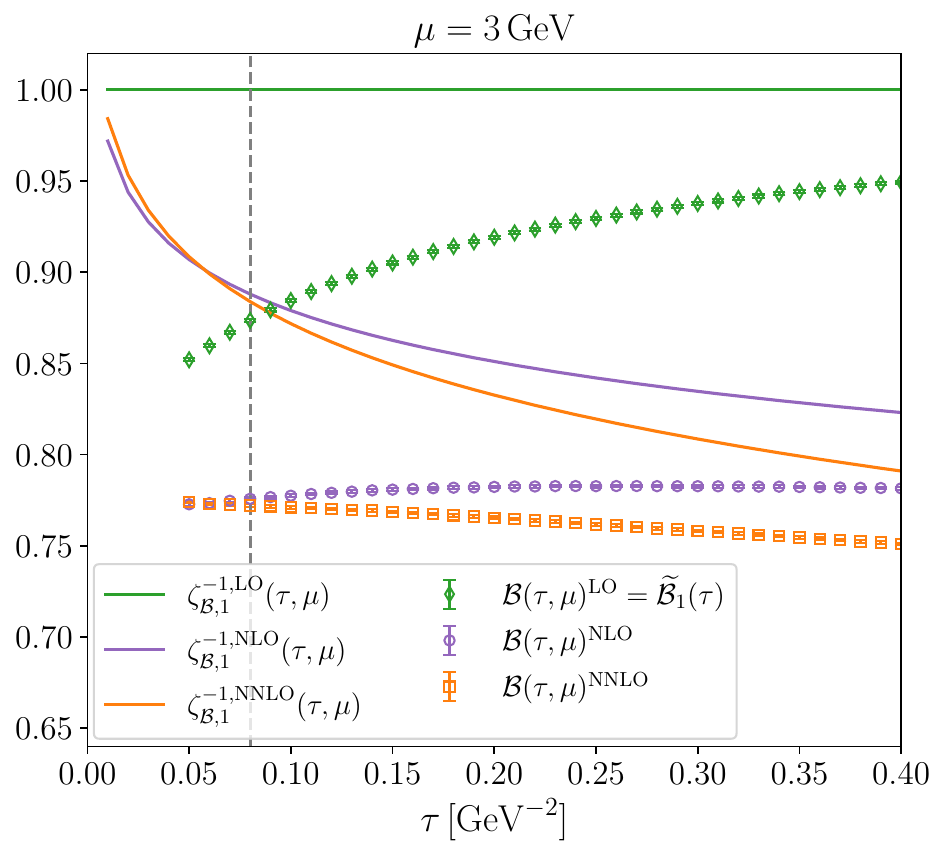}
    \caption{SFTX-matched bag parameter ${\cal B}_1(\tau,\mu=3\gev)$ at \LO\ (green), \NLO\ (purple), and \NNLO\ (orange). The associated matching coefficents $\zeta^{-1}_{{\cal B},1}$ are plotted in solid lines, where the \LO\ matching is 1, and the \NLO\ and \NNLO\ matching coefficients are given by~\cref{eq:zetapp}. he dashed line at $\tau=0.08\gev^{-2}$ indicates the appropriate $\tau_{\rm min}$ for our dataset. }\label{fig:D2B1_sketch}
\end{figure}

The \SFTX-matched bag parameters of \cref{eq:D2match,eq:D0match} -- which still depend on both the renormalization scale $\mu$ and the flow time $\tau$ -- are related to the \MSbar\ bag parameters by taking the $\tau\to0$ extrapolation, i.e.
\begin{equation}\label{eq:limtau0}
    {\cal B}_1^{\MSbar}(\mu) = \lim_{\tau\to0} {\cal B}_1(\tau,\mu),
\end{equation}
and similarly for the lifetime bag parameters $B_i,\,\epsilon_i$.
The limit of $\tau\to0$ is not trivial and involves identifying the correct region from which such an extrapolation can be performed which takes into account the constraints outlined in \cref{sec:sftx}.
\Cref{fig:D2B1_sketch} illustrates the situation for the example of the mixing bag parameter ${\cal B}_1$ by displaying the individual terms of \cref{eq:D2match}. The markers show data for ${\cal B}_1(\tau,\mu)$ obtained by using the matching coefficient $\zeta^{-1}_{{\cal B},1}$ at \LO\ (green diamonds), \NLO\ (purple circles), and \NNLO\ (orange squares). Since $\zeta^{-1,\text{\LO}}_{{\cal B},1}=1$, the green diamonds are equivalent to the pure lattice data $\tilde{\cal B}_1(\tau)$.

The matching coefficients are also shown separately as solid curves. They suggest reliable perturbative convergence over the full $\tau$ range of the figure, with the \NLO\ corrections amounting to less than 20\%, and the \NNLO\ effects adding less than another $\pm 5\%$. At first sight, this might seem remarkable considering that the flow time reaches up to relatively large values. However, it is important to note that we perform the matching at $\mu=3\,\gev$, which sets the scale for the strong coupling $\alphas(\mu)$, while the flow time occurs only via $\lmut$; cf.~\cref{eq:lmut}. At $\mu=3\,\gev$ and $\tau=0.4\,\gev^{-2}$, its numerical value is $\lmut\approx 2.55$, which apparently is sufficiently small for the matching coefficients to remain perturbatively stable.

The plot clearly exhibits the basic features expected from the \SFTX. While both
the lattice data for $\widetilde{\cal B}_1(\tau)$ as well as the higher-order
matching coefficients are logarithmically divergent as $\tau\to 0$, these
divergences formally cancel in the ratio ${\cal B}_1(\tau,\mu)$. Obviously, in
practice this cancellation is incomplete due to the finite perturbative order of
the matching coefficients as well as discretization effects when
$\sqrt{8\tau}\lesssim a$. Nevertheless, the (partial) cancellation is effective
over quite a large range of $\tau$, resulting in an approximately linear
behavior of ${\cal B}_1(\tau,\mu)$ in $\tau$ due to the truncation of these
terms in the \SFTX; see \cref{eq:sftx}. Moreover, the \NLO\ and \NNLO\ values
for ${\cal B}_1(\tau,\mu)$ are reasonably compatible within the displayed region
of $\tau$.

Based on these qualitative observations, it is the purpose of this section to work out a robust procedure for the extrapolation of the quantities defined in \cref{eq:D2match,eq:D0match} to $\tau=0$ which reliably takes into account the possible sources of systematic uncertainties.

Our goal is to derive the bag parameters in the \MSbar\ scheme at a certain reference value $\mu=\mu_0$. In principle, we can do this by inserting the (inverse) matching matrices at this value, $\zeta^{-1}(\tau,\mu_0)$, into \cref{eq:D2match,eq:D0match}.
However, it is equally valid to start from a different scale $\mu$ and evaluate the matching matrix at $\mu_0$ according to its renormalization-group (\RG) evolution:
\begin{equation}\label{eq:regRGE}
  \begin{aligned}
    \mu^2\frac{\rm d}{{\rm d}\mu^2}\zeta(\tau,\mu) = \zeta(\tau,\mu)\gamma(\alphas(\mu))\,,
  \end{aligned}
\end{equation}
with $\gamma$ the anomalous dimension in the \MSbar\ scheme of
$B_X$~\cite{Buras:1989xd,Ciuchini:1997bw,Buras:2000if,Aebischer:2025hsx},
provided in \cref{app:ren}. 
We denote the resulting matching matrix as $\zeta(\tau,\mu\to\mu_0)$.
Different choices of the initial value of $\mu$ typically change the $\tau$ dependence of $\zeta^{-1}(\tau,\mu\to\mu_0)\widetilde{B}(\tau)$, and therefore the $\tau\to 0$ limit, but critically not the value at $\tau=0$ itself.
Therefore we can better inform the $\tau\to0$ limit by simultaneously fitting to data for multiple values of $\mu$, where the flow-time dependence of each $\mu$ value can be parametrized separately.
In practice, we choose $\mu_0=3\gev$ and set $\nf=4$ as the number of active flavors at this scale.
For the set of $\mu\in\{2.5,3.0,3.5,4.0,4.5\}\gev$
we then perform a combined $\tau\to0$ extrapolation.
Our final results for ${\cal B}_1(\mu_0)$, $B_i(\mu_0)$, and $\epsilon_i(\mu_0)$ are obtained by performing a single global fit to $B(\mu\to\mu_0)$ for all $\mu$ using the functional form
\begin{equation}\label{eq:zerotau}
  \begin{aligned}
    f(\tau,\mu) = c_0 + \tau \left(c_1^{(\mu)} + c_2^{(\mu)}\ln (\mu^2\tau)\right)\,,
  \end{aligned}
\end{equation}
where the $c_{1,2}^{(\mu)}$ can be different for each value of $\mu$, while $c_0$ is a parameter common to all fit functions. This ensures a solution for which the perturbative \RG\ evolution is consistent across the chosen range of $\mu$.

We evaluate all possible fit intervals fixing $\tau_{\rm min}=0.08\gev^{-2}$ (cf.~\cref{sec:continuum}) and ranging up to at most $\tau_{\rm max}=0.4\gev^{-2}$, which is chosen to suppress $O(\tau)$ effects beyond the leading assumptions of the \SFTX{}.\footnote{The value of $\tau_{\rm min}$ is a stricter limit than $\tau_{\rm max}$ and is required by both the lattice spacings and the fixed-order perturbation theory used. Using finer lattice spacings and/or additional perturbative orders will push $\tau_{\rm min}$ to smaller values.}
To further account for bias and/or systematic effects in the choice of $\tau_{\rm min}$, we repeat this procedure for minimum values $\tau_{\rm min}+0.01,0.02,0.03,0.04\gev^{-2}$.
The results are given by identifying the spread of all fit results for $c_0$ with $p$-value $>0.05$ as the symmetrized uncertainty interval for the bag parameters at scale $\mu_0$.

This interval accounts for both statistical and systematical uncertainties of the $\tau\to0$ extrapolation, the latter arising from the finite perturbative order of the matching coefficients as well as the constraints on the allowed values of $\tau$; see \cref{sec:sftx}.

\begin{figure}[t]
    \centering
    \includegraphics[width=0.98\columnwidth]{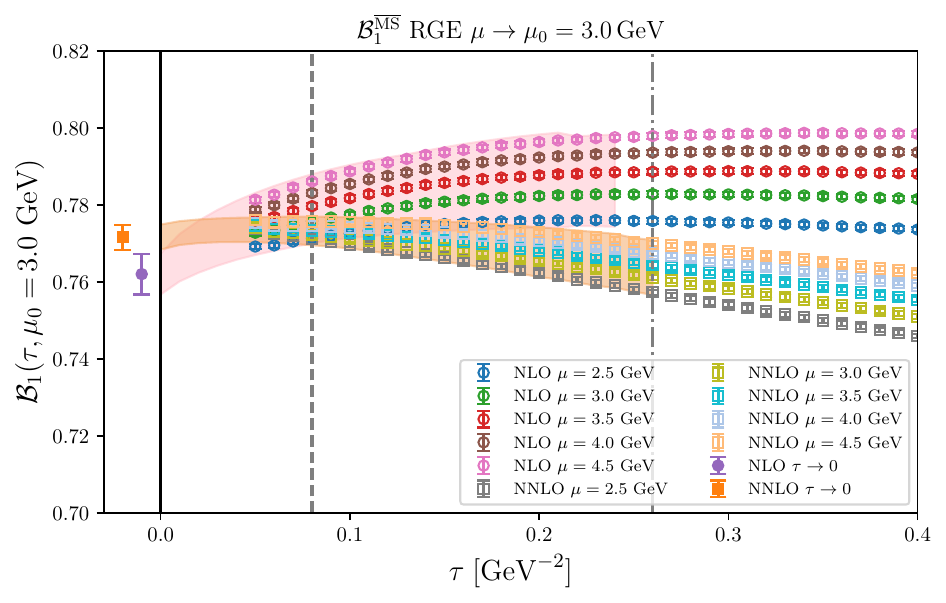}
    \caption{Zero-flow-time extrapolation of the matched flow-time data for the $\Delta Q=2$ ${\cal B}_1$ parameter. The matching to the \MSbar\ scheme is performed at both \NLO\ and \NNLO\ using $\mu\in\{2.5,3.0,3.5,4.0,4.5\}\gev$ before running back to $\mu_0=3\gev$. The pink and orange shaded areas indicate the spread of all acceptable fits to the data at \NLO\ and \NNLO, resulting in the purple and orange data points in the left panel respectively. The dashed line at $\tau=0.08\gev^{-2}$ indicates the appropriate $\tau_{\rm min}$ for our dataset, and the dot-dashed lines indicate the maximum flow time which enters the set of acceptable fits for each operator.}
    \label{fig:Q1_sftx}
\end{figure}

\begin{figure*}[th]
    \centering
    \includegraphics[width=0.48\textwidth]{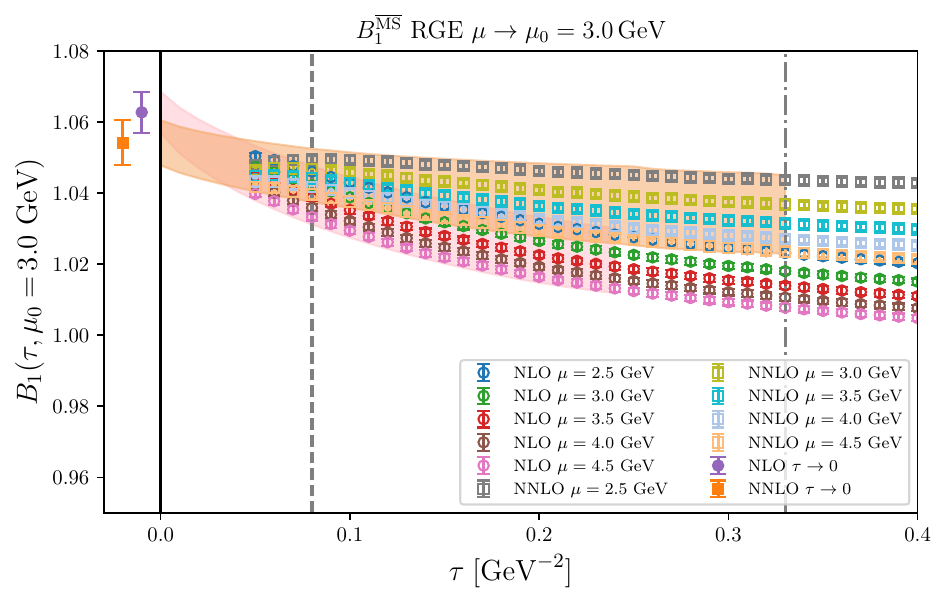}\hfill
    \includegraphics[width=0.48\textwidth]{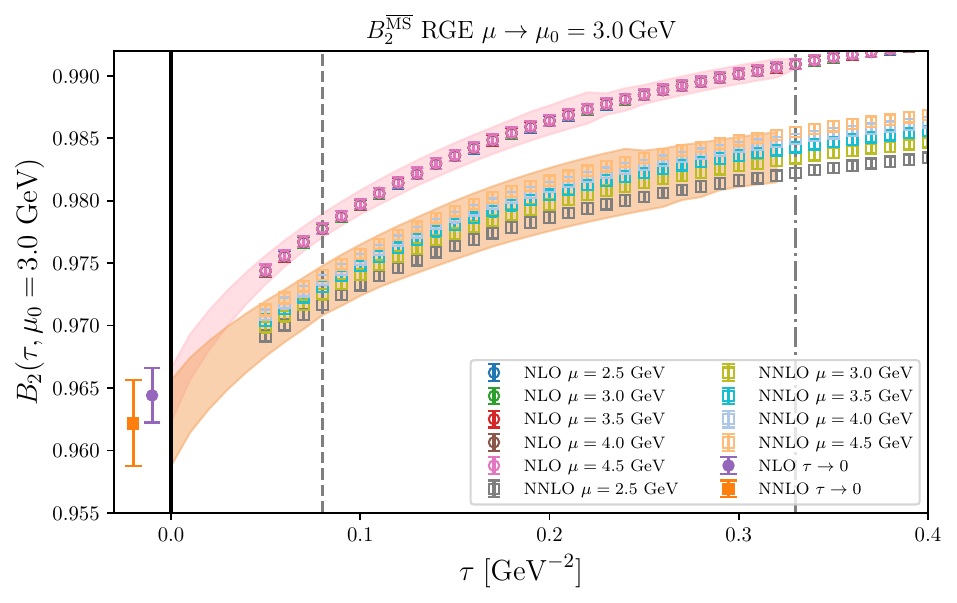}\\
    \includegraphics[width=0.48\textwidth]{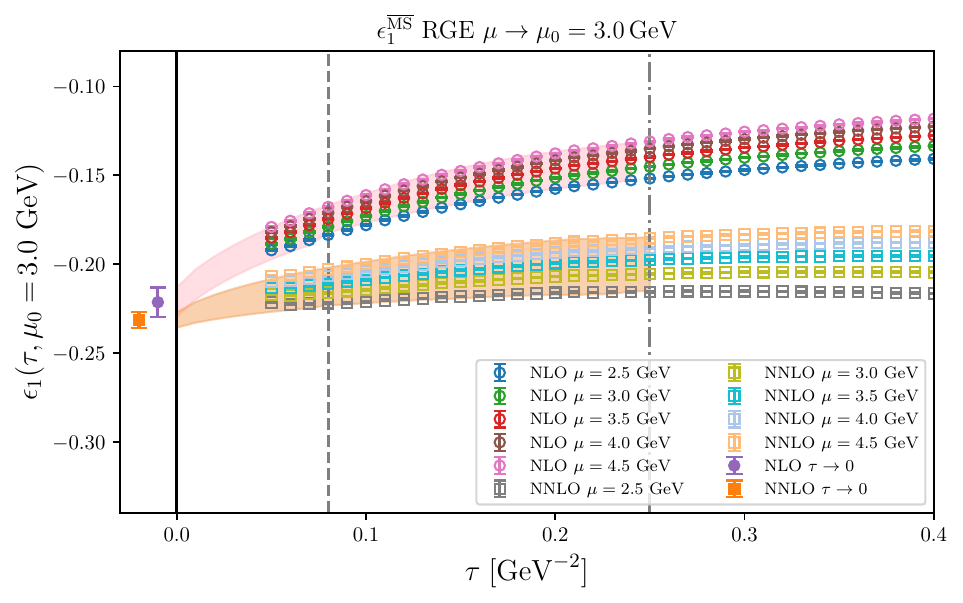}\hfill
    \includegraphics[width=0.48\textwidth]{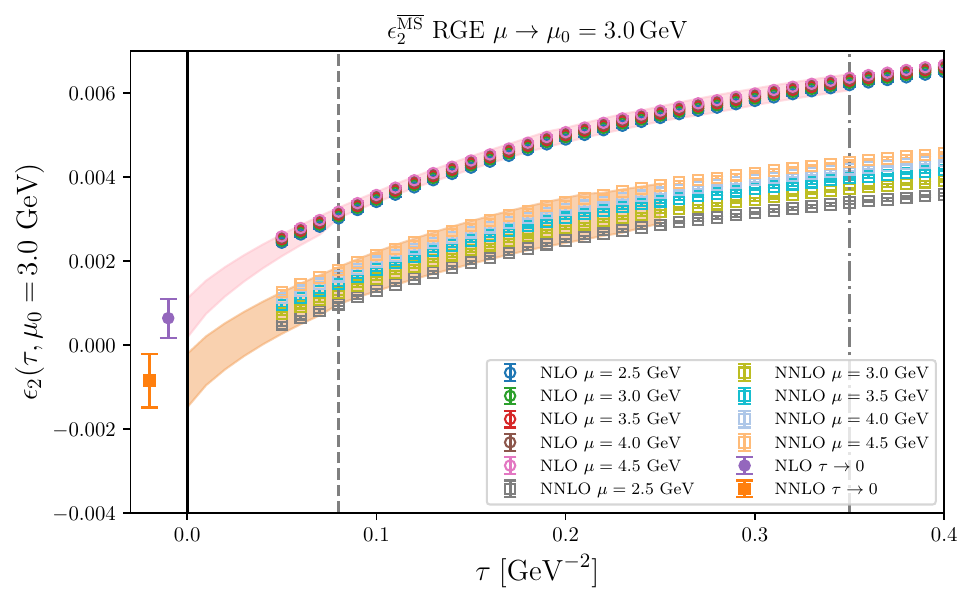}
    \caption{Zero-flow-time extrapolations of the matched flow-time data for the $\Delta Q=0$ bag parameters -- $B_1$ (top left), $B_2$ (top right), $\epsilon_1$ (bottom left), $\epsilon_2$ (bottom right). The matching to the \MSbar\ scheme is performed at both \NLO\ and \NNLO\ using $\mu\in\{2.5,3.0,3.5,4.0,4.5\}\gev$ before running back to $\mu_0=3\gev$. The pink and orange shaded areas indicate the spread of all acceptable fits to the data at \NLO\ and \NNLO\, resulting in the purple and orange data points in the left panel respectively. The dashed line at $\tau=0.08\gev^{-2}$ indicates the appropriate $\tau_{\rm min}$ for our dataset, and the dot-dashed lines indicate the maximum flow time which enters the set of acceptable fits for each operator.}
    \label{fig:DB0_sftx}
\end{figure*}

The $\tau\to0$ extrapolations for the $\Delta Q=2$ and $\Delta Q=0$ operators are shown in~\cref{fig:Q1_sftx,fig:DB0_sftx} respectively.
Next we discuss how we may improve upon this \SFTX{} matching procedure by taking advantage of RG running.

\subsection{Renormalization-group improvement}\label{sec:gammatilde}

The dependence of the flowed operators on the flow time can be parametrized by the \textit{flowed anomalous dimensions} $\widetilde{\gamma}$, which allows for the resummation of their logarithmic $\tau$ dependence. They are defined in Ref.~\cite{Harlander:2020duo} as
\begin{equation}\label{eq:flowRGE}
\tau\frac{{\rm d}}{{\rm d}\tau}\widetilde{\mathcal{O}}(\tau) = \widetilde{\gamma}\widetilde{\mathcal{O}}(\tau)\,,\quad\text{with}\quad\widetilde{\gamma}=\left(\tau\frac{{\rm d}}{{\rm d}\tau}\zeta\right)\zeta^{-1}\,.
\end{equation}
For the operators in~\cref{eq:DB0_ops}, we define the flow-time renormalization-group equation (\RGE) as
\begin{equation}\label{eq:resu:irra}
  \begin{aligned}
  \tau\frac{{\rm d}}{{\rm d}\tau}
    \begin{pmatrix}
      \widetilde{\mathcal{O}}_i(\tau)\\
      \widetilde{T}_i(\tau)
    \end{pmatrix} &=
    \widetilde{\gamma}_i
    \begin{pmatrix}
      \widetilde{\mathcal{O}}_i(\tau)\\
      \widetilde{T}_i(\tau)
    \end{pmatrix}\,,\qquad i\in \{1,2\}\,.
  \end{aligned}
\end{equation}
and then find the anomalous dimensions $\widetilde{\gamma}_i$ to be
\begin{align}
(\widetilde{\gamma}_1)_{11}=&\api^2\,\Bigl[\frac{5}{8}-\frac{1}{18}\nf
  +\left(-\frac{22}{3}+\frac{4}{9}\nf\right)\ln2 \nonumber \\
  &\eqindent{1}+\left(-\frac{11}{2}+\frac{1}{3}\nf\right)\ln3\Bigr]+\mathcal{O}(\api^3)\,, \nonumber \\
(\widetilde{\gamma}_1)_{12}=&\frac{3}{2}\api+\api^2\,\Bigl[\frac{249}{32}-\frac{5}{12}\nf
  +\left(\frac{33}{8}-\frac{1}{4}\nf\right)\lmut\Bigr] \nonumber \\
  &\eqindent{1.7}+\mathcal{O}(\api^3)\,, \nonumber \\
(\widetilde{\gamma}_1)_{21}=&\frac{1}{3}\api+\api^2\,\Bigl[\frac{83}{48}-\frac{5}{54}\nf
  +\left(\frac{11}{12}-\frac{1}{18}\nf\right)\lmut\Bigr] \nonumber \\
  &\eqindent{1.7}+\mathcal{O}(\api^3)\,, \nonumber \\
(\widetilde{\gamma}_1)_{22}=&-\frac{1}{2}\api+\api^2\,\Bigl[-\frac{63}{32}+\frac{1}{12}\nf
  +\left(-\frac{22}{3}+\frac{4}{9}\nf\right)\ln2 \nonumber \\
  &\eqindent{1}+\left(-\frac{11}{2}+\frac{1}{3}\nf\right)\ln3 \nonumber \\
  &\eqindent{1}+\left(-\frac{11}{8}+\frac{1}{12}\nf\right)\lmut\Bigr]+\mathcal{O}(\api^3)\,,
\end{align}
and
\begin{align}
(\widetilde{\gamma}_2)_{11}=&-2\api+\api^2\,\Bigl[-\frac{71}{6}+\frac{1}{2}\nf
  +\left(-\frac{22}{3}+\frac{4}{9}\nf\right)\ln2 \nonumber \\
  &\eqindent{1}+\left(-\frac{11}{2}+\frac{1}{3}\nf\right)\ln3 \nonumber \\
  &\eqindent{1}+\left(-\frac{11}{2}+\frac{1}{3}\nf\right)\lmut\Bigr]+\mathcal{O}(\api^3)\,, \nonumber \\
(\widetilde{\gamma}_2)_{12}=&\frac{15}{32}\api^2+\mathcal{O}(\api^3)\,, \nonumber \\
(\widetilde{\gamma}_2)_{21}=&\frac{5}{48}\api^2+\mathcal{O}(\api^3)\,, \nonumber \\
(\widetilde{\gamma}_2)_{22}=&\frac{1}{4}\api+\api^2\,\Bigl[\frac{389}{192}-\frac{1}{8}\nf
  +\left(-\frac{22}{3}+\frac{4}{9}\nf\right)\ln2 \nonumber \\
  &\eqindent{1}+\left(-\frac{11}{2}+\frac{1}{3}\nf\right)\ln3 \nonumber \\
  &\eqindent{1}+\left(\frac{11}{16}-\frac{1}{24}\nf\right)\lmut\Bigr]+\mathcal{O}(\api^3)\,.
\end{align}
The dependence on the scheme of evanescent operators drops out in $\widetilde{\gamma}$, which also means that the flowed anomalous dimension for the $\Delta Q=2$ operator is $\tilde{\gamma}_{++} = \tilde{\gamma}_{11}$.

In the \SFTX{} matching procedure used above, flow-time logarithms (\cref{eq:lmut}) -- which are naturally present in both the all-orders nonperturbative lattice data and the fixed-order perturbative matching coefficients -- will only cancel up to the available perturbative order.  We therefore expect higher-order logarithms to manifest as curvature at smaller flow times in~\cref{fig:Q1_sftx,fig:DB0_sftx}, and ultimately lead to divergences as $\tau\to0$, which is one of the limiting factors in pushing $\tau_{\rm min}$ in our fits to smaller values; see \cref{sec:sftx}. Our fit function in~\cref{eq:zerotau} is designed to capture part of this logarithmic behavior. Indeed, from inspecting the data we can see that residual logarithmic effects are small, but they may be non-negligible at the precision of our analysis.

One option to estimate these effects is to resum logarithmic terms with the help of \cref{eq:flowRGE}~\cite{Black:2025gft}. This equation enables us to run operators defined at some flow time $\tau$ to a different value $\tau_0$, which can be chosen to improve perturbative behavior of the operators and limit the effects of logarithms. A particularly appealing choice is $\tau_0 = {\rm e}^{-\EulerGamma}/2\mu_0^2$, since this would exactly set the logarithmic terms defined in~\cref{eq:lmut} to zero~\cite{Harlander:2018zpi}.
For the case of the $\Delta Q=2$ parameter ${\cal B}_1$, where the matching matrix is a single number, we can explicitly write this flow-time evolution as
\begin{equation}
    \widetilde{\cal B}_1(\tau\to\tau_0) = \exp\left[-\int_0^{L_{\mu_0\tau}}\widetilde{\gamma}(a_s(\mu_0),L)\,{\rm d}L\right]\,\widetilde{\cal B}(\tau),
\end{equation}
where $\widetilde{\cal B}_1(\tau\to \tau_0)$ is the set of values for $\widetilde{\cal B}_1(\tau_0)$ calculated for each $\tau$.
Having run our continuum results from each $\tau$ to $\tau_0$, we can combine these data with the matching coefficient at $\tau_0$, i.e.
\begin{align}\label{eq:flowRGtau0}
    {\cal B}(\mu_0) &= \lim_{\tau\to0}\zeta^{-1}(\tau_0,\mu_0)\widetilde{\cal B}_1(\tau\to \tau_0)
\end{align}
where we expect decreased flow-time dependence and thus an easier $\tau\to0$ extrapolation.

This procedure defines the \textit{\RG-improved} \SFTX{} and $\tau\to0$ extrapolation. The \RG\ improvement of the $\Delta Q=0$ operators follows analogously, however since the flowed anomalous dimensions are now matrices, we do not write an analytic expression for the evolution to $\tau_0$ but directly integrate~\cref{eq:flowRGE} numerically.

\begin{figure}[t]
    \centering
    \includegraphics[width=0.98\columnwidth]{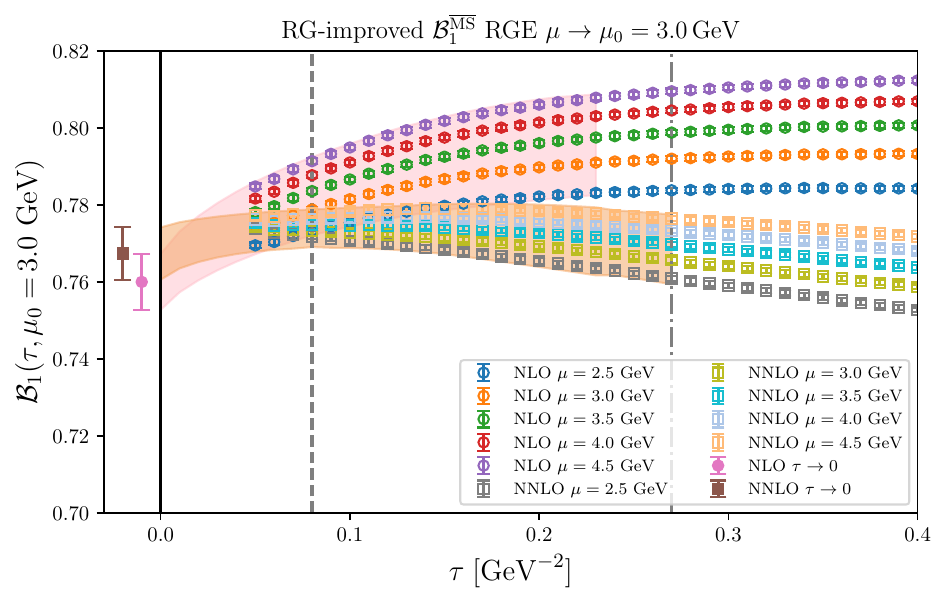}
    \caption{Zero-flow-time extrapolation of the \RG-improved matched flow-time data for the $\Delta Q=2$ ${\cal B}_1$ parameter. The matching to \MSbar\ is performed at both NLO and NNLO using $\mu\in\{2.5,3.0,3.5,4.0,4.5\}\gev$ before running back to $\mu_0=3\gev$. The flow-time \RGE\ is used to set flow-time logarithms to zero and improve the extrapolations. The pink and orange shaded areas indicate the spread of all acceptable fits to the data at NLO and NNLO, resulting in the pink and brown data points in the left panel. The dashed line at $\tau=0.08\gev^{-2}$ indicates the appropriate $\tau_{\rm min}$ for our dataset, and the dot-dashed lines indicate the maximum flow time which enters the set of acceptable fits for each operator.}
    \label{fig:Q1_sftx_RG}
\end{figure}

\begin{figure*}[th]
    \centering
    \includegraphics[width=0.48\textwidth]{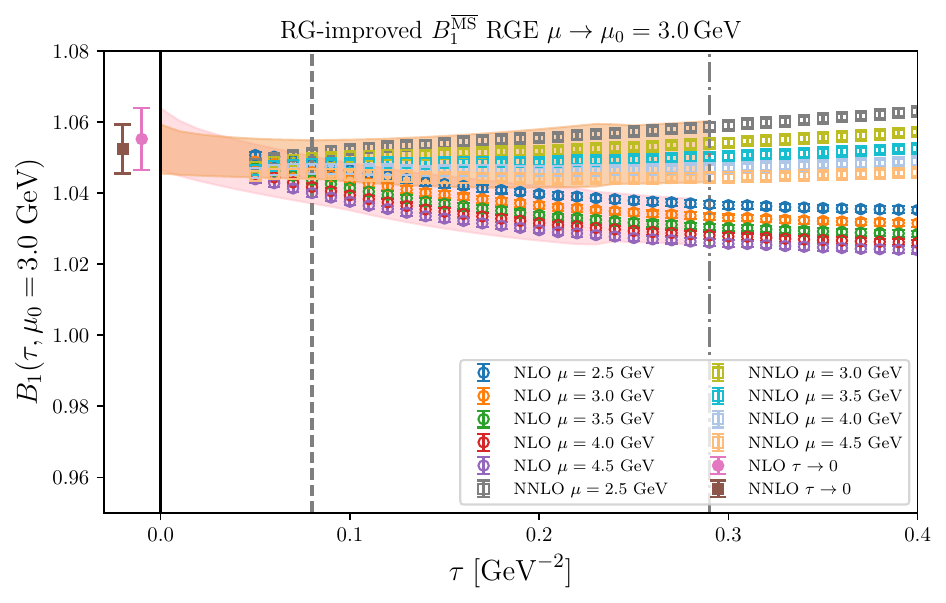}\hfill
    \includegraphics[width=0.48\textwidth]{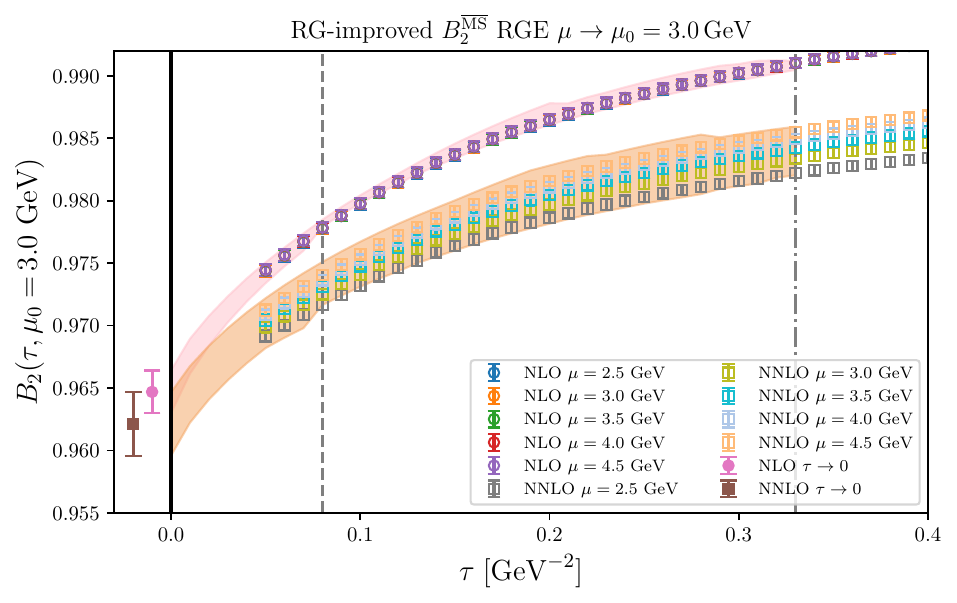}\\
    \includegraphics[width=0.48\textwidth]{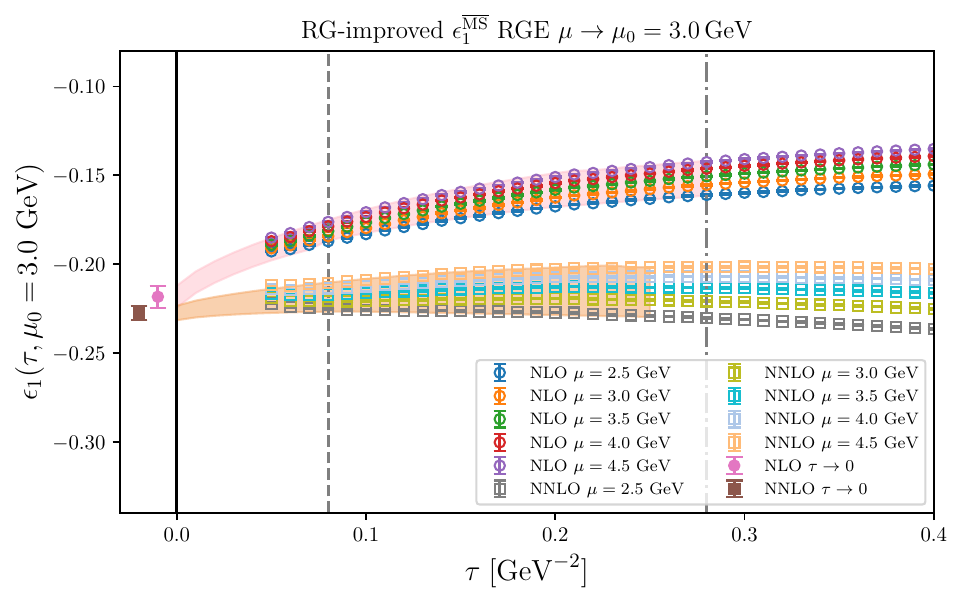}\hfill
    \includegraphics[width=0.48\textwidth]{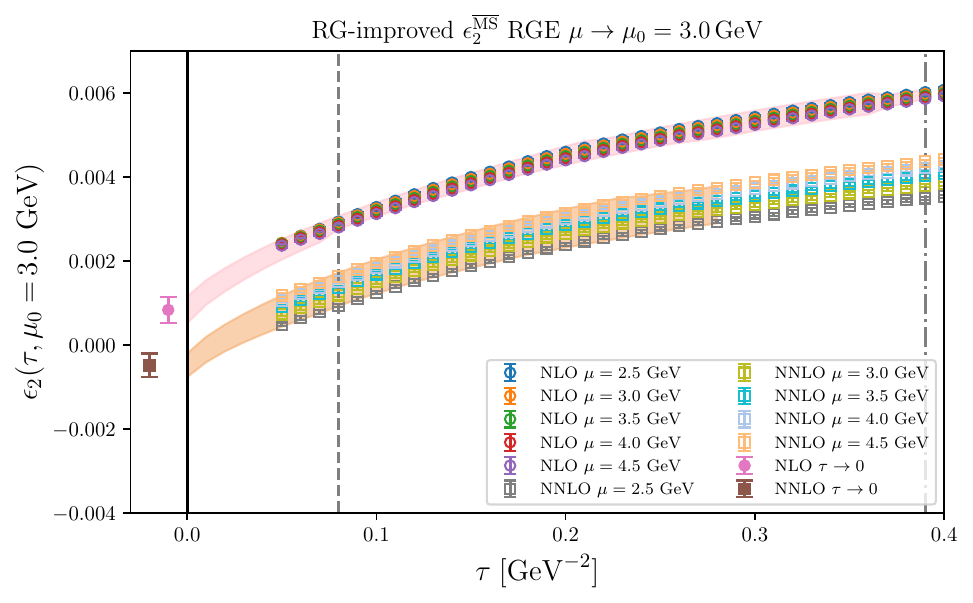}
    \caption{Zero-flow-time extrapolation of the \RG-improved matched flow-time data for the $\Delta Q=0$ bag parameters -- $B_1$ (top left), $B_2$ (top right), $\epsilon_1$ (bottom left), $\epsilon_2$ (bottom right). The matching to \MSbar\ is performed at both NLO and NNLO using $\mu\in\{2.5,3.0,3.5,4.0,4.5\}\gev$ before running back to $\mu_0=3\gev$. The flow-time \RGE\ is used to set flow-time logarithms to zero and improve the extrapolations. The pink and orange shaded areas indicate the spread of all acceptable fits to the data at NLO and NNLO, resulting in the pink and brown data points in the left panel. 
    The dashed lines indicate $\tau_{\rm min}=0.08\gev^{-2}$, whereas dash-dotted lines mark the maximum flow time included in our fit results.}
    \label{fig:DB0_sftx_RG}
\end{figure*}

In the following we use the \RG-improved \SFTX{} as our preferred method to obtain final results for the bag parameters in the \MSbar\ scheme.
A good validation of this choice is seen by comparing~\cref{fig:Q1_sftx,fig:DB0_sftx} to~\cref{fig:Q1_sftx_RG,fig:DB0_sftx_RG} where the final results with and without \RG\ improvement are very compatible with one another.
Furthermore, for all four $\Delta Q=0$ bag parameters, the \RG-improved matching
leads to more controlled extrapolations and thus somewhat smaller errors at
$\tau=0$. Hence, adopting our default choice of evanescent operators, our final results for the $\Delta Q=2$ and the four $\Delta
Q=0$ bag parameters are
\begin{align}
\begin{aligned}
   {\cal B}_1^{\MSbar}(3\gev) &=   \phantom{-}0.7673(68)_{\rm GF}, \\
    B_1^{\MSbar}(3\gev) &= \phantom{-}1.0524(69)_{\rm GF}, \\
    B_2^{\MSbar}(3\gev) &= \phantom{-}0.9621(26)_{\rm GF},\\
    \epsilon_1^{\MSbar}(3\gev) &= -0.2275(41)_{\rm GF},\\
    \epsilon_2^{\MSbar}(3\gev) &= -0.0005(\phantom{0}3)_{\rm GF},
\end{aligned}
\end{align}
where the uncertainty labeled `GF' combines statistical errors as well as systematic effects of the $\tau\to 0$ extrapolation from our analysis procedure which also directly propagates uncertainties of $a^{-1}$ quoted in Tab.~\ref{tab:confs}. We will now estimate additional sources of systematic uncertainties in our calculation.

\subsection{Systematic Uncertainties}\label{sec:syserror}

\begin{figure}[t]
    \centering
    \includegraphics[width=0.98\columnwidth]{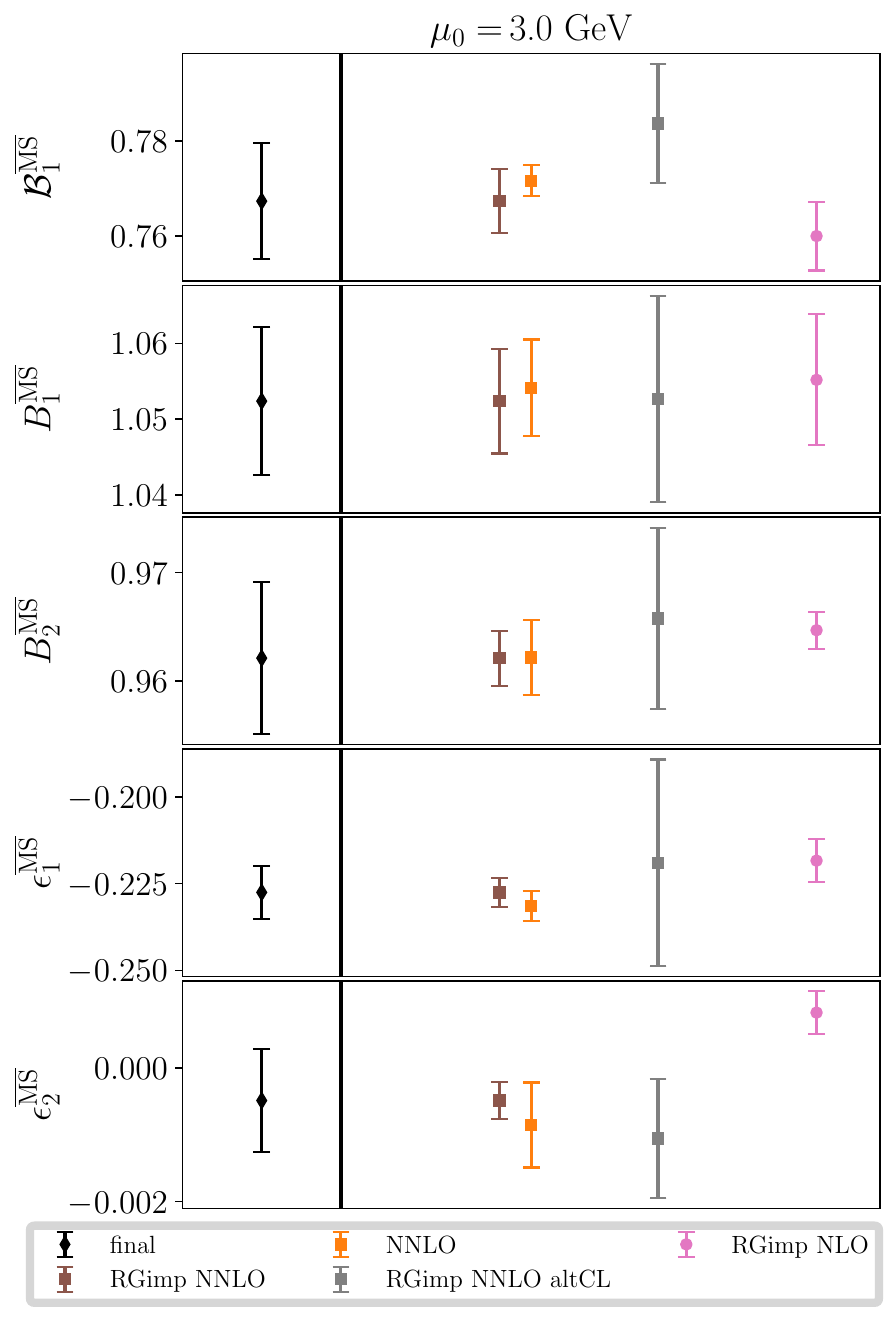}
    \caption{Final results for the $\Delta Q=2$ (top) and $\Delta Q=0$ bag parameters in the \MSbar\ scheme at $\mu_0=3\gev$ are shown by the black data points. The central value is taken as the NNLO result (brown) whose error is combined with half the differences to the alternative NNLO (gray) and NLO (pink) results as well as an additional 0.5\% quark mass mistuning estimate in quadrature. The data result from $\tau\to0$ extrapolations using the flow-time \RGE\ to improve the \SFTX{} as described in the text. The NNLO result without the flow-time \RGE\ is also plotted in orange to show consistency.}
    \label{fig:Bvals}
\end{figure}

Our values above are obtained at \NNLO\ with the improvement provided by the flow-time \RGE\ following~\cref{eq:flowRGtau0}.
In Fig.~\ref{fig:Bvals} we show these final values using brown squares and they are fully consistent with the orange squares obtained by carrying out the same analysis but without the flowed \RG\ improvement.
Subsequently, we discuss other possible sources of uncertainty and also estimate their impact on our results.

\subsubsection{Continuum-limit extrapolation}\label{sec:sys-CL} 
\begin{figure}[t]
    \centering
    \includegraphics[width=0.9\columnwidth]{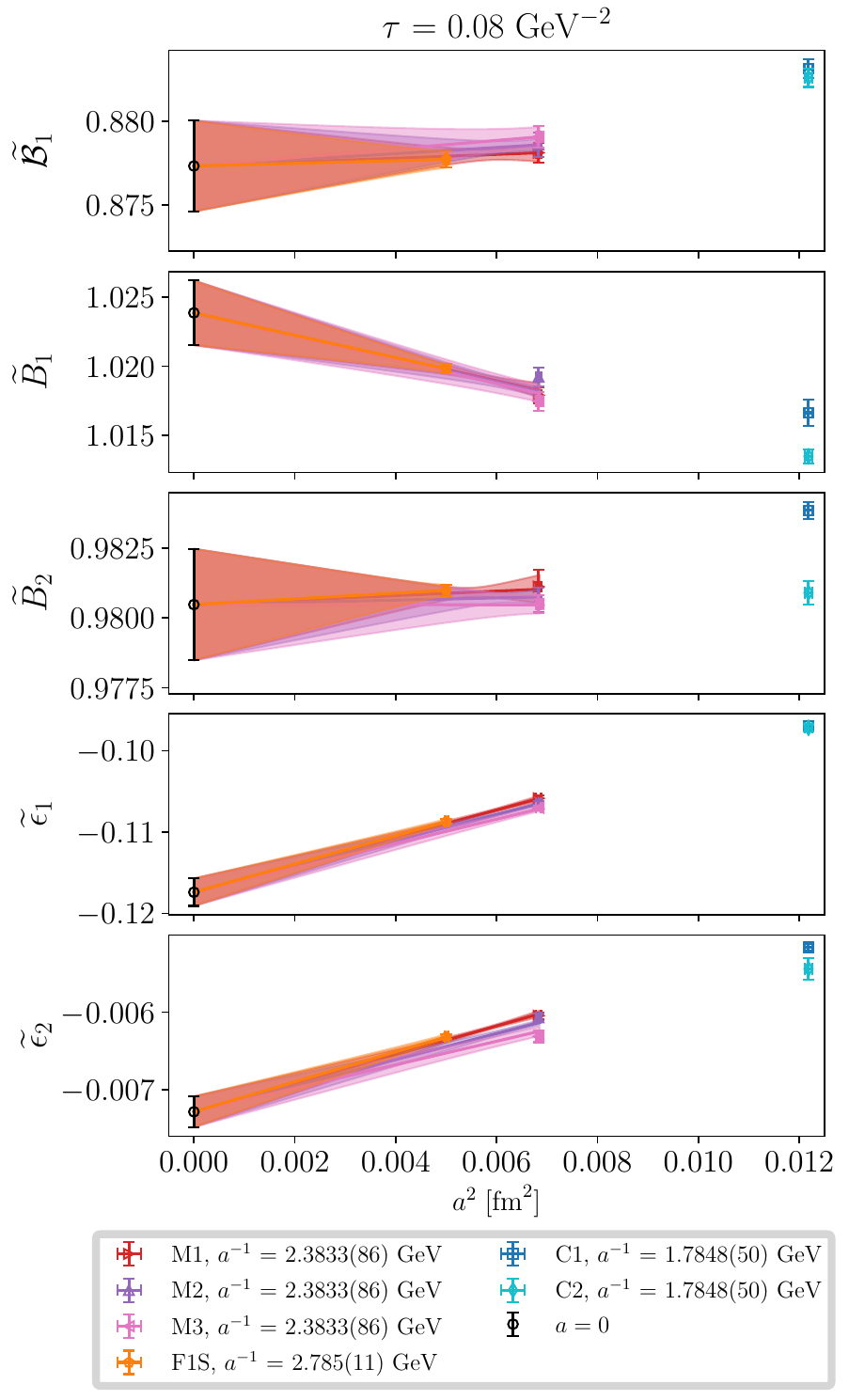}
    \caption{\label{fig:continuum1_alt} Alternative chiral-continuum limit fits for each bag parameter at $\tau=0.08\,{\rm GeV}^{-2}$.
        The black data point is the continuum-extrapolated value at physical pion mass, and the colored bands are the linear $a^2$ relations at each ensemble's pion mass after fitting to~\cref{eq:chiral-cont}.}
\end{figure}

\begin{figure}[t]
    \includegraphics[width=0.9\columnwidth]{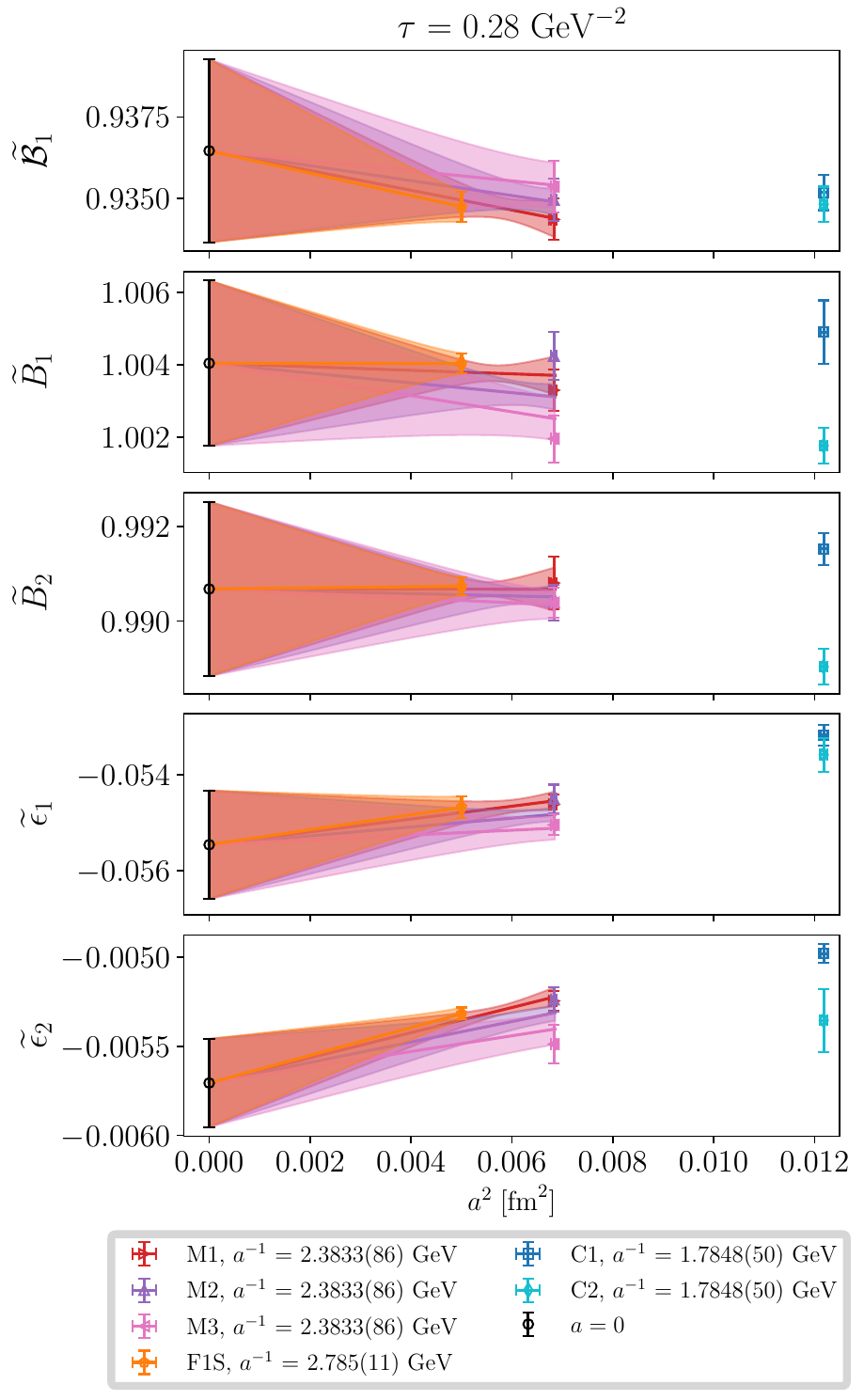}
    \caption{\label{fig:continuum2_alt} Chiral-continuum limit fits for each bag parameter at $\tau=0.28\,{\rm GeV}^{-2}$.
        The black data point is the continuum-extrapolated value at physical pion mass, and the colored bands are the linear $a^2$ relations at each ensemble's pion mass after fitting to~\cref{eq:chiral-cont}.}
\end{figure}

In order to test
for systematic effects in our continuum extrapolation, we repeat the full
analysis discarding the coarse C1 and C2 ensembles. This affects the analysis in
several ways:
\begin{itemize}
    \item Discretization effects are reduced because the smallest value of $a^{-1}$ increases from $1.78\gev$ to $2.38\gev$.
    \item Heavy-quark discretization effects are
    reduced since the heaviest bare charm-quark mass simulated decreases from
    $am_c=0.64$ to $0.45$.
    \item The uncertainties in
    the $\tau\to 0$ extrapolation increase, because the $a\to 0$ extrapolation
    is performed with two instead of three values of the lattice spacing, and thus, the continuum data have increased uncertainty.
\end{itemize}
We show examples of the resulting continuum limit extrapolations excluding the coarse ensembles in \cref{fig:continuum1_alt,fig:continuum2_alt}.
The outcome of these alternative continuum limits is compatible with our
previously obtained central results, albeit with larger
uncertainties, which is an expected effect of discarding $1/3$ of our ensembles.
We propagate these alternative continuum limits through the analysis to estimate the effect on our final bag parameters in the
\MSbar{} scheme.

For all five bag parameters considered, we find at most a 1.8\% change of
the central values and the obtained bag parameter values are shown in gray in
\cref{fig:Bvals}. To quantify the effect of these alternative continuum limits,
we assign half the difference of the central values as an additional uncertainty
to our final results in~\cref{eq:bag_vals} and use the label `\abbrev{CL}' to
indicate this continuum limit uncertainty.

\subsubsection{Perturbative truncation}
The systematics due to the truncation of the perturbative series are partly
taken into account through the variation of the renormalization scale $\mu$ as
discussed in the context of \cref{eq:regRGE}.  However, estimates based on scale
variations are known to be unreliable for capturing the full truncation effects.
This is illustrated by the pink circles in~\cref{fig:Bvals}, which show the
outcome of the \RG-improved analysis based on the \NLO\ rather than the \NNLO\
expressions for the matching coefficients. While they are consistent with the
\NNLO\ results for $\mathcal{B}_1^\text{\MSbar}$, $B_1^\text{\MSbar}$,
$B_2^\text{\MSbar}$, and $\epsilon_1^\text{\MSbar}$, the perturbative
uncertainty still seems to be underestimated for $\epsilon^\text{\MSbar}_2$. We
therefore include an additional perturbative truncation uncertainty by taking
half the difference of the central values at \NLO\ and \NNLO, referring to it as
`\abbrev{PT}' in the final results.

\subsubsection{Quark-mass tuning}
\begin{figure}[t]
    \centering
    \includegraphics[width=0.9\columnwidth]{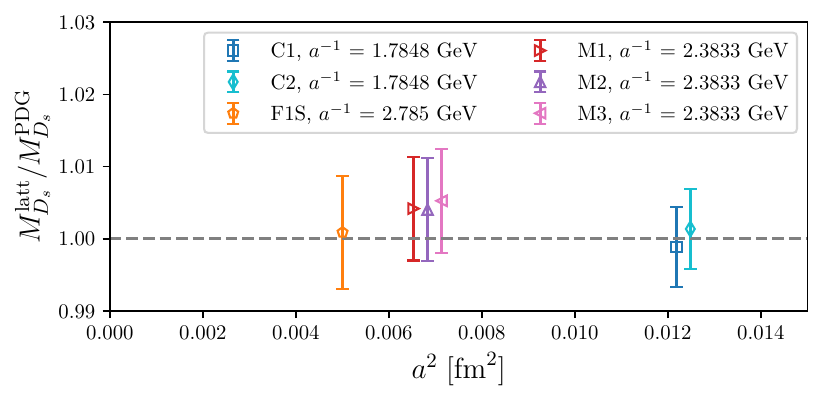}
    \caption{Ratios of $M_{D_s}^{\rm latt}/M_{D_s}^{\rm PDG}$ for each lattice ensemble used.}
    \label{fig:lat_Dmes}
\end{figure}

Starting from the physical values of the strange quark mass in lattice units reported by the \abbrev{RBC/UKQCD}  collaboration \cite{Allton:2008pn,Aoki:2010dy,Blum:2014tka,Boyle:2018knm}, we tune the charm-quark masses listed in \cref{tab:confs} such that the value of the $D_s$ meson mass matches the \abbrev{PDG} value~\cite{PDG:2024cfk} in physical units using the quoted values of the inverse lattice spacings.
In~\cref{fig:lat_Dmes} we show the relative deviations from the \abbrev{PDG}. Although all data are consistent within uncertainties, they have an associated uncertainty in the matching. Since the dataset of our exploratory calculation does not allow for a refined estimate of this effect on our bag parameters, we simply assign a $0.5\%$ uncertainty due to the matching in our final results.

\subsubsection{Gluon, light-, and strange-quark discretization}
Dominant light-quark and gluon discretization errors arise from the used actions and are of order ${\cal O}((a\Lambda_\text{QCD})^2)$. Using $\Lambda_\text{QCD}=300\,\mev$, this amounts to 1\% at our finest lattice spacing. We account and correct for that by our chiral-continuum extrapolation including terms linear in $a^2$. Remaining effects start at ${\cal O}((a \Lambda_\text{QCD})^4)\sim0.01\%$, which is negligible at our current precision. 
An indicator for discretization errors in domain-wall fermions is the residual mass $am_\text{res}$ which parametrizes the residual chiral-symmetry breaking as an additive mass term. In our calculation, these are of ${\cal O}(am_\text{res})\sim 0.4\%$; see Ref.~\cite{Black:2025gft}.

We add the quark-mass tuning (0.5\%) and residual chiral-symmetry breaking (0.4\%) systematic uncertainties in quadrature, and assign an additional $0.64\%$ uncertainty, labeled `OS' for other systematics, to the final results in~\cref{eq:bag_vals}.

\subsubsection{Heavy-quark discretization}
Charm quarks are simulated using stout-smeared~\cite{Morningstar:2003gk} M\"obius domain-wall fermions~\cite{Brower:2012vk} which are optimized for heavy flavors \cite{Cho:2015ffa} and allow the simulation of bare charm quark masses with $am_c\lesssim0.7$ while discretization effects remain small and under control~\cite{Cho:2015ffa,Boyle:2018knm,Aoki:2023qpa}.
This condition is fulfilled on all lattice ensembles in~\cref{tab:confs} and, hence, we expect discretization errors to be small.
As shown in Appendix C of Ref.~\cite{Black:2025gft}, the residual masses for charm quarks remains small ($am_{\rm res}\lesssim10^{-4}$) for all ensembles used and the long plateaus demonstrate that the domain-wall mechanism works well. We therefore conclude heavy quark discretization effects to be negligibly small which is in agreement with the small changes observed when performing alternative continuum limits discarding the coarse ensembles and reducing the largest bare charm quark mass from 0.64 to 0.45.

\subsubsection{Chiral extrapolation}
Although the valence charm and strange quarks are tuned to their physical values, our ensembles differ in the light sea-quark mass, which we have parametrized in the chiral-continuum ansatz of~\cref{eq:chiral-cont}.
The lattice-spacing dependence is the dominant component of the chiral-continuum extrapolations, while the chiral term is much smaller and not clearly resolved.
Therefore we do not assign a separate uncertainty due to the chiral extrapolation alone.

\subsubsection{Finite volume}
Focusing on a system with only charm and strange quark masses in the valence sector on ensembles with $M_\pi\cdot L > 4$, we consider finite-volume effects (\FVE{}s) to be negligible since heavy-light mesons have a short coherence length and the lightest excited state contaminations will scale with masses heavier than pions.
While there is no specific evaluation of \FVE{}s in the matrix elements of these operators for charm-strange mesons in the literature, we can interpolate from results for other systems.
Reference~\cite{Lee:2014ada} finds that the $\Delta B=0$ operator basis for $B_s$ mesons have \FVE{}s $\lesssim0.05\%$, while Ref.~\cite{Boyle:2024gge} estimates \FVE{}s of $\sim0.1\%$ for the bag parameters of $\Delta Q=2$ operators. 
We infer that the \FVE{}s of the bag parameters of charm-strange mesons lie between $0.1\%$ and $0.05\%$; this is a negligible uncertainty at our current precision.
We therefore do not account for finite-volume effects separately.

\subsubsection{Isospin-breaking and QED corrections}
Furthermore, corrections due to isospin breaking for the light quarks in the sea sector may arise as well as effects due to accounting for electric charges both in the sea and valence sector. Putting the focus on exploring the concept of using \GFSFTX, we refrain from estimating these effects and, hence, corresponding uncertainties are not accounted for.

\section{Phenomenological results}\label{sec:phenores}
Using a setup with pseudoscalar $D_s$ meson states in pure \QCD{} without
electric charges, we determined the bag parameter contributing to the mass
difference of a non-physical charm-strange meson at short distance given by a
$\Delta Q=2$ four-quark operator. Accounting for all systematic effects
discussed in the previous section, we find in the \MSbar{} scheme at
$\mu_0=3\gev$ and adopting our default choice of evanescent operators
\begin{align}
    {\cal B}_1^{\MSbar}(3\gev) &= 0.7673(68)_{\rm GF}(82)_{\rm CL}(37)_{\rm PT}(49)_{\rm OS} \nonumber \\
                            &= 0.7673(123) \label{eq:Q1_val}.
\end{align}
Based on our knowledge from the $B$ meson sector, we may assume spectator
effects to be small and that most of the difference in the hadronic matrix
element between using a strange and a light spectator is captured through the
decay constant~\cite{Aoki:2014nga,FLAG:2024oxs}.  We therefore infer that the
bag parameter calculated in~\cref{eq:Q1_val} is largely compatible with
literature results for short-distance contributions to $D^0$ mixing.  Lattice
results exist for the bag parameters of the $D^0$ meson from both the
\abbrev{ETM} and \abbrev{FNAL/MILC} collaborations.  The former find ${\cal
B}_1(3\gev)=0.757(27)$~\cite{Carrasco:2015pra} and the latter ${\cal
B}_1(3\gev)=0.784(56)$~\cite{Bazavov:2017weg}.\footnote{We convert the quoted
matrix element using $f_D^{N_f=2+1}=210.4(1.5)\mev$~\cite{FLAG:2024oxs} and
$M_{D^{0}}=1864.84(5)\mev$~\cite{PDG:2024cfk}.}
These results are in excellent
agreement with our result in~\cref{eq:Q1_val} albeit with larger uncertainties
which are however also expected when extrapolating to/working at physical light
quark masses.  The bag parameters parametrizing short-distance $D^0$ mixing
have also been calculated using \HQET{} sum rules predicting ${\cal
B}_1(3\gev)=0.654^{+0.060}_{-0.052}$ \cite{Kirk:2017juj}. This result exhibits a
noticeable tension to the lattice determinations. However, 
\HQET{} is less suitable for charm quarks than for bottom quarks, and
Ref.~\cite{Kirk:2017juj} does not include $O(1/m_Q^n)$ corrections in the
\HQET{}-\QCD{} matching procedure.
Note that the results of Refs.~\cite{Kirk:2017juj,Carrasco:2015pra,Bazavov:2017weg}, based on \NLO\ matching, employ choices of evanescent operators~\cite{Buras:2000if,Beneke:1998sy} which differ from ours. Nevertheless, our final results can be compared consistently with these because the differences occur only at \NNLO\ for $\mathcal{B}_1^\text{\MSbar}$.

For the bag parameters parametrizing $\Delta Q=0$ operator matrix elements
describing the lifetime ratio of $D_s/D^0$ mesons, we obtain, using our
default choice of evanescent operators,
\begin{align}
    B_1^{\MSbar}(3\gev) &= \phantom{-}1.0524(69)_{\rm GF}(\phantom{0}1)_{\rm CL}(14)_{\rm PT}(67)_{\rm OS} \notag \\
                     &= \phantom{-}1.0524(97), \notag \\
    B_2^{\MSbar}(3\gev) &= \phantom{-}0.9621(25)_{\rm GF}(18)_{\rm CL}(13)_{\rm PT}(62)_{\rm OS} \notag \\
                     &= \phantom{-}0.9621(70), \notag \\
    \epsilon_1^{\MSbar}(3\gev) &= -0.2275(41)_{\rm GF}(42)_{\rm CL}(46)_{\rm PT}(14)_{\rm OS} \notag \\
                            &= -0.2275(76), \notag \\
    \epsilon_2^{\MSbar}(3\gev) &= -0.0005(\phantom{0}3)_{\rm GF}(\phantom{0}0)_{\rm CL}(\phantom{0}7)_{\rm PT}(\phantom{0}0)_{\rm OS} \notag  \\
                            &= -0.0005(\phantom{0}8). \label{eq:bag_vals}
\end{align}
These are the first lattice results for the bag parameters of $\Delta Q=0$ operators with a full error budget, and represent a significant step towards more precise predictions of heavy meson lifetimes and lifetime ratios using lattice \QCD{}.
Although our focus is on establishing the use of \GFSFTX{} to renormalize four-quark operators, we obtain percent-level results for $B_1^{\MSbar}$, $B_2^{\MSbar}$, and $\epsilon_1^{\MSbar}$, while $\epsilon_2^{\MSbar}$ is much smaller and vanishes within uncertainties. The latter is however an artifact of a particular choice of evanescent operators (cf.~Appendix \ref{app:evscheme}).

Although SU(3)$_{\rm F}$-breaking corrections to the $\Delta Q=0$ bag parameters are calculated in Ref.~\cite{King:2021jsq} within the \HQET{} sum-rules framework, the \HQET{}-\QCD{} matching for this scenario has not been calculated and, hence, we cannot directly compare results.
From Tables 3 and 4 of Ref.~\cite{King:2021jsq}, we see, however, that the
difference between $B_d$ and $B_s$ bag parameters is very small, implying
mostly-negligible spectator effects. Therefore we expect again that the results
calculated here using $D_s$ mesons are largely compatible with $D^0$
calculations.
The most up-to-date values of the $\Delta Q=0$ bag parameters in \QCD{} from sum
rules for $D$ mesons are presented in Ref.~\cite{Black:2024bus}, where a
different basis of evanescent operators has been used than here.
Performing the basis transformation described in~\cref{app:evscheme}, we express our results of~\cref{eq:bag_vals} in the scheme used in Ref.~\cite{Black:2024bus} and obtain
\begin{equation}
  \begin{aligned}
    B_1^{\prime\,\MSbar}(3\gev) &= \phantom{-}1.103(10),  \\
    B_2^{\prime\,\MSbar}(3\gev) &= \phantom{-}0.9754(71),  \\
    \epsilon_1^{\prime\,\MSbar}(3\gev) &= -0.2427(71),  \\
    \epsilon_2^{\prime\,\MSbar}(3\gev) &= -0.0289(8). \label{eq:bag_vals_NLOFierz}
  \end{aligned}
\end{equation}
As another consistency check, we also performed the $\tau\to0$ extrapolations by
directly using the matching coefficients defined for this evanescent scheme and
find fully compatible results.  In comparison to Eq.~(6.27) of
Ref.~\cite{Black:2024bus} our results are an order of magnitude more precise. We
note, however, that the uncertainties in Ref.~\cite{Black:2024bus} are dominated
by the \HQET{}-\QCD{} matching procedure. Furthermore, our values in particular
for $B_1^{\MSbar}$ and $\epsilon_1^{\MSbar}$ differ significantly from those
of Ref.~\cite{Black:2024bus}. This may again be due to the \HQET{}-\QCD{}
matching procedure, which may
require including $O(1/m_Q^n)$ corrections to more accurately describe $D$ meson.

From a phenomenological perspective, our value for $\epsilon_1^{\MSbar}$ is particularly interesting, because in the \HQE\ the Wilson coefficient multiplying this parameter is rather large~\cite{King:2021xqp}.
Hence, the observed difference of about a factor of two compared to the
existing literature may lead to a significant effect in the lifetime predictions for $D$
mesons. An independent verification of our findings would therefore be
desirable.

In addition to lifetimes and lifetime ratios, the $\Delta Q=0$ four-quark operators also contribute to other processes described by the \HQE, such as rare semileptonic $B$-meson decays and are in fact the dominant uncertainties at high $q^2$~\cite{Huber:2020vup,Fael:2025xmi}. 
A future high-precision determination of these quantities for $B$ mesons is therefore advantageous to the broader flavor physics community.
In particular, the lifetime-ratio operator basis is immediately relevant to the decay-rate ratios of the currents $b\to s\ell\ell$ and $b\to u\ell\nu$ as discussed in Ref.~\cite{Huber:2020vup} while the future application of our procedure to the eye diagrams will be highly beneficial to $b\to s\ell\ell$ processes themselves.

\section{Conclusion}\label{sec:concl}
Using six \abbrev{RBC/UKQCD} $2+1$-flavor domain-wall ensembles and performing a valence sector measurement with charm and strange quarks tuned to their physical values, we calculate in a heavy-light system $\Delta Q=2$ operators describing neutral meson mixing as well as $\Delta Q=0$ operators related to bag parameters entering the determination of heavy meson lifetime ratios. Renormalization of operators is performed using fermionic \GF\ in combination with the \SFTX\ to match our results to the \MSbar{} scheme. This matching is performed at both \NLO\ and \NNLO\ at a scale of $\mu_0=3\,\gev$. Taking advantage of perturbatively calculated anomalous dimensions, we improve the $\tau\to 0$ extrapolation by running the matching scale to higher values and cancelling flow-time logarithms using the flowed \RGE. In the future this may further be improved by nonperturbative determinations of the corresponding anomalous dimensions~\cite{Hasenfratz:2022wll,Carosso:2018bmz,Hasenfratz:2019hpg}.

Although the focus of this work lies on the method to calculate and renormalize four-quark operators using \GFSFTX, we further achieve percent-level results for the bag parameters indicating that \GFSFTX\ is also a competitive method w.r.t.~overall uncertainties.
While our results have a full error budget, we do stress, however, that in this study we choose to simplify the problem by neglecting contributions from eye diagrams which are required to obtain absolute lifetimes and $\text{SU(3)}_{\rm F}$-breaking effects.
The challenge of the eye diagrams arises due to their poor signal-to-noise ratio demanding different numerical techniques to simulate and, moreover, they also induce mixing with operators of lower mass-dimension under renormalization. In the case of $\epsilon_2^{\MSbar}$, we observe sensitivity to the choice of evanescent operator scheme and provide formulae to convert to a broad class of such choices.

The successful application of the \GFSFTX\ framework demonstrated here and also
in other works~\cite{Black:2025gft,Francis:2025pgf,Francis:2025rya,Iritani:2018idk}
motivates an extended program 
in the future. It will be
interesting to incorporate the full $\Delta Q=0$ operator basis and handle
eye diagrams with the associated mixing of lower mass-dimension operators in the
\SFTX.
Furthermore, it looks promising to choose a larger set of ensembles
in order to improve control of all systematic effects in a high-precision study extending this calculation to use both physical light quarks as well as ``heavier-than-charm'' quarks to extrapolate towards bottom. 
This will allow us to obtain a more complete picture of lifetimes and lifetime ratios for heavy-light mesons including $D^0,\,D^+,\,D^+_s,\,B^0,\,B^+,\,B_s$ as well as heavy baryons.

\textit{Note added ---}
  While this paper was under review, Ref.\,\cite{Moretti:2026waq} presented
  the \NNLO{} Wilson coefficients for heavy-meson lifetime ratios and combined
  them with the bag parameters determined here.  It finds that our lattice
  results significantly change the charm lifetime-ratio predictions compared
  to the \HQET{} sum-rule results~\cite{Black:2024bus}, reduce the associated
  uncertainties, and improve the agreement with experiment.

\section*{Acknowledgments}
We thank the \abbrev{RBC/UKQCD} Collaboration for generating and making their
gauge field ensembles publicly available. Special thanks is given to Felix
Erben, Ryan Hill, and J.~Tobias Tsang for assistance in setting up the
simulation code, and we thank Anna Hasenfratz and Martin Lang for constructive
discussions.
Fruitful input and encouragement from the members of the \abbrev{CRC~TRR}~257 is
highly appreciated. Special thanks in this respect go to Alexander Lenz, Uli
Nierste, and Thomas Mannel. We are also grateful to the authors of
Ref.~\cite{Moretti:2026waq}, especially Francesco Moretti, for comparing the anomalous dimensions of the general basis of evanescent operators given in \cref{app:ren}, including the non-physical entries.
Similarly, we thank the authors of Ref.~\cite{Aebischer:2025hsx}, especially Pol Morell, for help comparing to our anomalous dimensions.

These computations used resources provided by the \abbrev{OMNI} cluster at the University of Siegen, the \abbrev{HAWK} cluster at the High-Performance Computing Center Stuttgart, the DeiC Large Memory \abbrev{HPC} Type3/Hippo System managed by the eScience Center at the University of Southern Denmark, and \abbrev{LUMI-G} at the \abbrev{CSC} data center Finland (DeiC National \abbrev{HPC} g.a.~DEIC-SDU-L5-13 and DEIC-SDU-N5-2024053).

This research was supported by Deutsche Forschungsgemeinschaft 
(\abbrev{DFG}, German Research Foundation) through grant
396021762 - TRR~257 ``Particle Physics Phenomenology after the Higgs
Discovery'', through grant 513989149, Germany’s Excellence Strategy – EXC 3107 – Project 533766364,
\abbrev{UK STFC} grant ST/X000494/1, the Swiss National Science Foundation (\abbrev{SNSF}) under contract TMSGI2\_211209, the
National Science Foundation under grant PHY-2209185, and the \abbrev{DOE} Topical
Collaboration ``Nuclear Theory for New Physics'' award No.\ DE-SC0023663.

\section*{Data Availability}
The data that support the findings of this article are openly available~\cite{DatasetGF}.

\appendix

\section{Renormalization constants}
\label{app:ren}
The bare gauge coupling is related to the \MSbar-renormalized coupling by
\begin{align}
\label{eq:gren}
    g_0=\left(\frac{\mu^2{\rm e}^{\EulerGamma}}{4\pi}\right)^{\ep/2}
    Z_g\left(\api(\mu)\right)g(\mu)\,,
\end{align}
where $\mu$ denotes the renormalization scale and $Z_g$ is the renormalization
constant of the coupling,
\begin{equation}
\label{eq:Zg}
  \begin{aligned}
    Z_g(\api) &= 1 -
    \api\frac{\beta_0}{2\ep}  +
    \api^2\left(\frac{3\beta_0^2}{8\epsilon^2}
    -\frac{\beta_1}{4\epsilon}\right)  +
\mathcal{O}(\api^3)\,,
  \end{aligned}
\end{equation}
with the coefficients
\begin{equation}
\label{eq:b01}
  \begin{aligned}
    \beta_0 &= \frac{1}{4}\left(11
    -\frac{2}{3}\nf\right)\,,\\
    \beta_1 &= \frac{1}{16}\left(102-\frac{38}{3}\nf\right)\,,\\
  \end{aligned}
\end{equation}
of the \QCD{} beta function
\begin{equation}
\label{eq:beta}
  \begin{aligned}
    \beta(\api) = -\ep-\api\sum_{n=0}^\infty\beta_n\api^n\,.
  \end{aligned}
\end{equation}
The \NNLO\ relation between a generic renormalization constant and the corresponding anomalous dimension, defined by
\begin{equation}
\label{eq:ADM-def}
\frac{\rm d}{{\rm d}\ln\mu^2}Z_{ij}=-\gamma_{ik} Z_{kj}\,,
\end{equation}
and
\begin{equation}
    \label{eq:gamma}
    \gamma_{ij}=-\api\sum_{n=0}^\infty\api^n\gamma_{n,ij}\,,
\end{equation}
 is
\begin{align}
    Z_{ij}=\delta_{ij}&+\api\left(
    -\frac{1}{\epsilon}\gamma_{0,ij}+Z^{10}_{ij}\right)
    +\api^2\Bigl(\frac{1}{2\epsilon^2}[\gamma_{0,ik}\gamma_{0,kj} \nonumber \\
    &\eqindent{1}+\beta_0\gamma_{0,ij}]
    -\frac{1}{2\epsilon}[\gamma_{1,ij}+\beta_0Z^{10}_{ij}
    +\gamma_{0,ik}Z^{10}_{kj} \nonumber \\
    &\eqindent{2}+Z^{10}_{ik}\gamma_{0,kj}]+Z_{ij}^{20}\Bigr)+\mathcal{O}(\api^3)\,. \label{eq:Zgamma}
\end{align}
The finite parts of the renormalization constants, $Z_{ij}^{n0}$ are only
non-zero if $i\in{\mathcal{E}}$ and $j\in\mathcal{O}$ as described in
\cref{sec:evops}.

As mentioned in \cref{sec:pt_results}, the two sets of operators $\mathcal{O}^{(1)}\equiv\{\mathcal{O}_1,\mathcal{T}_1\}$ and $\mathcal{O}^{(2)}\equiv\{\mathcal{O}_2,\mathcal{T}_2\}$ do not mix with each other.
The result for the physical submatrix of the anomalous dimension is \cite{Buras:1989xd,Ciuchini:1997bw,Buras:2000if}
\begin{equation}
\label{eq:gamma01}
    \gamma_0^{(1)}=
    \begin{pmatrix}
    0&-\frac{3}{2}\\
    -\frac{1}{3}&\frac{1}{2}
    \end{pmatrix}\,,
\end{equation}
\begin{equation}
\label{eq:gamma02}
    \gamma_0^{(2)}=
    \begin{pmatrix}
    2&0\\
    0&-\frac{1}{4}
\end{pmatrix}\,,
\end{equation}
and
\begin{equation}
\label{eq:gamma11}
    \gamma_1^{(1)}=
    \begin{pmatrix}
    \frac{7}{24}&-\frac{111}{32}+\frac{7}{24}\nf\\
    -\frac{85}{48}+\frac{7}{108}\nf&-\frac{185}{96}+\frac{11}{72}\nf\\
\end{pmatrix}\,,
\end{equation}
\begin{equation}
\label{eq:gamma12}
    \gamma_1^{(2)}=
    \begin{pmatrix}
    \frac{49}{6}-\frac{5}{18}\nf&-\frac{95}{32}+\frac{1}{12}\nf\\
    -\frac{23}{144}+\frac{1}{54}\nf&-\frac{269}{64}+\frac{37}{144}\nf
\end{pmatrix}\,.
\end{equation}
In order to be able to rotate to a more general scheme (see
\cref{app:evscheme}), we also need unphysical parts of the
renormalization matrix, which we find to be\footnote{These results have been
obtained independently in Ref.\,\cite{Moretti:2026waq}}.
\begin{align}
\gamma^{(1)}_{0,\mathcal{O}\mathcal{E}}&=
\begin{pmatrix}
      0&\frac{1}{4}&0&0\\
      \frac{1}{18}&\frac{5}{48}&0&0\\
\end{pmatrix}\,,\quad
\gamma^{(2)}_{0,\mathcal{O}\mathcal{E}}=
\begin{pmatrix}
      0&\frac{1}{4}&0&0\\
      \frac{1}{18}&\frac{5}{48}&0&0\\
\end{pmatrix}\,,\notag\\
\gamma^{(1)}_{1,\mathcal{O}\mathcal{E}}&=
\begin{pmatrix}
      -\frac{17}{48}&\frac{11}{192}-\frac{1}{144}\nf&\frac{7}{576}&\frac{35}{1536}\\
      -\frac{169}{864}-\frac{1}{648}\nf&\frac{317}{1152}-\frac{5}{1728}\nf&\frac{35}{6912}&-\frac{1}{3072}\\
\end{pmatrix}\,,\notag\\
\gamma^{(2)}_{1,\mathcal{O}\mathcal{E}}&=
\begin{pmatrix}
      -\frac{11}{36}&\frac{7}{24}-\frac{1}{144}\nf&\frac{7}{576}&\frac{35}{1536}\\
      -\frac{31}{216}-\frac{1}{648}\nf&\frac{283}{1152}-\frac{5}{1728}\nf&\frac{35}{6912}&-\frac{1}{3072}\\
\end{pmatrix}\,,\notag\\
\gamma^{(1)}_{0,\mathcal{E}\mathcal{E}}&=
\begin{pmatrix}
      0&-\frac{7}{2}&0&\frac{1}{4}\\
      -\frac{7}{9}&-\frac{1}{3}&\frac{1}{18}&\frac{5}{48}\\
      *&*&*&*\\
      *&*&*&*\\
\end{pmatrix}\,,\notag\\
\gamma^{(2)}_{0,\mathcal{E}\mathcal{E}}&=
\begin{pmatrix}
      -\frac{2}{3}&-4&0&\frac{1}{4}\\
      -\frac{8}{9}&-\frac{1}{12}&\frac{1}{18}&\frac{5}{48}\\
      *&*&*&*\\
      *&*&*&*\\
\end{pmatrix}\,,
\label{eq:gammanonphys}
\end{align}
where the entries denoted by $*$ in $\gamma_{0,\mathcal{E}\mathcal{E}}$ do not
contribute to our final result.  They require including the evanescent operators
of the third generation and were thus not computed.  The two-loop anomalous
dimensions $\gamma_1$ are scheme dependent, meaning that their values are
specified for the choice of evanescent operators of \cref{sec:evops}. The result
in a general scheme can be found in the ancillary files (see \cref{app:anc} and Ref.~\cite{SupplementalMaterial}).

For the $\Delta Q=2$ operator, we find
\begin{equation}
\label{eq:gammamix}
\gamma^{(0)}_{{\cal Q}_1}=-\frac{1}{2}\,,\qquad
\gamma^{(1)}_{{\cal Q}_1}=\frac{7}{32}-\frac{1}{72}\nf\,.
\end{equation}
The \MSbar\ renormalization constant of the flowed-quark field $Z^{\MSbar}_\chi$, introduced in \cref{eq:gff:gapy}, is related by \cref{eq:Zgamma} to the anomalous dimensions given by\,\cite{Luscher:2013cpa,Harlander:2018zpi}
\begin{equation}
\label{eq:gammachi}
  \begin{aligned}
    \gamma_{\chi,0} = &-1\,,\\
    \gamma_{\chi,1} = &-\frac{73}{8}+\frac{11}{36}\nf+\frac{26}{9}\ln 2\,.
  \end{aligned}
\end{equation}
The finite renormalization $\zeta_\chi$ relating the ringed scheme to the \MSbar\ scheme,
\begin{equation}
\mathring{Z}_\chi=\zeta_\chi(\tau,\mu)Z^{\MSbar}_\chi\,,
\end{equation}
is\,\cite{Makino:2014taa,Artz:2019bpr}
\begin{align}
    \zeta_\chi(\tau,\mu) = 1&
    -\api\left(\gamma_{\chi,0}\lmut
    +\ln 3+\frac{4}{3}\ln 2\right) \nonumber \\
    &+\api^2\Bigg\{\frac{\gamma_{\chi,0}}{2}\left(\gamma_{\chi,0}-\beta_0
    \right)\lmut^2 \nonumber \\
    &+\Big[\gamma_{\chi,0}\left(\beta_0
      -\gamma_{\chi,0}\right)\ln3
      +\frac{4}{3}\gamma_{\chi,0}\left(\beta_0
      -\gamma_{\chi,0}\right)\ln 2 \nonumber \\
      &-\gamma_{\chi,1}\Big]\lmut
    +\frac{c_\chi^{(2)}}{16}\Bigg\}+\mathcal{O}(\api^3)\,, \label{eq:ringedfin}
\end{align}
where
\begin{equation}
\label{eq:chi2}
  \begin{aligned}
    c_\chi^{(2)} &= 4\, c_{\chi,\text{A}}
    +\frac{16}{9}\, c_{\chi,\text{F}}+\frac{2}{3}\nf\, c_{\chi,\text{R}}\,.
  \end{aligned}
\end{equation}
The coefficients have been determined in Ref.\,\cite{Artz:2019bpr}:
\begin{align}
    c_{\chi,\text{A}} &= -23.7947,\qquad c_{\chi,\text{F}}= 30.3914,\qquad \nonumber \\
    c_{\chi,\text{R}} &=
    -\frac{131}{18}+\frac{46}{3}\zeta(2)
    +\frac{944}{9}\ln 2+\frac{160}{3}\ln^2 2
    -\frac{172}{3}\ln 3 \nonumber \\
    &+\frac{104}{3}\ln2 \ln 3
    -\frac{178}{3}\ln^2 3+\frac{8}{3}\text{Li}_2(1/9) \label{eq:chi2coeffs} \\
    &-\frac{400}{3}\text{Li}_2(1/3)+\frac{112}{3}\text{Li}_2(3/4)
    =-3.92255\ldots\,. \nonumber
\end{align}
Only digits which are not affected by the numerical uncertainty are quoted in \cref{eq:chi2coeffs}.

\section{Scheme dependence}
\label{app:evscheme}
The definition of the scheme for the evanescent operators
in~\cref{eq:evopsmin1,eq:evopsmin2} includes an arbitrariness; in particular,
one is free to add terms which vanish in $D=4$. Such additional terms, however,
change the result of our calculation and correspond to choosing a different
renormalization scheme. In a more general scheme, the evanescent operators of the first generation are
\begin{align}
\mathcal{E}_{{\cal O}_1}^{\prime(1)}=&\left(\bar{\psi}_\alpha\gamma_{\mu\nu\rho}(1-\gamma_5) \psi_\beta\right)\left(\bar{\psi}_\gamma\gamma_{\rho\nu\mu}(1-\gamma_5) \psi_\delta\right) \nonumber \\
&+E^{(1)}_1\,{\cal O}_1\,, \nonumber \\
\mathcal{E}_{T_1}^{\prime(1)}=&\left(\bar{\psi}_\alpha\gamma_{\mu\nu\rho}(1-\gamma_5)T^A \psi_\beta\right)\left(\bar{\psi}_\gamma\gamma_{\rho\nu\mu}(1-\gamma_5)T^A \psi_\delta\right) \nonumber \\
&+E^{(1)}_2\,T_1\,, \nonumber \\
\mathcal{E}_{{\cal O}_2}^{\prime(1)}=&\left(\bar{\psi}_\alpha\gamma_{\mu\nu}(1-\gamma_5) \psi_\beta\right)\left(\bar{\psi}_\gamma\gamma_{\nu\mu}(1+\gamma_5) \psi_\delta\right) \nonumber \\
&+E^{(1)}_3\,{\cal O}_2\,, \nonumber \\
\mathcal{E}_{T_2}^{\prime(1)}=&\left(\bar{\psi}_\alpha\gamma_{\mu\nu}(1-\gamma_5)T^A \psi_\beta\right)\left(\bar{\psi}_\gamma\gamma_{\nu\mu}(1+\gamma_5)T^A \psi_\delta\right) \nonumber \\
&+E^{(1)}_4\,T_2\,, \label{eq:evops1}
\end{align}
where
\begin{equation}
\begin{aligned}
E_1^{(1)}&=-4+E_{1,1}^{1}\varepsilon+E_{1,2}^{1}\varepsilon^2\,,\\
E_2^{(1)}&=-4+E_{2,1}^{1}\varepsilon+E_{2,2}^{1}\varepsilon^2\,,\\
E_3^{(1)}&=-4+E_{3,1}^{1}\varepsilon+E_{3,2}^{1}\varepsilon^2\,,\\
E_4^{(1)}&=-4+E_{4,1}^{1}\varepsilon+E_{4,2}^{1}\varepsilon^2\,.\\
\end{aligned}
\label{eq:evopscoeffs1}
\end{equation}
For the second generation, we work with the operators
\begin{align}
\mathcal{E}_{{\cal O}_1}^{\prime(2)}=&\left(\bar{\psi}_\alpha\gamma_{\mu\nu\rho\sigma\tau}(1-\gamma_5) \psi_\beta\right)\left(\bar{\psi}_\gamma\gamma_{\tau\sigma\rho\nu\mu}(1-\gamma_5) \psi_\delta\right) \nonumber \\
&+E^{(2)}_1\,{\cal O}_1+B_1\mathcal{E}_{{\cal O}_1}^{(1)}\,, \nonumber \\
\mathcal{E}_{T_1}^{\prime(2)}=&\left(\bar{\psi}_\alpha\gamma_{\mu\nu\rho\sigma\tau}(1-\gamma_5)T^A \psi_\beta\right)\left(\bar{\psi}_\gamma\gamma_{\tau\sigma\rho\nu\mu}(1-\gamma_5)T^A \psi_\delta\right) \nonumber \\
&+E^{(2)}_2\,T_1+B_2\mathcal{E}_{T_1}^{(1)}\,, \nonumber \\
\mathcal{E}_{{\cal O}_2}^{\prime(2)}=&\left(\bar{\psi}_\alpha\gamma_{\mu\nu\rho\sigma}(1-\gamma_5) \psi_\beta\right)\left(\bar{\psi}_\gamma\gamma_{\sigma\rho\nu\mu}(1+\gamma_5) \psi_\delta\right) \nonumber \\
&+E^{(2)}_3{\cal O}_2+B_3\mathcal{E}_{{\cal O}_2}^{(1)}\,, \nonumber \\
\mathcal{E}_{T_2}^{\prime(2)}=&\left(\bar{\psi}_\alpha\gamma_{\mu\nu\rho\sigma}(1-\gamma_5)T^A \psi_\beta\right)\left(\bar{\psi}_\gamma\gamma_{\sigma\rho\nu\mu}(1+\gamma_5)T^A \psi_\delta\right) \nonumber \\
&+E^{(2)}_4\,T_2+B_4\mathcal{E}_{T_2}^{(1)}\,, \label{eq:evops2}
\end{align}
where
\begin{equation}
\begin{aligned}
E_1^{(2)}&=-16+E_{1,1}^{2}\varepsilon+E_{1,2}^{2}\varepsilon^2\,,\\
B_1&=B_{1,0}+B_{1,1}\varepsilon\,,\\
E_2^{(2)}&=-16+E_{2,1}^{2}\varepsilon+E_{2,2}^{2}\varepsilon^2\,,\\
B_2&=B_{2,0}+B_{2,1}\varepsilon\,,\\
E_3^{(2)}&=-16+E_{3,1}^{2}\varepsilon+E_{3,2}^{2}\varepsilon^2\,,\\
B_3&=B_{3,0}+B_{3,1}\varepsilon\,,\\
E_4^{(2)}&=-16+E_{4,1}^{2}\varepsilon+E_{4,2}^{2}\varepsilon^2\,,\\
B_4&=B_{4,0}+B_{4,1}\varepsilon\,.
\end{aligned}
\label{eq:evopscoeffs2}
\end{equation}
The numerically specified coefficients were obtained by Fierz relations and are
chosen such that the operators are evanescent; all other coefficients are part
of the definition of the basis of evanescent operators.

In the case of the $\Delta Q=2$ operators, we use the non-mixing
basis which refers to the
physical operators $\{\mathcal{O}_{+},\mathcal{O}_{-}\}$ and the evanescent
operators $\{\mathcal{E}_{{\cal O}_1}^{\prime(1)},\mathcal{E}_{{\cal
T}_1}^{\prime(1)},\mathcal{E}_{{\cal O}_1}^{\prime(2)},\mathcal{E}_{{\cal
T}_1}^{\prime(2)}\}$ given in \cref{eq:evops1,eq:evops2} with $\alpha=\gamma=Q$,
$\beta=\delta=q$ and the coefficients of \cref{eq:evopscoeffs1},
\cref{eq:evopscoeffs2} given by
\cite{Buras:1989xd,Buras:2006gb}
\begin{equation}
\begin{aligned}
\label{eq:evopscoeffsnonmixing}
E_{1,1}^{1}&=E_{2,1}^{1}=8\,,\\
E_{1,2}^{1}&=E_{2,2}^{1}=-4\,,\\
E_{1,1}^{2}&=E_{2,1}^{2}=64\,,\\
E_{1,2}^{2}&=-\frac{1688}{25}\,,E_{2,2}^{2}=\frac{2632}{25}\,,\\
B_{1,0}&=B_{2,0}=-40\,,\\
B_{1,1}&=B_{2,1}=40\,.\\
\end{aligned}
\end{equation}
This choice of scheme respects Fierz symmetry and ensures that there is no
mixing among $\mathcal{O}_{+}$ and $\mathcal{O}_{-}$ such that there are no
off-diagonal entries in the physical submatrices of both the matching matrix as
well as the anomalous dimensions.

In the ancillary file, we provide the matching matrix and the anomalous
dimension matrix in a general scheme as a function of the coefficients
$E_{i,j}^{k}$.\footnote{The choice of $B_{4,1}$ neither impacts the matching
matrix nor the physical part of the anomalous dimension, which can be seen from
the fact that in the rotation to a general scheme it only appears in the matrix
describing the rotation among the evanescent operators $M_i$.} The results in a
general scheme can be obtained by performing a rotation of the evanescent
operators from the minimal basis of~\cref{eq:evopsmin1,eq:evopsmin2} to the more
general basis of~\cref{eq:evops1,eq:evopscoeffs1,eq:evops2,eq:evopscoeffs2}~\cite{Gorbahn:2004my,Buras:2006gb}
\begin{equation}
\mathcal{E}^{\prime}_i=M_i(\mathcal{E}_i+\varepsilon U_i{\cal O}_i+\varepsilon^2 V_i{\cal O}_i)\,,
\end{equation}
where
\begin{equation}
\begin{aligned}
M_1&=
\begin{pmatrix}
      1&0&0&0\\
      0&1&0&0\\
      B_{1,0}+\varepsilon B_{1,1}&1&1&0\\
      0&B_{2,0}+\varepsilon B_{2,1}&0&1\\
\end{pmatrix}\,,\\
M_2&=
\begin{pmatrix}
      1&0&0&0\\
      0&1&0&0\\
      B_{3,0}+\varepsilon B_{3,1}&1&1&0\\
      0&B_{4,0}+\varepsilon B_{4,1}&0&1\\
\end{pmatrix}\,,
\end{aligned}
\end{equation}
and
\begin{equation}
\begin{aligned}
U_1&=
\begin{pmatrix}
      E_{1,1}^{1}&0\\
      0&E_{2,1}^{1}\\
      E_{1,1}^{2}&0\\
      0&E_{2,1}^{2}\\
\end{pmatrix}\,,\qquad
U_2&&=
\begin{pmatrix}
      E_{3,1}^{1}&0\\
      0&E_{4,1}^{1}\\
      E_{3,1}^{2}&0\\
      0&E_{4,1}^{2}\\
\end{pmatrix}\,,\\
V_1&=
\begin{pmatrix}
      E_{1,2}^{1}&0\\
      0&E_{2,2}^{1}\\
      E_{1,2}^{2}&0\\
      0&E_{2,2}^{2}\\
\end{pmatrix}\,,\qquad
V_2&&=
\begin{pmatrix}
      E_{3,2}^{1}&0\\
      0&E_{4,2}^{1}\\
      E_{3,2}^{2}&0\\
      0&E_{4,2}^{2}\\
\end{pmatrix}\,.
\end{aligned}
\end{equation}
We can now obtain the matching matrix in a general scheme by applying the transformation to the bare results before performing the renormalization. Alternatively, the renormalized result in a general scheme can be obtained by multiplying the renormalized matching matrix with a finite renormalization induced by the rotation matrices $U,V$
\begin{equation}
\begin{aligned}
\label{eq:Zfinform}
Z_{\text{fin},1/2}&=1+\api Z^{1}_{\text{fin},1/2}+\api^2 Z^{2}_{\text{fin},1/2}+\mathcal{O}(\api^3)\,,\\
Z^{1}_{\text{fin},1/2}&=\gamma^{(1/2)}_{0,\mathcal{O}\mathcal{E}}U_{1/2}\,,\\
Z^{2}_{\text{fin},1/2}&=\frac{1}{2}\gamma^{(1/2)}_{1,\mathcal{O}\mathcal{E}}U_{1/2}+\frac{1}{2}\left(\gamma^{(1/2)}_{0,\mathcal{O},k}\gamma^{(1/2)}_{0,k\mathcal{E}}+\beta_0\gamma^{(1/2)}_{0,\mathcal{O}\mathcal{E}}\right)V_{1/2}\\
&\eqindent{1}-\gamma^{(1/2)}_{0,\mathcal{O}\mathcal{E}}V_{1/2}\gamma^{(1/2)}_0\,.
\end{aligned}
\end{equation}
We then get the result for the matching matrix in the general scheme $\zeta^{\prime-1}_{1/2}$ by
\begin{equation}
\label{eq:transformationgeneralscheme}
\zeta^{\prime-1}_{1/2}=Z^{-1}_{\text{fin},1/2}\zeta^{-1}_{1/2}\,.
\end{equation}
A third way to determine the general scheme dependence is the direct calculation
of the matching matrix using the evanescent operators displayed
in~\cref{eq:evops1,eq:evopscoeffs1,eq:evops2,eq:evopscoeffs2}. We find agreement
between all three methods which provides a strong check on our result.
The finite renormalization constant in~\cref{eq:Zfinform} can also be found in the ancillary file; it can be used to convert the results of the bag parameters from~\cref{eq:bag_vals} to any scheme captured by~\cref{app:evscheme}.

\section{Ancillary files}
\label{app:anc}
We provide the perturbative results of this paper in computer-readable form as ancillary files in both \textit{Mathematica} and \textit{Python} format in the files \texttt{Zeta.m} and \texttt{Zeta.py}, respectively~\cite{SupplementalMaterial}. The provided quantities and their notation in either format are listed in~\cref{tab:ancillarycontent}, while the variables they depend on can be found in~\cref{tab:ancvars}.

We provide the matching coefficients in two schemes for the flowed quarks, where this scheme can be chosen by setting the variable \texttt{Xzetachi} to 0 or 1 to get the \MSbar\ or the ringed scheme (see \cref{eq:gff:gapy}). The coefficients $c_\chi^{(2)}$ are denoted by \texttt{C2} and the function \texttt{chi2(nf)} in \texttt{Zeta.m} and \texttt{Zeta.py},
respectively. Using the replacement rule \texttt{ReplaceC2}, $c_\chi^{(2)}$ can be substituted by the values in \cref{eq:chi2}.

Additionally, the scheme related to the evanescent operators can be chosen by setting $E^i_{j,k}=$ \texttt{Eijepk} to any value. In particular, using the replacement rules \texttt{ReplaceScheme1}, the default scheme choice of this paper, which is used in both \cref{eq:bag_vals} as well as~\cref{eq:resu:irus,eq:resu:itys} (for the definition of this scheme see~\cref{sec:evops}) can be substituted. We also provide the explicit coefficients of the scheme used in Ref.~\cite{Black:2024bus} and \cref{eq:bag_vals_NLOFierz} which can be replaced by \texttt{ReplaceScheme2}.
Our results in the ancillary files are expressed in dependence of $\ctr=\frac{1}{2}$.
\begin{table*}[th]
  \begin{center}
    \begin{tabular}{llll}
      \hline\hline
      \texttt{Zeta.m} & \texttt{Zeta.py} & meaning & reference\\\hline
      \verb$ZfinInv1$ & \verb$ZfinInv1(alpha_s,nf)$ & $Z_{fin,1}^{-1}$ & \cref{eq:Zfinform}\\
      \verb$ZfinInv2$ & \verb$ZfinInv2(alpha_s,nf)$ & $Z_{fin,2}^{-1}$ & \cref{eq:Zfinform}\\
      \verb$zetaInv1$ & \verb$zetaInv1(alpha_s,nf,mu,t)$ & $\zeta_1^{\prime-1}$ & \cref{eq:zetas,eq:resu:irus,eq:transformationgeneralscheme}\\
      \verb$zetaInv2$ & \verb$zetaInv2(alpha_s,nf,mu,t)$ & $\zeta_2^{\prime-1}$ & \cref{eq:zetas,eq:resu:itys,eq:transformationgeneralscheme}\\
      \verb$zetaInvMixing$ & \verb$zetaInvMixing(alpha_s,nf,mu,t)$ & $\zeta_{{\cal Q}_1}^{-1}$ & \cref{eq:zetas,eq:zetapp}\\
      \verb$zetaInv1NormA$ & \verb$zetaInv1NormA(alpha_s,nf,mu,t)$ & $\zeta^{\prime-1}_{B,1}$ & \cref{eq:zetas,eq:transformationgeneralscheme}\\
      \verb$zetaInv2NormP$ & \verb$zetaInv2NormP(alpha_s,nf,mu,t)$ & $\zeta^{\prime-1}_{B,2}$ & \cref{eq:zetas,eq:transformationgeneralscheme}\\
      \verb$zetaInvMixingNormA$ & \verb$zetaInvMixingNormA(alpha_s,nf,mu,t)$ & $\zeta^{-1}_{{\cal B},1}$ & \cref{eq:zetas}\\
      \verb$gamma1$ & \verb$gamma1(alpha_s,nf)$ & $\gamma^{\prime(1)}$ & \cref{eq:gamma,eq:gamma01,eq:gamma11,eq:gammanonphys,eq:transformationgeneralscheme} \\
      \verb$gamma2$ & \verb$gamma2(alpha_s,nf)$ & $\gamma^{\prime(2)}$ & \cref{eq:gamma,eq:gamma02,eq:gamma12,eq:gammanonphys,eq:transformationgeneralscheme} \\
      \verb$gammaMixing$ & \verb$gammaMixing(alpha_s,nf)$ & $\gamma_{{\cal Q}_1}$ & \cref{eq:gamma,eq:gammamix}\\
      \hline\hline
    \end{tabular}
  \end{center}
  \caption{\label{tab:ancillarycontent} The expressions of the ancillary
    files \texttt{Zeta.m} and \texttt{Zeta.py}
    that encode the main results of this paper. The notation of the
    variables should be self-explanatory; more details are given in the
    header of these files.}
\end{table*}

\begin{table*}[th]
  \begin{center}
    \begin{tabular}{llll}
    \hline\hline
      \texttt{Zeta.m} & \texttt{Zeta.py} & meaning & reference\\\hline
      \verb$nc$ &\verb$NC$ &  $\nc$ &  \cref{sec:GFR} \\
      \verb$tr$ &\verb$TR$ &  $\ctr$ &   \cref{app:anc}\\
      \verb$Lmut$&\verb$Lmut$ &  $\lmut$ &  \cref{eq:lmut}\\
      \verb$as$ &\verb$a_s$ &$\api$ &  \cref{sec:pt_results}\\
      \verb$nf$ &\verb$nf$ &  $\nf$ &  \cref{sec:pt_results}\\
      \verb$z2$ &\verb$zeta2$ & $\zeta(2)$ &  \cref{eq:transcendentals}\\
      \hline\hline
    \end{tabular}
  \end{center}
  \caption{\label{tab:ancvars}Notation for the variables in the
      ancillary files.}
\end{table*}

\section{Correlator Fits}
We now show the results of fits to~\cref{eq:ratio} for all five bag parameters studied on each ensemble listed in~\cref{tab:confs} at select flow times, with $p$-values.
The fits combine multiple source separations $\Delta T$ as described in~\cref{sec:data} and are fully correlated in both independent analyses.

Fits on the C1 and C2 ensembles are collected in~\cref{tab:Cfits}, then on the M1, M2, and M3 ensembles in~\cref{tab:Mfits}, and finally for the F1S ensemble in~\cref{tab:Ffits}.

Moreover, the fit results of the chiral-continuum limits for each bag parameter at a set of physical flow times are shown in~\cref{tab:a0fits}.
\newpage
\onecolumngrid
\begin{table*}[th]
    \centering
    \begin{tabular}{c@{~~}||c@{~~}|@{~~}c@{~~}|@{~~}c@{~~}c@{~~}|@{~~}c@{~~}c@{~~}|@{~~}c@{~~}c@{~~}|@{~~}c@{~~}c@{~~}|@{~~}c@{~~}c@{~~}}
        \hline\hline
        Ens & $\tau\,[{\rm GeV}^{-2}]$ & $\tau/a^2$ & ${\cal B}_1^{\rm GF}$ & $p$ & $B_1^{\rm GF}$ & $p$ & $B_2^{\rm GF}$ & $p$ & $\epsilon_1^{\rm GF}$ & $p$ & $\epsilon_2^{\rm GF}$ & $p$ \\
        \hline\hline
        \multirow{8}{*}{C1} & 0.094 & 0.30 & 0.8894(6) & 0.67 & 1.0149(10) & 0.85 & 0.9846(3) & 0.94 & -0.0919(2) & 0.51 & -0.00527(4) & 0.95 \\
                            & 0.157 & 0.50 & 0.9099(6) & 0.80 & 1.0097(9) & 0.86 & 0.9875(3) & 0.93 & -0.0749(2) & 0.54 & -0.00538(4) & 0.98 \\
                            & 0.220 & 0.70 & 0.9247(5) & 0.70 & 1.0067(9) & 0.87 & 0.9898(3) & 0.87 & -0.0623(2) & 0.48 & -0.00524(5) & 0.98 \\
                            & 0.251 & 0.80 & 0.9305(5) & 0.71 & 1.0057(9) & 0.87 & 0.9908(3) & 0.85 & -0.0573(2) & 0.51 & -0.00512(5) & 0.98 \\
                            & 0.314 & 1.00 & 0.9400(5) & 0.86 & 1.0043(9) & 0.89 & 0.9924(3) & 0.92 & -0.0489(2) & 0.50 & -0.00481(5) & 0.98 \\
                            & 0.345 & 1.10 & 0.9440(5) & 0.75 & 1.0038(9) & 0.90 & 0.9931(3) & 0.89 & -0.0454(2) & 0.55 & -0.00464(6) & 0.95 \\
                            & 0.408 & 1.30 & 0.9509(5) & 0.63 & 1.0031(8) & 0.91 & 0.9944(3) & 0.94 & -0.0395(2) & 0.66 & -0.00429(6) & 0.96 \\
        \hline
        \multirow{8}{*}{C2} & 0.094 & 0.30 & 0.8889(6) & 0.72 & 1.0117(5) & 0.92 & 0.9817(4) & 0.95 & -0.0921(4) & 0.49 & -0.00556(15) & 0.46 \\
                            & 0.157 & 0.50 & 0.9097(6) & 0.64 & 1.0065(5) & 0.94 & 0.9847(4) & 0.95 & -0.0752(4) & 0.50 & -0.0057(2) & 0.43 \\
                            & 0.220 & 0.70 & 0.9243(6) & 0.62 & 1.0035(5) & 0.92 & 0.9871(4) & 0.95 & -0.0627(4) & 0.64 & -0.0056(2) & 0.32 \\
                            & 0.251 & 0.80 & 0.9303(6) & 0.54 & 1.0025(5) & 0.91 & 0.9882(4) & 0.96 & -0.0577(4) & 0.63 & -0.0055(2) & 0.32 \\
                            & 0.314 & 1.00 & 0.9396(5) & 0.63 & 1.0011(4) & 0.90 & 0.9900(4) & 0.99 & -0.0493(4) & 0.71 & -0.0052(2) & 0.42 \\
                            & 0.345 & 1.10 & 0.9436(5) & 0.62 & 1.0007(4) & 0.89 & 0.9907(4) & 1.00 & -0.0458(3) & 0.75 & -0.0050(2) & 0.30 \\
                            & 0.408 & 1.30 & 0.9505(5) & 0.42 & 1.0000(4) & 0.87 & 0.9920(4) & 0.97 & -0.0398(3) & 0.86 & -0.0047(2) & 0.42 \\
        \hline\hline
    \end{tabular}
    \caption{\label{tab:Cfits} Bag parameters extracted from ratio fits for the coarse ensembles at select flow times with $p$-values.}
\end{table*}

\begin{table*}[th]
    \centering
    \begin{tabular}{c@{~~}||c@{~~}|@{~~}c@{~~}|@{~~}c@{~~}c@{~~}|@{~~}c@{~~}c@{~~}|@{~~}c@{~~}c@{~~}|@{~~}c@{~~}c@{~~}|@{~~}c@{~~}c@{~~}}
        \hline\hline
        Ens & $\tau\,[{\rm GeV}^{-2}]$ & $\tau/a^2$ & ${\cal B}_1^{\rm GF}$ & $p$ & $B_1^{\rm GF}$ & $p$ & $B_2^{\rm GF}$ & $p$ & $\epsilon_1^{\rm GF}$ & $p$ & $\epsilon_2^{\rm GF}$ & $p$ \\
        \hline\hline
        \multirow{8}{*}{M1} & 0.106 & 0.60 & 0.8901(6) & 0.90 & 1.0139(6) & 0.50 & 0.9830(6) & 0.78 & -0.0951(4) & 0.93 & -0.00610(5) & 0.18 \\
                            & 0.158 & 0.90 & 0.9085(6) & 0.88 & 1.0088(6) & 0.57 & 0.9861(6) & 0.82 & -0.0783(4) & 0.96 & -0.00595(5) & 0.10 \\
                            & 0.211 & 1.20 & 0.9218(7) & 0.86 & 1.0057(6) & 0.49 & 0.9884(7) & 0.73 & -0.0663(2) & 0.24 & -0.00565(6) & 0.39 \\
                            & 0.246 & 1.40 & 0.9288(7) & 0.82 & 1.0043(6) & 0.60 & 0.9898(6) & 0.86 & -0.0599(2) & 0.19 & -0.00545(5) & 0.43 \\
                            & 0.299 & 1.70 & 0.9373(6) & 0.89 & 1.0028(6) & 0.65 & 0.9914(6) & 0.88 & -0.0519(2) & 0.12 & -0.00513(6) & 0.51 \\
                            & 0.352 & 2.00 & 0.9442(6) & 0.78 & 1.0018(6) & 0.67 & 0.9926(5) & 0.97 & -0.0455(2) & 0.15 & -0.00479(6) & 0.68 \\
                            & 0.405 & 2.30 & 0.9500(6) & 0.68 & 1.0012(6) & 0.70 & 0.9937(5) & 0.98 & -0.0403(2) & 0.18 & -0.00447(6) & 0.74 \\
        \hline
        \multirow{8}{*}{M2} & 0.106 & 0.60 & 0.8906(7) & 0.28 & 1.0152(7) & 0.20 & 0.9825(4) & 0.54 & -0.0957(3) & 0.90 & -0.00610(5) & 0.76 \\
                            & 0.158 & 0.90 & 0.9089(7) & 0.36 & 1.0099(6) & 0.20 & 0.9856(4) & 0.52 & -0.0787(3) & 0.86 & -0.00595(6) & 0.78 \\
                            & 0.211 & 1.20 & 0.9223(6) & 0.43 & 1.0067(6) & 0.14 & 0.9880(4) & 0.49 & -0.0665(3) & 0.85 & -0.00566(6) & 0.79 \\
                            & 0.246 & 1.40 & 0.9293(6) & 0.43 & 1.0053(7) & 0.12 & 0.9893(4) & 0.47 & -0.0599(3) & 0.85 & -0.00545(7) & 0.80 \\
                            & 0.299 & 1.70 & 0.9378(6) & 0.48 & 1.0038(7) & 0.13 & 0.9909(4) & 0.39 & -0.0518(2) & 0.38 & -0.00511(7) & 0.84 \\
                            & 0.352 & 2.00 & 0.9448(6) & 0.39 & 1.0027(6) & 0.15 & 0.9923(4) & 0.38 & -0.0455(2) & 0.38 & -0.00479(7) & 0.75 \\
                            & 0.405 & 2.30 & 0.9503(7) & 0.47 & 1.0020(5) & 0.06 & 0.9933(4) & 0.44 & -0.0402(3) & 0.38 & -0.00444(8) & 0.82 \\
        \hline
        \multirow{8}{*}{M3} & 0.106 & 0.60 & 0.8910(7) & 0.20 & 1.0133(7) & 0.50 & 0.9824(3) & 0.97 & -0.0962(3) & 0.27 & -0.00634(8) & 0.51 \\
                            & 0.158 & 0.90 & 0.9094(8) & 0.24 & 1.0079(6) & 0.50 & 0.9855(3) & 0.99 & -0.0793(3) & 0.23 & -0.00618(8) & 0.26 \\
                            & 0.211 & 1.20 & 0.9227(7) & 0.27 & 1.0044(7) & 0.64 & 0.9880(3) & 0.99 & -0.0667(3) & 0.52 & -0.00584(9) & 0.15 \\
                            & 0.246 & 1.40 & 0.9296(8) & 0.26 & 1.0030(7) & 0.65 & 0.9893(3) & 0.99 & -0.0603(2) & 0.30 & -0.00562(9) & 0.15 \\
                            & 0.299 & 1.70 & 0.9383(8) & 0.30 & 1.0015(6) & 0.66 & 0.9910(3) & 0.96 & -0.0523(2) & 0.31 & -0.00537(11) & 0.51 \\
                            & 0.352 & 2.00 & 0.9451(8) & 0.34 & 1.0005(6) & 0.68 & 0.9924(3) & 0.92 & -0.0458(2) & 0.41 & -0.00498(11) & 0.37 \\
                            & 0.405 & 2.30 & 0.9508(7) & 0.39 & 0.9999(6) & 0.66 & 0.9935(3) & 0.88 & -0.0405(2) & 0.44 & -0.00470(12) & 0.68 \\
        \hline\hline
    \end{tabular}
    \caption{\label{tab:Mfits} Bag parameters extracted from ratio fits for the medium ensembles at select flow times with $p$-values.}
\end{table*}

\begin{table*}[th]
    \centering
    \begin{tabular}{c@{~~}||c@{~~}|@{~~}c@{~~}|@{~~}c@{~~}c@{~~}|@{~~}c@{~~}c@{~~}|@{~~}c@{~~}c@{~~}|@{~~}c@{~~}c@{~~}|@{~~}c@{~~}c@{~~}}
        \hline\hline
        Ens & $\tau\,[{\rm GeV}^{-2}]$ & $\tau/a^2$ & ${\cal B}_1^{\rm GF}$ & $p$ & $B_1^{\rm GF}$ & $p$ & $B_2^{\rm GF}$ & $p$ & $\epsilon_1^{\rm GF}$ & $p$ & $\epsilon_2^{\rm GF}$ & $p$ \\
        \hline\hline
        \multirow{8}{*}{F1S} & 0.103 & 0.80 & 0.8888(5) & 0.11 & 1.0156(3) & 0.12 & 0.9828(2) & 0.56 & -0.0981(3) & 0.84 & -0.00629(3) & 0.75 \\
                             & 0.155 & 1.20 & 0.9080(4) & 0.05 & 1.0099(3) & 0.11 & 0.9859(2) & 0.51 & -0.0804(3) & 0.68 & -0.00608(3) & 0.76 \\
                             & 0.206 & 1.60 & 0.9211(4) & 0.07 & 1.0066(3) & 0.10 & 0.9883(2) & 0.45 & -0.0677(2) & 0.83 & -0.00578(4) & 0.67 \\
                             & 0.258 & 2.00 & 0.9311(4) & 0.60 & 1.0044(3) & 0.11 & 0.9901(2) & 0.46 & -0.0581(2) & 0.73 & -0.00546(3) & 0.86 \\
                             & 0.309 & 2.40 & 0.9391(3) & 0.61 & 1.0030(2) & 0.12 & 0.9915(2) & 0.47 & -0.0506(2) & 0.73 & -0.00512(4) & 0.63 \\
                             & 0.348 & 2.70 & 0.9440(4) & 0.63 & 1.0023(2) & 0.12 & 0.9925(2) & 0.44 & -0.0459(2) & 0.72 & -0.00486(5) & 0.73 \\
                             & 0.400 & 3.10 & 0.9496(4) & 0.64 & 1.0016(2) & 0.21 & 0.9935(2) & 0.44 & -0.0407(2) & 0.71 & -0.00451(4) & 0.74 \\
        \hline\hline
    \end{tabular}
    \caption{\label{tab:Ffits} Bag parameters extracted from ratio fits for the F1S ensemble at select flow times with $p$-values.}
\end{table*}

\begin{table*}[th]
    \centering
    \begin{tabular}{c@{~~}||c@{~~}|@{~~}c@{~~}c@{~~}|@{~~}c@{~~}c@{~~}|@{~~}c@{~~}c@{~~}|@{~~}c@{~~}c@{~~}|@{~~}c@{~~}c@{~~}}
        \hline\hline
        Ens & $\tau\,[{\rm GeV}^{-2}]$ & ${\cal B}_1^{\rm GF}$ & $p$ & $B_1^{\rm GF}$ & $p$ & $B_2^{\rm GF}$ & $p$ & $\epsilon_1^{\rm GF}$ & $p$ & $\epsilon_2^{\rm GF}$ & $p$ \\
        \hline\hline
        \multirow{8}{*}{$a=0$} & 0.100 & 0.8845(8) & 0.54 & 1.0172(10) & 0.12 & 0.9800(4) & 0.36 & -0.1072(4) & 0.05 & -0.00716(6) & 0.41 \\
                               & 0.150 & 0.9051(8) & 0.64 & 1.0101(9) & 0.15 & 0.9839(4) & 0.27 & -0.0866(4) & 0.17 & -0.00678(7) & 0.45 \\
                               & 0.200 & 0.9192(8) & 0.77 & 1.0059(9) & 0.16 & 0.9867(4) & 0.25 & -0.0720(3) & 0.24 & -0.00630(7) & 0.79 \\
                               & 0.250 & 0.9295(8) & 0.81 & 1.0033(9) & 0.19 & 0.9887(4) & 0.16 & -0.0617(3) & 0.26 & -0.00589(7) & 0.73 \\
                               & 0.300 & 0.9377(9) & 0.73 & 1.0016(9) & 0.19 & 0.9904(4) & 0.10 & -0.0533(3) & 0.42 & -0.00553(8) & 0.53 \\
                               & 0.350 & 0.9442(8) & 0.68 & 1.0006(9) & 0.17 & 0.9917(4) & 0.07 & -0.0467(3) & 0.64 & -0.00512(9) & 0.71 \\
                               & 0.400 & 0.9495(8) & 0.72 & 0.9998(9) & 0.17 & 0.9927(4) & 0.05 & -0.0414(3) & 0.70 & -0.00477(9) & 0.57 \\
        \hline\hline
    \end{tabular}
    \caption{\label{tab:a0fits} Chiral-continuum limit extrapolations of the bag parameters at select flow times with $p$-values.}
\end{table*}

\twocolumngrid

\bibliography{lit.bib}

\end{document}